\RequirePackage{fix-cm}
\documentclass[smallextended]{svjour3}       
\smartqed  
\usepackage{alltt                                    
	, multirow
	,subfig
	, booktabs
	, listings
	, graphicx
	,float
	,cite
	,verbatim
	,mathtools
	,url
}

\usepackage[table]{xcolor}
\usepackage{xcolor, soul}
\usepackage[round]{natbib}     
\usepackage{syntax}
\usepackage{algorithmic, algorithm}
\usepackage{threeparttable}
\usepackage{framed}
\usepackage{expl3}
\usepackage{hyperref}
\usepackage{filecontents}

\usepackage{pgfplots}
\pgfplotsset{compat=1.16}
\usepackage{pgfplotstable}

\usepackage{scalefnt}
\usepackage{tikz}

\usepackage{tabularx}
\usepackage{lipsum}
\usepackage{array}

\usepackage{tcolorbox}
\usepackage{multirow}

\usetikzlibrary{shapes.gates.logic.US,trees,positioning,arrows}
\usetikzlibrary{shapes.multipart}
\usetikzlibrary{shapes,arrows}

\usepackage{tikz}

\tikzstyle{vertex}=[ellipse,fill=black!25,minimum size=20pt, inner sep=0pt]
\tikzstyle{edge} = [draw,thin,-]
\tikzstyle{glabel} = [text width=1cm,text centered,font=\bf]
\pgfdeclarelayer{bg}    
\pgfsetlayers{bg,main} 

\ExplSyntaxOn
\newcommand\latinabbrev[1]{
  \peek_meaning:NTF . {
    #1\@}%
  { \peek_catcode:NTF a {
      #1., \@ }%
    {#1., \@}}}
\ExplSyntaxOff

\usepackage{tikz}
\usepackage{verbatim}
\usetikzlibrary{arrows,shapes,backgrounds}

\tikzstyle{vertex}=[ellipse,fill=black!25,minimum size=20pt, inner sep=0pt]
\tikzstyle{edge} = [draw,thin,-]
\tikzstyle{glabel} = [text width=1cm,text centered,font=\bf]
\pgfdeclarelayer{bg}    
\pgfsetlayers{bg,main}  

\usepackage{xcolor}
\usepackage{soul}
\usepackage{pifont}
\usepackage{booktabs,makecell}
\usepackage[export]{adjustbox}
\usepackage{balance}
\usepackage{paralist}

\newcommand{\CASE}[1]{\STATE \textbf{case} #1\textbf{:} \begin{ALC@g}}
\newcommand{\ENDCASE}{\end{ALC@g}}

\newcommand{\DEFAULT}{\STATE \textbf{default:} \begin{ALC@g}}
\newcommand{\ENDDEFAULT}{\end{ALC@g}}
\newcommand{\DEFAULTLINE}[1]{\STATE \textbf{default:} }

\newsavebox{\supbox}
\newcommand{\bsup}{\begin{lrbox}{\supbox}$\tt\scriptstyle}
\newcommand{\esup}{$\end{lrbox}{}^{\usebox{\supbox}}}
\def\eg{\latinabbrev{e.g}}
\def\ie{\latinabbrev{i.e}}

\definecolor{lightpurple}{rgb}{0.8,0.8,1}
\definecolor{codebg}{RGB}{255,255,255}
\definecolor{commentcolor}{RGB}{11,140,11}
\usepackage[normalem]{ulem} 
\usepackage{xcolor}

\usepackage{ifthen}
\usepackage{amssymb}
\newboolean{showcomments}

\setboolean{showcomments}{false}

\ifthenelse{\boolean{showcomments}}
{\newcommand{\nbc}[3]{
 {\colorbox{#3}{\bfseries\sffamily\scriptsize\textcolor{white}{#1}}}
 {\textcolor{#3}{\sf\small$\blacktriangleright$\textit{#2}$\blacktriangleleft$}}
 }
}
{\newcommand{\nbc}[3]{}
 \renewcommand{\hl}[1]{#1} 
 }
 
 \AtBeginDocument{%
	\paperwidth=\dimexpr
	1in + \oddsidemargin
	+ \textwidth
	+ 1in + \oddsidemargin
	\relax
	\paperheight=\dimexpr
	1in + \topmargin
	+ \headheight + \headsep
	+ \textheight
	+ 1in + \topmargin
	\relax
	\usepackage[pass]{geometry}\relax
}

\usepackage{amsmath}
\usepackage{enumitem}

\begin{document}

\title{The Reproducibility of Programming-Related Issues in Stack Overflow Questions 
}


\author{Saikat Mondal         \and
        Mohammad Masudur Rahman \and
        Chanchal K. Roy \and
        Kevin Schneider 
}


\institute{Saikat Mondal \at
              Software Research Lab, Department of Computer Science\\ University of Saskatchewan, Canada \\
              \email{saikat.mondal@usask.ca}           
           \and
           Mohammad Masudur Rahman \at
           Faculty of Computer Science, Dalhousie University, Canada \\
           \email{masud.rahman@dal.ca}
           \and
           Chanchal K. Roy \at
           Software Research Lab, Department of Computer Science\\ University of Saskatchewan, Canada \\
          \email{chanchal.roy@usask.ca}
          \and
           Kevin Schneider \at
           Software Research Lab, Department of Computer Science\\ University of Saskatchewan, Canada \\
          \email{kevin.schneider@usask.ca}
}

\date{Received: date / Accepted: date}

\maketitle

\begin{abstract}
Software developers often look for solutions to their code-level problems using the Stack Overflow Q\&A website. To receive help, developers frequently submit questions that contain sample code segments along with the description of the programming issue. Unfortunately, it is not always possible to reproduce the issues from the code segments they provide. Issues that are not easily reproducible may impede questions from receiving prompt and appropriate solutions. We conducted an exploratory study on the reproducibility of issues discussed 400 Java and 400 Python questions. We parsed, compiled, executed, and carefully examined the code segments from these questions to reproduce the reported programming issues, expending 300 person-hours of effort. 
The outcomes of our study are three-fold.
First, we can reproduce the issues for approximately 68\% of Java and 71\% of Python code segments. In contrast, we were unable to reproduce approximately 22\% of Java and 19\% of Python issues. Of the reproducible issues, approximately 67\% of the Java and 20\% of the Python code segments required minor or major modifications to reproduce the issues.
Second, we carefully investigated why programming issues could not be reproduced and provided evidence-based guidelines to write effective code examples for Stack Overflow questions.
Third, we investigated the correlation between the issue reproducibility status of questions and the corresponding answer meta-data, such as the presence of an accepted answer. According to our analysis, a reproducible question has at least two times higher chance of receiving an accepted answer than an irreproducible question. Besides, the median time delay in receiving accepted answers is double if the issues reported in questions could not be reproduced. We also investigated the confounding factors (e.g., user reputation) that can affect questions receiving answers besides reproducibility. We found that such factors do not hurt the correlation between reproducibility status and answer meta-data.


\end{abstract}

\section{Introduction}
\label{introductio}

Stack Overflow has become the most popular programming Q\&A website for software developers \citep{understanding2014}. The number of users and the questions posted on Stack Overflow is increasing exponentially \citep{datadumpapi}.
A large number of these questions contain code segments that have programming issues (e.g., errors, unexpected behaviors) \citep{treude2011programmers}.
Once a question is submitted, users of Stack Overflow often attempt to \emph{reproduce} the programming issues discussed in these questions and submit solutions. 
In this paper, we follow the definition of reproducibility of issues by \citet{crashdroid}. According to them, reproducibility is the complete agreement between the reported issues and the investigated issues. Unfortunately, it is not always possible to reproduce the programming issues 
reported by other users \citep{crashscope, guided-ga-crash}, and this might prevent the questions from getting prompt, appropriate answers. This scenario might also explain the 47.04\% unresolved and 13.36\%  unanswered questions on Stack Overflow \citep{rahman2015insight,asaduzzaman2013answering}. Given these major challenges in question answering, a detailed investigation is warranted on why programming issues reported in Stack Overflow questions cannot be reproduced from their code segments.

Several existing studies \citep{querytousablecode, gistable} investigated the usability and executability of the code segments posted at Q\&A sites. \citet{querytousablecode} extract 914,974 code segments from the accepted answers of Stack Overflow and analyzed their parsability and compilability. However, their analysis was completely automatic, and they did not address the challenges of issue reproducibility. \citet{gistable} analyzed the executability of the Python code segments found on the GitHub Gist system. They identify several issues (e.g., import error, syntax errors, indentation error) that can prevent the execution of their code segments. However, 
a successful execution of the code might not necessarily guarantee the reproduction of the issues discussed in Stack Overflow questions. Thus, their approach also might fail to address the reproducibility challenges that we are dealing with. In short, all the challenges of reproducibility cannot be resolved using only automated analysis. 
Thus, there is a marked lack of research that (1) carefully investigates the challenges of issue reproducibility (of Stack Overflow questions) and (2) develops automated approaches to help overcome such challenges.

In this paper, we report on an exploratory study on the \emph{reproducibility} of programming issues discussed in 400 Java and 400 Python questions from Stack Overflow. For each of these questions, we follow three steps. First, we attempt to understand the issue of a question by analyzing its code segment and the associated textual description. Second, we copy/paste the code segment into our integrated development environment (\texttt{IDE}) and attempt to reproduce the reported issue. Third, we record our findings on the reproducibility of the issues from each of the questions. If we are unable to reproduce an issue, we investigate why the issue could not be reproduced. We spent a total of 300 person-hours in our analysis for this study. Our findings from this study are three-fold. First, we find that 68\% of issues reported in Java questions can be reproduced, although only 32\% of these can be reproduced using the verbatim code from Stack Overflow without modification. Unfortunately, a total of about 22\% of code segments completely fail to reproduce the issues reported by developers.
We also analyze Python code segments and found a similar pattern. For Python, 71\% of code segments can reproduce the issues reported in the Stack Overflow questions. Two-thirds of them require no modification to reproduce the issues. The remaining code segments require either minor or major modifications to reproduce the issues.
Second, we investigate the challenges that prevent reproducibility of issues using the code segments included in the questions. We also provide evidence-based guidelines to write effective code examples for Stack Overflow questions.  
Third, we investigate the relationship between reproducibility status of the issues from questions and answer meta-data (\eg\ accepted answer). We found that a question with reproducible issues has at least a two times higher chance of getting an acceptable answer than a question with an irreproducible issue. Our in-depth investigation and findings not only establish \emph{issue reproducibility} as a question quality paradigm but also encourage automated tool supports for improving the code segments submitted on Stack Overflow. 
We also investigate the confounding factors besides reproducibility of issues that can affect questions receiving answers. In particular, we analyze three factors -- \emph{user reputation}, \emph{number of posts} (i.e., questions and answers) and \emph{question submission time} (e.g., day/night). According to our analysis, questions have a higher chance of receiving accepted answers when users with a higher reputation submitted them. On the contrary, submitting a higher number of posts can reduce the chance of receiving accepted answers. However, question submission time has almost no impact on questions receiving answers.
We answer four research questions and make four main contributions in this paper as follows.

\vspace{1mm}
    $\bullet$ \textbf{RQ\textsubscript1:} \textbf{What are the challenges in reproducing the issues reported in Stack Overflow questions? How can issue reproducibility be determined?}
    We conduct extensive manual analysis and investigate which types of actions are needed to reproduce the issues reported in Stack Overflow questions. We find that several common editing actions (e.g., library inclusion, method invocation, code migration, instrumentation) 
    should be performed on both Java and Python code segments to reproduce the issues reported in questions. We also identify a few more editing actions required to reproduce issues reported in questions that are specific to either Java (e.g., addition of classes and methods) or Python code segments (e.g., fix indentation). 
    However, several challenges prevent the reproduction of issues
    using the code segments such as class/interface/method not found, an important part of code missing, external library not found, database/file dependency.
    
    We 
    classify the reproducibility status into two major categories: \emph{reproducible} and \emph{irreproducible}. We then divide the reproducible status into three more sub-categories based on the level of human effort and time required to reproduce the issue. They are: \emph{none -- reproducible without modification}; \emph{minor -- reproducible with minor modification}; and, \emph{major -- reproducible with major modification}.

    \vspace{1.5mm}
    $\bullet$ \textbf{RQ\textsubscript2:} \textbf{What proportion of reported issues in Stack Overflow questions can be reproduced successfully?}
    We conduct a detailed statistical analysis to determine what percentage of the issues can be reproduced and what percentage cannot be reproduced even after performing major modifications to their corresponding code segments.
    According to our analysis, about 68\% of the Java and 71\% of Python code segments can reproduce the issues reported in Stack Overflow questions. Among them, only 32\% of Java code segments do not require any modifications to reproduce the issues. Such percentage is 78\% for Python code segments. However, the remaining code segments require minor or major modifications to reproduce the reported issues. Unfortunately, a total of about 22\% of Java and 19\% of Python code segments completely fail to reproduce their issues even after major modifications.
    
    \vspace{1.5mm}
    $\bullet$ \textbf{RQ\textsubscript3:} \textbf{Does reproducibility of issues reported in Stack Overflow questions help them receive high-quality responses including the acceptable answers?}
    We conduct a detailed investigation and determine the correlation between issue reproducibility (of questions) and corresponding answer meta-data such as presence of an accepted answer, time delay between posting of a question and accepted answer, and the number of answers.
    According to our investigation, a question whose code segment can reproduce the reported issue has at least \emph{two} times higher chance of receiving an acceptable answer than a question with irreproducible issues. Furthermore, the median time delay of receiving an accepted answer is about \emph{double} for the questions with irreproducible issues than those with reproducible issues. For example, half of the Java questions with reproducible issues receive accepted answers in less than five minutes. On the contrary, it takes about ten minutes for questions with irreproducible issues.
    
    \vspace{1.5mm}
    $\bullet$ \textbf{RQ\textsubscript4:} \textbf{What factors can affect questions receiving answers besides reproducibility of issues?}
    We attempt to investigate the confounding factors (e.g., user reputation) that could affect questions receiving accepted answers, time delay in receiving accepted answers and number of answers. We also see whether such factors hurt the correlation between reproducibility status and answer meta-data.According to our investigation, users with a higher reputation score receive more acceptable answers to their questions. On the contrary, the chance of receiving acceptable answers decreases with submitting a higher number of posts. However, question submission time (e.g., day, night) does not affect questions receiving answers. Regardless of the confounding factors, questions with reproducible issues have a higher chance of receiving accepted answers with minimum time delay and encourage more answers.
    

\vspace{3mm}    


\noindent \textbf{Contribution.} 
This paper is a significantly extended version of our previous study \citep{Mondal-SOIssueReproducability-MSR2019} that investigated 400 questions of Stack Overflow related to Java programming problems. Previously, we answered three research questions and reported four key findings that could help one write effective code examples for Stack Overflow questions.
This study extends our previous study in several aspects. 
First, we 
extend our investigation by adding
400 questions discussing
Python programming problems to our dataset. 
Second, we extend our earlier analysis and answer four research questions ( i.e., as opposed to three questions of previous study). 
In particular, we attempt to reproduce our previous findings using these Python-related questions.
Third, we extend the literature part with the studies that investigate the reproducibility on diverse domains (e.g., computational reproducibility, issues in reproducing research results). 
Fourth, we report {\color{black}{eight}} key findings  ( i.e., as opposed to four in the previous study) that could help one write effective code examples when submitting questions to Stack Overflow.

\vspace{2mm}
\noindent \textbf{Replication Package.} Replication Package can be found in our online appendix \citep{replicationPackage}.

\vspace{2mm}
\noindent \textbf{Structure of the Article.} 
%
In Section \ref{motiExample}, we discuss two examples that motivate our idea of issue reproducibility. Section \ref{methodology} discusses our study methodology. In Section \ref{studyFindings}, we classify the reproducibility status of issues in Stack Overflow questions, explore the challenges behind issue reproducibility, investigate their correlation with answer meta-data, and then investigate what factors are associated with reproducibility that can affect questions receiving answers. Several key findings are discussed in Section \ref{discussion}. Section \ref{threat} focuses on the threats to validity, Section \ref{relatedwork} discusses the related work and finally, Section \ref{conclusion} concludes the paper with future work.

\begin{figure}
\centering
	\includegraphics[width=4in]{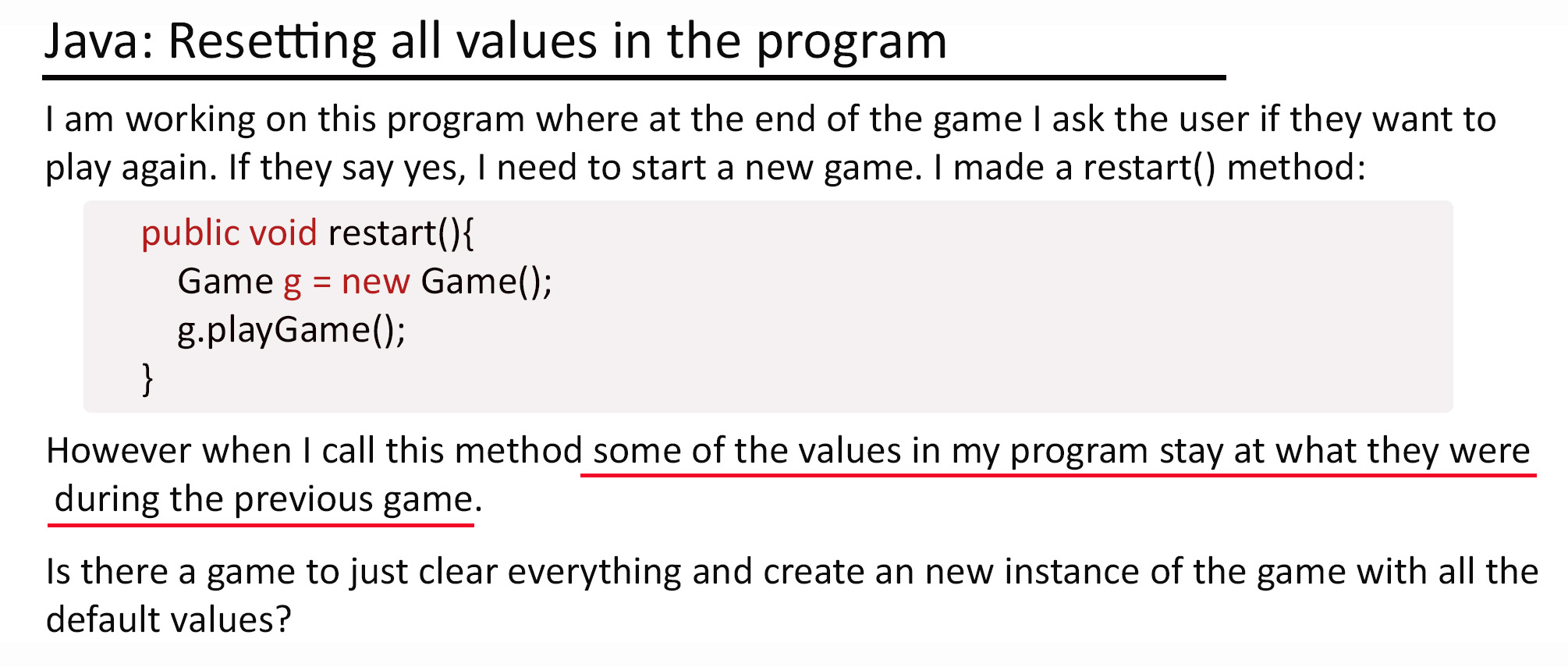}
	\caption{An example of a question with an irreproducible issue (\url{https://stackoverflow.com/questions/798184}).}
	\label{example-issue}
\end{figure}

\begin{figure}[!htb]
\centering
	\includegraphics[width=4in]{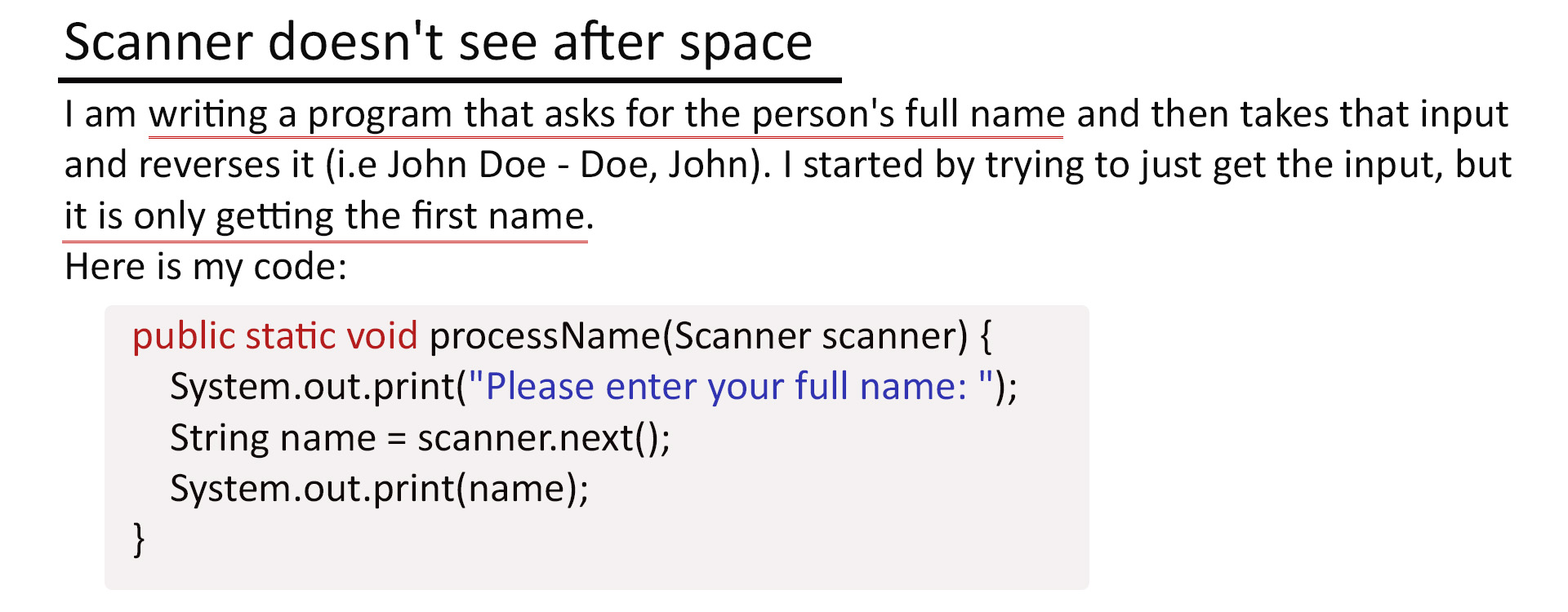}
	\caption{An example of a question with a reproducible issue (\url{https://stackoverflow.com/questions/19509647}).}
	\label{example-issue-reproducible}
\end{figure}

\section{Motivating Examples} 
\label{motiExample}

Code segments submitted as a part of Stack Overflow questions might not always be sufficient enough to reproduce the reported issues. Let us consider the example question in Fig.~\ref{example-issue}. Here, the user attempts to reset all the variables while starting a new game. They discover that the code is not working as expected and some of the variables are retaining their old values. Unfortunately, this issue cannot be reproduced since essential parts of the code are missing. For example, one user commented -- \emph{``Can you post more code? The Game class? The class that contains the restart() method?"} -- while attempting to answer the question. In Stack Overflow, this question has failed to receive a precise response. Even though the code (i.e., Fig.~\ref{example-issue}) could be made parsable, compilable, and executable with all the necessary editing, the reported issue could not be reproduced easily due to its complex nature. 
Thus, the automated analyses done by the earlier studies \citep{gistable,querytousablecode} is likely not sufficient to overcome all the challenges of reproducing this issue.


Let us consider another example question as shown in Fig.~\ref{example-issue-reproducible}. 
Suppose Alice, a software developer, wants to answer this question. First, Alice attempts to identify the issue and soon understands that the user is attempting to capture a person's full name (\eg\ John Doe). Unfortunately, the user is getting only the first part of the given name (\eg\ John). They invoked the \texttt{next()} method of the \texttt{Scanner} class to obtain the input. In order to provide an answer, Alice first copies the code segment to their IDE and then finds that the code does not even parse. As a result, the IDE returns several parsing and compilation errors. Fortunately, by performing several edits (\eg\ addition of a demo class and main method) Alice is able to reproduce the stated issue. Similarly, this issue was also reproduced by other users of Stack Overflow, and the question received a high-quality answer within a couple of minutes. As the solution suggests, the user should have used the \texttt{nextLine()} method instead of the \texttt{next()} method to avoid the reported issue. 

\begin{figure}
	\centering
	\includegraphics[width=4.68in]{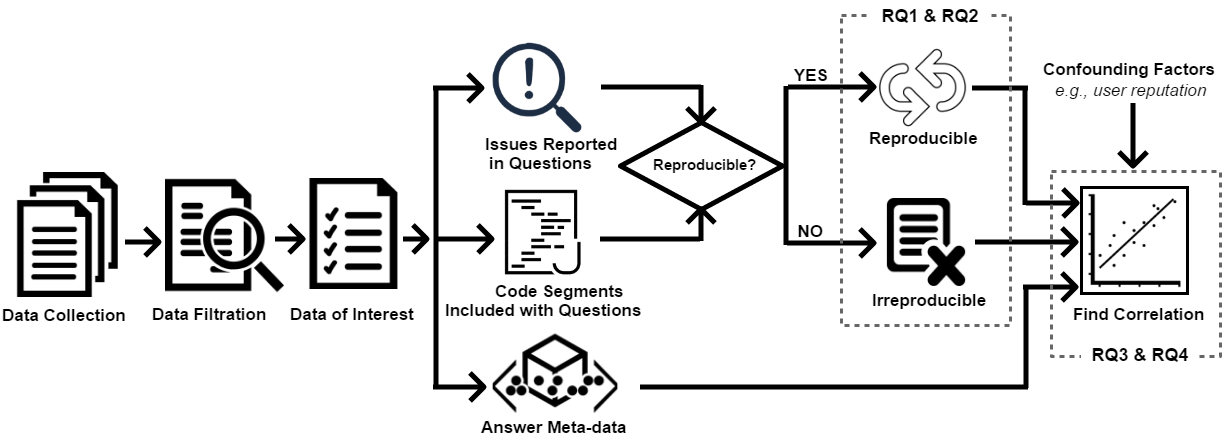}
	\caption{Schematic diagram of our exploratory study.}
	\label{fig:schematic-diagram}
\end{figure}

\section{Study Methodology}
\label{methodology}

Fig.~\ref{fig:schematic-diagram} shows the schematic diagram of our exploratory study. We first randomly select 800 questions related to Java and Python (400 from each) from Stack Overflow and then attempt to reproduce the discussed programming issues. In particular, we attempt to reproduce the reported issues using the submitted code segments by performing a series of edits as necessary. The following sections discuss the different steps in our methodology.

\begin{figure}
	\centering
	\includegraphics[width=4.6in]{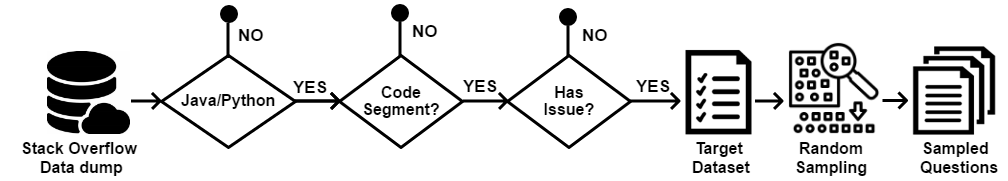}
	\caption{Selection of dataset for our study.}
	\label{fig:data-collection-process}
\end{figure}

\subsection{Dataset Preparation}
\label{dbCollection}
Fig. \ref{fig:data-collection-process} shows our data collection steps. We collected a December 2019 data dump of Stack Overflow from the Stack Exchange site \citep{datadumpapi}. In particular, we selected questions related to Java and Python programming languages since they are very popular and widely used. Stack Overflow data dump stores the latest version of questions and answers. Since we collected a December 2019 data dump of Stack Overflow, our selected questions were from the snapshot of December 2019. 
There were a total of 93,757 questions with \emph{$<$java$>$} or equivalent tags (e.g., \emph{$<$java-8$>$}, \emph{$<$java-11$>$}) and 105,137 questions with \emph{$<$python$>$} or equivalent tags (e.g., \emph{$<$python-2.x$>$}, \emph{$<$python-3.x$>$}) in the data dump. It should be noted that we imposed this restriction on the question tags to (1) choose purely Java or Python related questions and (2) keep our analysis simple. We then discarded questions without code segments. Since we are addressing reproducibility of reported issues, the presence of code segments in the question is necessary. We thus consider only such questions that have at least one line of code. According to our investigation, 75,669 out of 93,757 (\ie\ 80.71\%) of Java questions and 91,164 out of 105,137 (\ie\ 86.71\%) of Python questions had at least one line of code. We identified such questions using specialized HTML tags such as \texttt{<code>} within \texttt{<pre>}, and selected them for our study. However, a question can have multiple code segments (e.g., multiple source files). They were either combined (if required) or separately investigated to reproduce the issue.

\begin{table}[!t]
	\centering
	\captionsetup{justification=centering, labelsep=newline}
	\caption{A list of keywords to mine Stack Overflow questions that discuss programming-related issues}
	\label{table:list-of-keywords}
	\resizebox{3.8in}{!}{%
    \begin{tabular}{l} \toprule
    error, warning, issue, exception, fix, problem, wrong, fail \\ \bottomrule
    \end{tabular}
    }
\end{table}

\begin{figure}
 	\centering
	\includegraphics[width=4.5in]{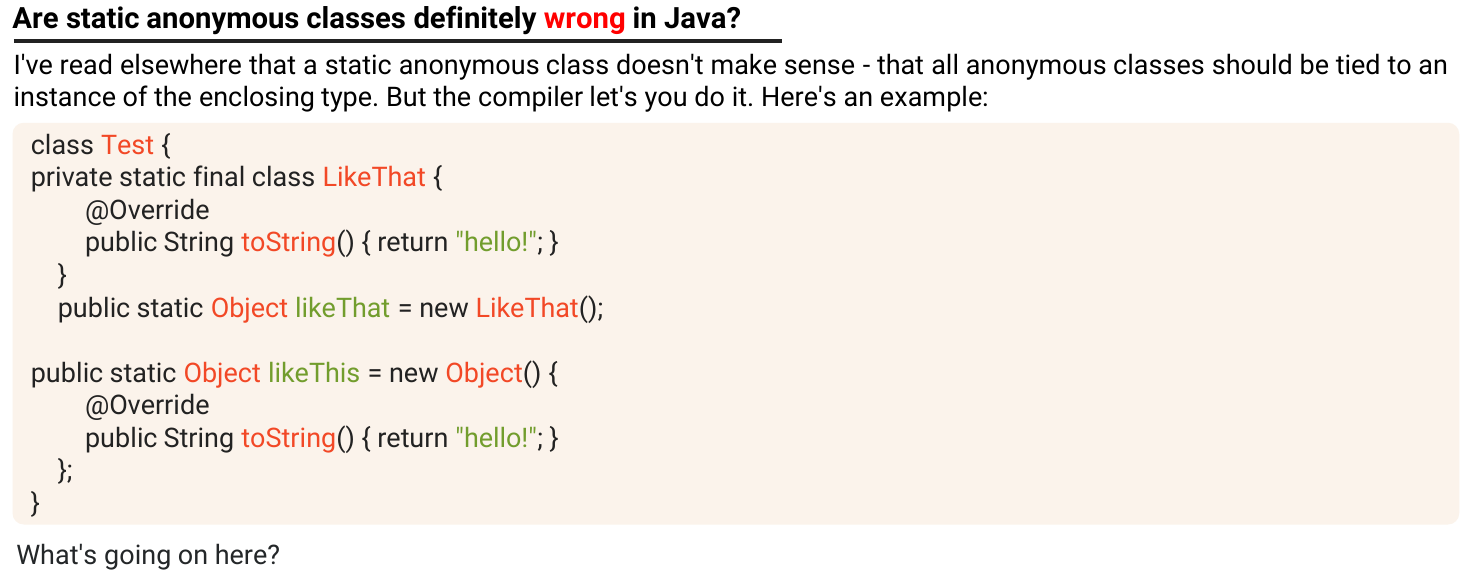}
	\caption{An example question contains a keyword (e.g., \emph{wrong}) from our selected list but does not discuss any programming-related issues. (\url{https://stackoverflow.com/questions/1243700}).}
	\label{fig:exam-false-positive}
\end{figure}

Questions containing code segments might not always have an issue that needs to be reproduced to submit an answer. Programmers might be seeking general help or asking for source code that is more efficient than the submitted code segment. However, in this study, we target only such questions that discuss at least one programming related issue and contain at least one code segment. We thus carefully analyze the questions to find the keywords that could identify questions that discuss programming-related issues. We identify several keywords (e.g., Table \ref{table:list-of-keywords}) and then look for such keywords in the text of the question. In this way, we identified a total of 29,180 Java and 34,569 Python questions that might have an issue. 
However, questions may not discuss a programming issue even if they contain one or more keywords mentioned in Table \ref{table:list-of-keywords}. They can be used in a different context. For example, Fig. \ref{fig:exam-false-positive} shows a question that contains the keyword \emph{wrong} but does not discuss any programming-related issues. Instead, it seeks opinions from others. 
We thus attempt to measure how accurately our selected keywords identify the questions that discuss programming-related issues.

To measure accuracy, we first randomly sampled 400 questions. Out of 400, we picked 200 such questions where we did not find a match with any of our selected keywords (i.e., identified as ``not discussing any issues''). The remaining 200 were picked from where we found a match with one or more keywords. To analyze the accuracy of the keywords, we manually label the sampled 400 questions as follows.
1) \textbf{Detected}: question discusses an issue or not as identified by the keywords. 
2) \textbf{Expected}: whether issues are actually discussed in questions based on our manual analysis. 
We then create a confusion matrix to analyze the performance of the selected keywords as follows.
i) True Positive (TP): \emph{detected} question discusses issue AND \emph{expected} question discusses issue. 
ii) False Positive (FP):  \emph{detected} question discusses issue BUT \emph{expected} question does not discuss issue. 
iii) True Negative (TN): \emph{detected} question does not discuss issue AND \emph{expected} question does not discuss issue. 
iv) False Negative (FN): \emph{detected} question does not discuss issue BUT \emph{expected} question discusses issue.
Table \ref{table:confusion-matrix} shows the confusion matrix. Using the matrix, we compute four standard metrics to evaluate the performance of our selected keywords as follows. Precision $P=82\%$, Recall $R=82.41\%$, F1-Score $F1=82.21\%$, and Accuracy $A=82.25\%$.

\begin{table}[!htb]
	\centering
	\captionsetup{justification=centering, labelsep=newline}
	\caption{Confusion matrix of measuring performance of our selected keywords}
	\label{table:confusion-matrix}
	\resizebox{2in}{!}{%
    \begin{tabular}{c|c|c}
    \multicolumn{1}{l|}{\textbf{N=400}}                                                   & \multicolumn{1}{c|}{\textbf{\begin{tabular}[c]{@{}c@{}}Expected\\ YES\end{tabular}}} & \multicolumn{1}{c}{\textbf{\begin{tabular}[c]{@{}c@{}}Expected\\ NO\end{tabular}}}         \\ \midrule
    \multicolumn{1}{c|}{\textbf{\begin{tabular}[c]{@{}c@{}}Detected\\ YES\end{tabular}}} & 164 & 36 \\ \midrule
    \multicolumn{1}{c|}{\textbf{\begin{tabular}[c]{@{}c@{}}Detected\\ NO\end{tabular}}}  & 35 & 165 \\ \midrule
    \end{tabular}
    }
\end{table}

We then randomly sampled 1,000 (out of 29,180) questions from Java and 1,000 (out of 34,569) questions from Python for initial analysis.
During the initial analysis, we found several duplicate questions. 
We retained the original questions with explicit issues and discarded the duplicate questions. Several questions containing one or more of the keywords mentioned in Table \ref{table:list-of-keywords} did not discuss a programming issue, which was simply discarded. We also discarded a few questions that were closed by Stack Overflow due to duplicate, off-topic or opinion-based \citep{Closed-Question-StackOverflow}. Our study analyzes the correlation between issue reproducibility status and answer meta-data (e.g., accepted answers). The closed status of a question might hurt the chance of receiving answers. Thus we discarded the closed questions. However, the closed questions were less than 1\% in our dataset.
After initial analysis, about 20\% of questions from each group were discarded due to the reasons mentioned above. Thus, we find about 800 questions (out of 1000) for each Java and Python that genuinely discuss programming-related issues. Finally, we randomly sampled 400 Java and 400 Python questions from them that met all of our selection criteria. This sample size is statistically significant with a 95\% confidence level and 5\% confidence interval. We then manually analyzed the randomly selected 800 questions and answered our research questions.

\subsection{Environment Setup}
\label{envSetup}

The environment setup to analyze Java and Python code segments is discussed as follows.

\textbf{Java.} We use \texttt{Eclipse Oxygen.3a Release (4.7.3a)}\footnote{\url{https://www.eclipse.org}} and \texttt{NetBeans 8.2}\footnote{\url{https://netbeans.org}} to execute the code segments and to reproduce the programming issues. Eclipse and NetBeans are two popular IDEs that are frequently used for Java programming. When the issues are related to compilation errors, we first employ \texttt{javac} to detect the compilation problems and then use Eclipse and NetBeans to reproduce the issues. In particular, we attempt to find an exact match between our error messages and the ones mentioned in the questions from Stack Overflow. We use \texttt{Java Development Kit(JDK)-1.8} as our development framework, \texttt{JavaParser}\footnote{\url{http://javaparser.org}} as our custom AST parser, and MySQL Workbench\footnote{\url{https://www.mysql.com/products/workbench}} 8.0 as our example database for this study. We use a desktop computer with a 64-bit Windows 10 Operating System (OS) and 16GB primary memory (i.e., RAM). We allocate 4GB as Java memory (Java heap for the IDE).

\textbf{Python.} We use \texttt{PyCharm 2019.2}\footnote{\url{https://www.jetbrains.com/pycharm}} professional
edition to execute the code segments and to reproduce the programming issues. PyCharm is widely used for Python programming. Like Java, here we also attempt to find an exact match between our error messages and the ones mentioned in the questions of Stack Overflow. We use \texttt{Python 2.7.17} and \texttt{Python 3.6.9} as Python interpreter\footnote{\url{https://www.python.org}}. 
Primarily, the \texttt{Python 2.7.17} interpreter is used to analyze the code segments and reproduce the issues. Upon failure, we use the Python \texttt{Python 3.6.9} interpreter. We use a desktop computer with a 64-bit Ubuntu 18.04.4 LTS Operating System (OS) and 16GB primary memory (i.e., RAM).

\begin{table}[!htb]
	\centering
	\captionsetup{justification=centering, labelsep=newline}
	\caption{List of Code Editing Actions to Reproduce the Issues Reported in Stack Overflow Questions}
	\label{table:list-of-actions}
	\resizebox{3.7in}{!}{%
    \begin{tabular}{l} \toprule
    \rowcolor{lightgray!50} \multicolumn{1}{c}{\textbf{Common Actions}} \\ 
    $-$ Instrumentation \\
    $-$ Code migration (outdated APIs)\\ 
    $-$ Invocation of methods\\
    $-$ Inclusion of native and external libraries\\ 
    $-$ External file, database, and dataset creation\\
    $-$ Deletion of redundant and erroneous statements\\
    $-$ Object creation, identifier declaration, and initialization\\

    \rowcolor{lightgray!50} \multicolumn{1}{c}{\textbf{Actions for Java}}\\ 
    $-$ Exception handling \\
    $-$ Addition of demo classes and methods\\ 
    
    \rowcolor{lightgray!50} \multicolumn{1}{c}{\textbf{Actions for Python}} \\ 
    $-$ Fix indentation \\
    $-$ Test case generation \\
    $-$ Create folder structure \\ \bottomrule
    \end{tabular}
    }
\end{table}

\subsection{Qualitative Analysis} 
\label{qAnalyis}

Two of the authors took part in the manual analysis of Java code segments and spent a total of 200 person-hours on the 400 Java-related questions. The author who took the lead responsibility has five, and another author has eight years of professional Java development experience. We hired two Python developers with more than four years of professional experience to manually analyze the 400 Python-related questions. They spent 100 person-hours on the analysis. We follow a two-step approach for reproducing the issues reported in the Stack Overflow questions. First, we attempt to clearly understand the reported issues and identify the key problem statements from the question description. We also gather supporting data such as input-output values and file formats from the question text. Second, using the code segment and supporting data, we attempt to reproduce the reported issues. We perform trivial, minor, and major edits on the code segments 
to reproduce the issues. Table \ref{table:list-of-actions} provides a list of our editing actions. In the following three sections we discuss each of the actions that were performed during the qualitative analysis to make the code segments reproduce the reported issues, namely: (i) common editing actions (both Java and Python), (ii) editing actions for Java, and (ii) editing actions for Python.

\subsubsection{Common editing actions}
This section discusses the editing actions that 
were applied to both
Java and Python code segments to reproduce programming issues.

\begin{inparaenum}[(a)]
    \item \emph{Inclusion of Native and External Libraries}: JDK comes with many libraries that help the developers accomplish many of the common tasks. However, developers frequently use external libraries for various specialized tasks. In Stack Overflow questions, code segments that use classes and methods from native or external libraries often miss the \emph{import} statements. We add the import statements associated with native libraries with the help of the IDE (e.g., Eclipse). In the case of external libraries, we look for relevant library references in the question texts. If such libraries were found, we include them in the IDE and then add necessary import statements.
    
    Like Java, Python code segments also miss the import statements targeting
    native or external libraries/packages that are required to execute the code. We thus add the import statements for the native libraries/packages. In the case of external libraries/packages, we first attempt to install them using the Python package manager (e.g., pip) and if installed, we then add the relevant import statements. In some cases, we need to use an updated version of the libraries/packages to execute the code.
    
    \item \emph{Object Creation, Identifier Declaration and Initialization}: The undeclared identifier is one of the common compilation errors for the Java code segments submitted to Stack Overflow. 
    We resolve the undeclared identifiers according to their inferred types and 
    initialize them with appropriate values. On the other hand, Python is a dynamically typed programming language. Thus, we did not need to declare the identifiers explicitly rather initialize them with appropriate values.
    
    \item \emph{Invocation of Methods}: We identify several code snippets that have user defined methods (or functions) but the methods were not invoked from anywhere. We thus add extra statements to call the methods if the issue reproduction warrants the execution of these methods. We also add appropriate parameters to call these methods.
        
    \item \emph{Deletion of Redundant and Erroneous Statements}: The inclusion of an erroneous code segment with a question is not a problem. Instead, we attempt to see if the error discussed in the texts can be reproduced from the given code segment. We find several code segments containing redundant and even erroneous code statements unrelated to the reported issues. We then attempt to eliminate the redundant and erroneous statements except for the discussed one. We thus delete the redundant statements or simply comment erroneous statements out so that compilers and interpreters ignore them. Our goal is to make the code segments as concise as possible that could reproduce the reported issue.
        
    \item \emph{Code Migration}: We identify several code snippets that use outdated APIs which are not compatible with our environment. We replace the outdated APIs with the equivalent updated APIs. In some cases, we also invoke extra APIs to make the code compilable/executable. 
        
    \item \emph{External File, Database and Dataset Creation}: We identify several code segments that accept input from \texttt{.txt} or \texttt{.csv} file. In such cases, we create one or more necessary files in our local drive and add the sample data from the question so that we can verify the program's correctness. Besides text files or spreadsheets, some programs require image files, especially that are related to User Interface (UI). In such cases, we create images with a specified format and dimension and then execute the program. We also create a demo relational database (\eg\ MySQL) and necessary tables when the Java code segments deal with database operations. Needless to say, we use the sample dataset provided by the question.
    
    \item \emph{Debugging}: We identify several code segments that warrant debugging to reproduce the issue. Sometimes programmers claim that they are getting unexpected behaviour from the code. For example, a developer reports that one of the \texttt{if} statements is never executed, e.g., the expression of the \texttt{if} has always been false. Then we debug the code using appropriate breakpoints and check whether the if statement is really executed. When programmers claim that they are not getting expected values from an identifier, we usually print the value or debug the code to check the run-time value of the identifier.
    
\end{inparaenum}

\subsubsection{Editing actions for Java code segments}
In this section, we discuss the editing actions that 
were performed to reproduce the issues using
Java code segments.

\begin{figure}
	\centering
	
	\subfloat[Example question with reproducibility issue]{\includegraphics[width=3.8in]{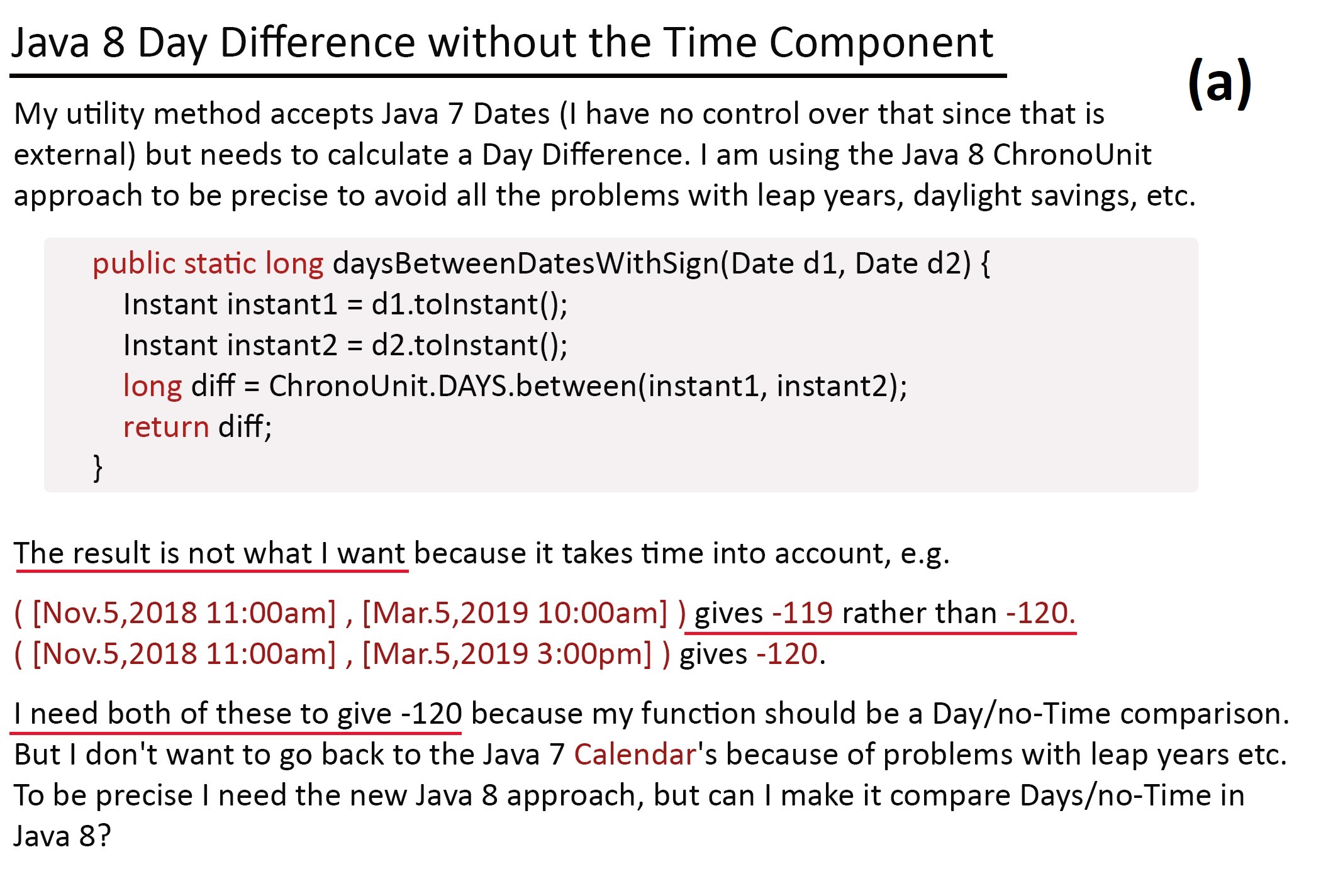}\label{fig:actionexample(a)}}

	\subfloat[Code after editing to reproduce the programming issue]
	{\includegraphics[width=3.8in]{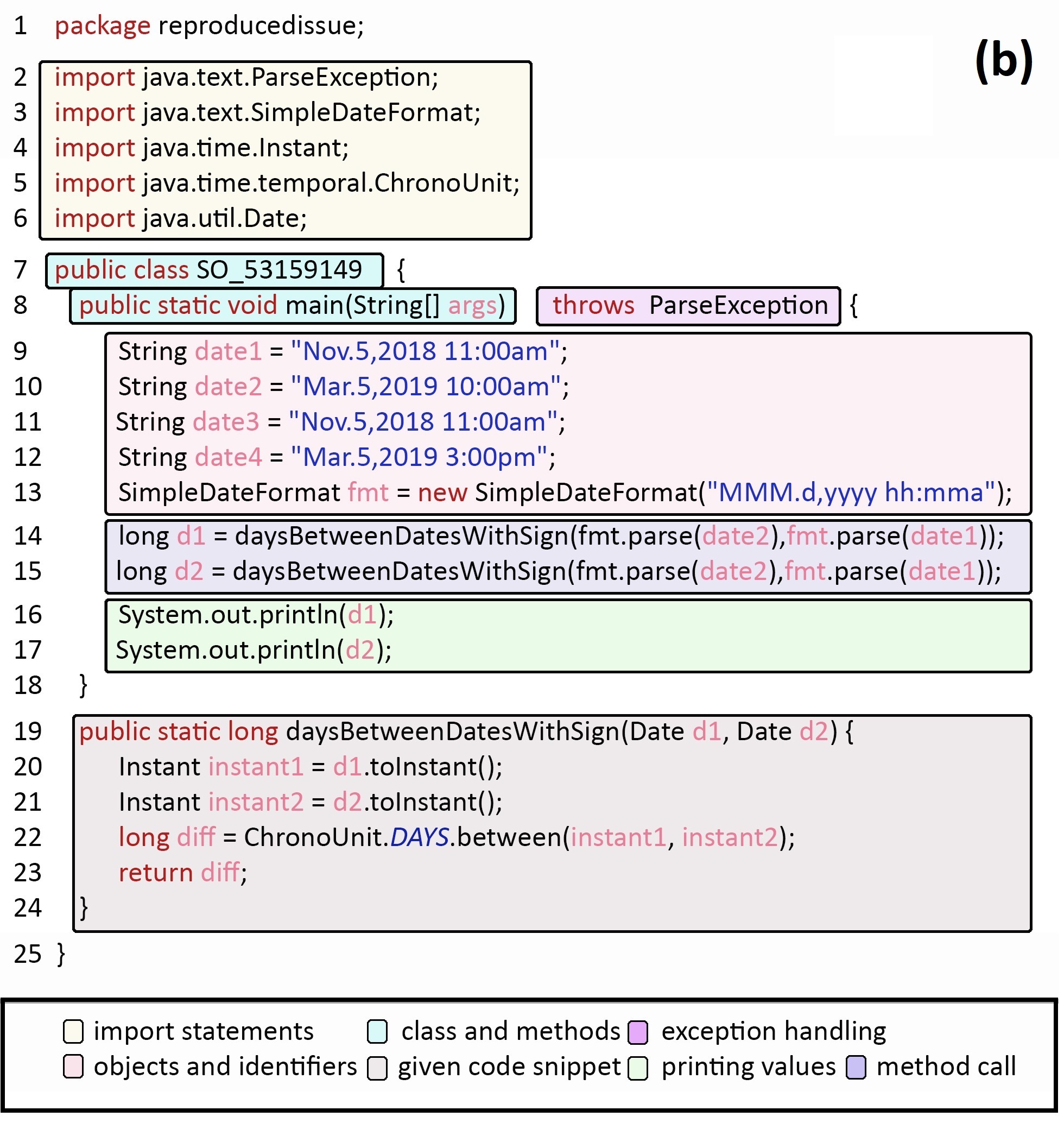}\label{fig:codeafteraction(b)}}
	
	\caption{An example question before and after editing the code to make it reproducible (\url{https://stackoverflow.com/questions/53159149}).}
	\label{fig:exam-question-to-perform-action}


\end{figure}

\begin{inparaenum}[(a)]

    \item \emph{Addition of Container Classes and Methods}: In Stack Overflow, users often submit bare minimum code examples with questions that are neither complete nor compilable. We enclose such code segments with a container Java class and place them under a main method. The main method acts as the entry point to the program. 
    
    In some cases, code segments contain statements related to the creation of an undefined object or
    the method invocation from undefined object.
    For instance, one of our code snippets has the following statements:
        \begin{tcolorbox}
            \texttt{A obj_A = new A(x,y);}
            
            \texttt{obj_A.add();}
        \end{tcolorbox}
    We see that the definition of the \texttt{class A} is absent. In order to make this code compilable, we define the class, add a constructor and also define the method add() within the class using available information from the question texts. Although the actual implementation of the method was often unavailable, such modifications helped us resolve the compilation errors.
    
    \item \emph{Exception Handling}: Developers often submit question claiming that their code throws unexpected exceptions. We find some questions where issues are not related to Java exceptions but their code throws one or more exceptions. In such cases, we resolve them using appropriate exception handling.
    Consider the example in Fig. \ref{fig:exam-question-to-perform-action}. We create an object of the \texttt{SimpleDateFormat} class (Fig. \ref{fig:codeafteraction(b)}, line No. 13) to invoke the parse API that converts the text values to the \texttt{Date} type. Since \texttt{SimpleDateFormat} throws \texttt{ParseException}, we had to handle the exception.
    
\end{inparaenum}

\subsubsection{Editing actions for Python code segments}
In this section, we discuss the editing actions that were performed on the Python code segments to reproduce the reported issues.

\begin{inparaenum}[(a)]

    \item \emph{Fix indentation:} Indentation error is one of the common errors in Python. Indentation refers to the spaces at the beginning of a line of code. While indentation is used for code readability in other programming languages,
    Python uses it to indicate a block of code. 
    Indentation error could occur when question submitters paste their code 
    on Stack Overflow question 
    editor. We fix the indentation errors carefully to execute the code segments and to reproduce the reported issues.
    
    \item \emph{Test case generation:} In Stack Overflow questions, users
    often discuss issues that are related to their 
    program inputs. For example, one user was getting an index out of range error for a list of data. However, 
    the provided code snippet
    did not provide any list values. In such cases, we generate sample inputs according to the problem discussions.
    
    \item \emph{Create folder structure:} We found several cases where users attempt to access a file from wrong locations.
    We found that there could be
    multiple files in different locations with the same name, and users often attempt to access the wrong file.
    We need to create the folder structures and choose the right path to reproduce the issues. 
    
\end{inparaenum}

\subsection{An Example of Issue Reproduction}
Let us consider the Java question shown in the Fig. \ref{fig:actionexample(a)}. First, we attempt to identify the issue and soon understand that the user is trying to calculate the time delay (in days) between two given dates. Unfortunately, the user did not get the expected results for several input values. We then copy the code segment in the IDE and find that the code does not even parse. As a result, the IDE returns several parsing and compilation errors. Fortunately, by performing several actions we can reproduce the findings.
First, we add a demo class (Fig. \ref{fig:codeafteraction(b)}, line No. 7) and place the method \texttt{daysBetweenDatesWithSignwith} within the class.
Second, we add a main method (Fig. \ref{fig:codeafteraction(b)}, line No. 8).
Third, we create four \texttt{String} objects (Fig. \ref{fig:codeafteraction(b)}, line No. 9-12) and initialize them with sample values according to the question description. 
Fourth, we then create an object \texttt{fmt} (Fig. \ref{fig:codeafteraction(b)}, line No. 13) of the class \texttt{SimpleDateFormat} to invoke the parse API that converts the text values stored in \texttt{date1, date2, date3} and \texttt{date4} to the type \texttt{Date}.
Since \texttt{SimpleDateFormat} throws \texttt{ParseException}, we also handle the exception (Fig. \ref{fig:codeafteraction(b)}, latter part of line No. 8). 
Afterward, we invoke the given method and keep the return values in \texttt{d1} and \texttt{d2} of type \texttt{long} (Fig. \ref{fig:codeafteraction(b)}, line No. 14-15). Two inline variable declarations are also required here.
We import the libraries for the classes \texttt{SimpleDateFormat, Date, ParseException, Instant} and also the enum \texttt{ChronoUnit} (Fig. \ref{fig:codeafteraction(b)}, line No. 2-6).
We then execute the modified code, print the outputs and then check whether the outputs match with the ones reported by the user. Interestingly, the issue was reproducible.


\section{Study Findings} \label{studyFindings}

We attempt to reproduce the issues reported in 400 Java and 400 Python question texts by executing their corresponding code segments. We also ask four research questions in this study and answer them carefully with the help of our empirical and qualitative findings as follows.

\subsection{Answering RQ$_1$}
\label{subsec:rq1}

\textbf{RQ$_1$(a): What are the challenges in reproducing the issues reported in Stack Overflow questions?}
We attempt to reproduce the reported issue discussed in the question using the given code segment and other supporting information (e.g., data, instruction). However, we fail to reproduce them due to several non-trivial challenges. The challenges that prevent reproducibility are discussed below.

\textbf{Java.} First, programmers often submit code segments that use methods from classes which are not defined in the code segment. Despite adding appropriate class and method definitions, many issues cannot be reproduced. Second, code segments often miss such statements that are essential to reproduce the issues. For instance, the reproducibility of an issue depends on the values of an array that are absent from the code segment. We could not reproduce such issues with the sample array values. Third, dependency on external libraries is another major challenge towards issue reproducibility from the submitted code. In many cases, we do not find any hints that point to the appropriate libraries. We also identify such code segments that are too short of reproducing the issue. Too short code is also identified as the reason for getting no answer to Stack Overflow questions \citep{asaduzzaman2013answering}. We also identify several code segments that could not reproduce the issues due to their complex interactions with UI elements, databases, and external files. We find a few segments containing outdated code (\eg\ deprecated API) that cannot be replaced by an alternative one. Several code segments are too long and noisy to reproduce the issues. 

Table \ref{table:reason-not-reproducible-java} shows the major challenges that prevent the reproduction of issues reported in Stack Overflow questions. A question might experience multiple challenges. We note that half of the irreproducible issues are plagued by undefined classes, interfaces, and methods. That is, they are used in the code segments without proper definitions. We also note that about 40\% of issues could not be reproduced due to their missing important statements and missing external libraries.

\begin{table*}[!htb]
	\centering
	\caption{Challenges Preventing Issue Reproduction for Java}
	\label{table:reason-not-reproducible-java}
    \centering
    \resizebox{3.2in}{!}{%
    \rowcolors{1}{}{lightgray!50}
    \begin{tabular}{l|c} \toprule
    \textbf{Identified Reason} & \textbf{Percentage} \\ \midrule
    Class/Interface/Method not found               & 51.00\% \\
    Important part of code missing                 & 22.99\% \\
    External library not found                     & 20.69\% \\
    Identifier/Object type not found               & 14.94\% \\
    Too short code snippet                         & 12.64\% \\
    Miscellaneous                                  & 6.90\% \\ 
    Database/File/UI dependency                    & 4.60\% \\
    Outdated code                                  & 1.15\% \\ \bottomrule
    \end{tabular}
    }
\end{table*}

\textbf{Python.} We find several common challenges between Java and Python, such as important part of code missing, method not found, library not found, database/file dependency, and too short code snippet. Apart from those, we find several additional challenges while attempting to reproduce the issues using Python code segments. First, some issues demand test cases (i.e., sample input-output) to reproduce. In some cases, issues are related to unexpected outputs. However, question submitters did not specify the inputs for which they get such outputs. We also could not guess from the question description. Second, we could not execute several code segments and thus fail to reproduce the reported issues. Such an execution failure is due to the incompatibility with our Python environment. The environment details (e.g., Python version) were required to execute the code segments that were often missed in the problem description. Some other challenges include API key missing, long, and buggy code segments.

Table \ref{table:reason-not-reproducible-python} shows the major challenges that prevent the reproduction of issues reported in Questions using Python code segments. A question might experience multiple challenges. We note that more than half of the issues could not be reproduced due to missing important statements of code. About 22\% of issues could not be reproduced due to their dependency on a database or external file.

\begin{table*}[!htb]
	\centering
	\caption{Challenges Preventing Issue Reproduction for Python}
	\label{table:reason-not-reproducible-python}
    \centering
    \resizebox{3in}{!}{%
    \rowcolors{1}{}{lightgray!50}
    \begin{tabular}{l|c} \toprule
    \textbf{Identified Reason} & \textbf{Percentage} \\ \midrule
    Important part of code missing                 & 54.05\% \\
    Database/File dependency                       & 21.62\% \\
    Missing test case                              & 8.11\%\\
    Incompatible version                           & 6.76\% \\
    Too short code snippet                         & 6.76\% \\
    Miscellaneous                                  & 6.76\% \\
    Library not found                              & 5.41\% \\ 
    Method not found                               & 1.35\% \\ \bottomrule
    \end{tabular}
    }
\end{table*}

		

	


\begin{figure}
 	\centering
	\includegraphics[width=3.8in]{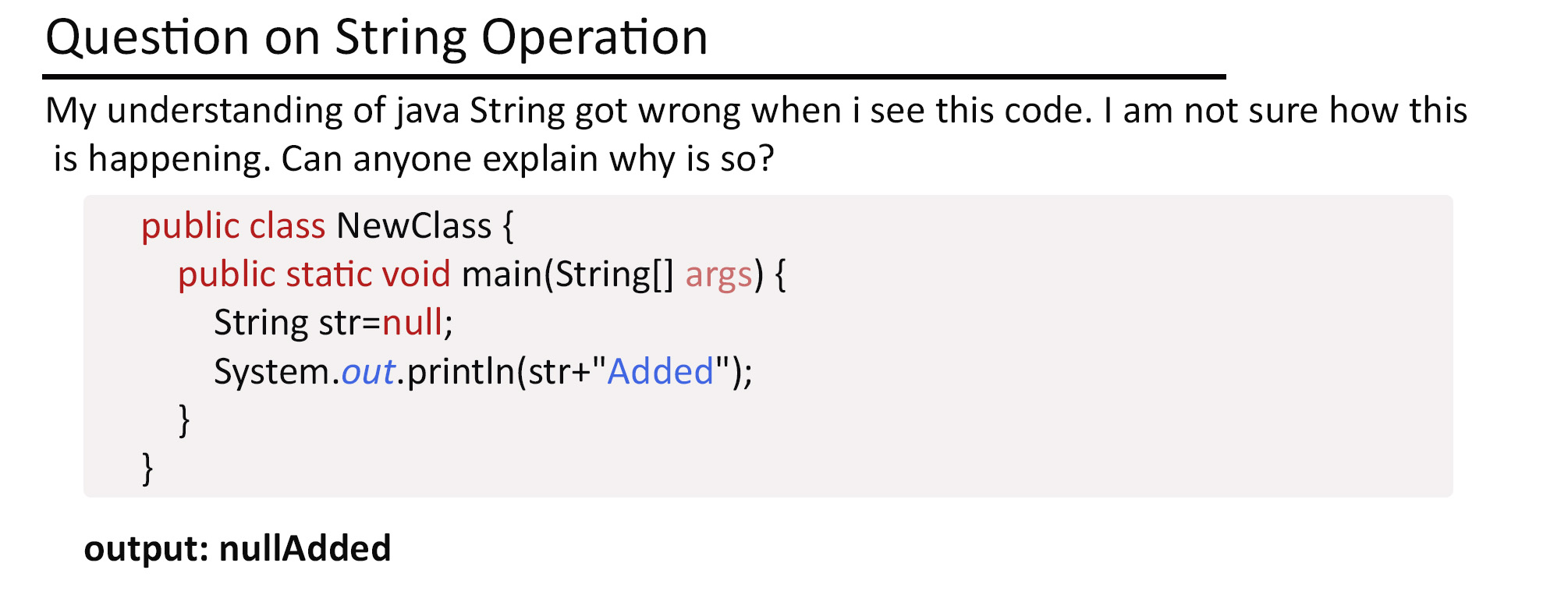}

	\caption{An example question that is reproducible without modification (\url{https://stackoverflow.com/questions/1715533}).}
	\label{fig:exam-without-modification}
\end{figure}

\textbf{RQ$_1$(b): How can issue reproducibility be determined?}
We divide the issue reproducibility status (of questions) using two different dimensions. They are the success status of reproducibility and level of editing efforts in reproducing the issues reported in Stack Overflow questions. Reproducibility status can be classified into \emph{two} major categories: \emph{reproducible} and \emph{irreproducible}.
In the following section, we discuss both categories in detail.

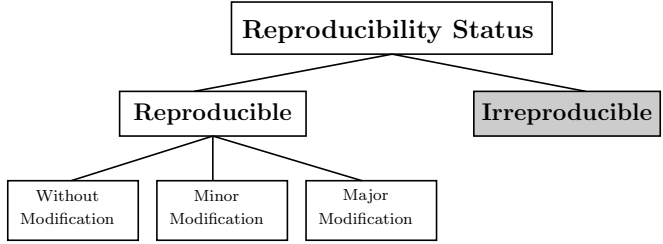
\begin{figure}
	\centering
	\resizebox{3.45in}{!}{
		\begin{tikzpicture}[scale=1.3, auto,swap]
		
		\begin{pgfonlayer}{bg}
		
		\node at (5.6, 2.4) (middle-root) {\fontsize{13}{16}\selectfont{\textbf{Reproducibility Status}}};
		\draw[thick] (3.5, 2.1) rectangle (7.8,2.8);

		\draw[thick] (5.655,2.1)--(3,1.6); 
		\draw[thick] (5.655,2.1)--(8.25,1.6); 
		
		\node at (3.25,1.3) (leftRec) {\fontsize{12}{14}\selectfont{\textbf{Reproducible}}};
		\draw[thick] (2, 1) rectangle (4.5,1.6);
		\draw[thick] (3.25,1)--(1.35,0.4); 
		\draw[thick] (3.25,1)--(3.25,0.4); 
		\draw[thick] (3.25,1)--(5.25,0.4); 
		
		\node at (1.3,0.2) (LM) {{Without}};
		\node at (1.3,0 -0.1) (LM) {{Modification}};
		\draw[thick] (0.5, -0.4) rectangle (2.25,0.4);
		
		\node at (3.3,0.2) (MID) {{Minor}};
		\node at (3.3,0 -0.1) (MID) {{Modification}};
		\draw[thick] (2.5, -0.4) rectangle (4.25,0.4);
		
		\node at (5.3,0.2) (RM) {{Major}};
		\node at (5.3,0 -0.1) (RM) {{Modification}};
		\draw[thick] (4.5, -0.4) rectangle (6.25,0.4);
	
		\draw[thick,fill=black!20] (6.75, 1) rectangle (9.25,1.6);
		\node at (8,1.3) (righRect) {\fontsize{12}{14}\selectfont{\textbf{Irreproducible}}};
		
		
		

		\end{pgfonlayer}
		\end{tikzpicture}
	}
	\caption{Classification of issue reproducibility status}
	\label{fig:issue-and-effort-classification}
\end{figure}

    \textbf{Reproducible (REP)}: Reproducibility of issues usually requires some modifications to the code segments. In some cases, issues can be reproduced from the given code segments without any modification. We therefore classify the \emph{reproducible} status into \emph{three} more sub-categories based on the complexity of the code level modifications, the level of human efforts spent and the time required to reproduce the issue. 
    We track each question's analysis time (in minutes) to reproduce the reported issue using the code segment. Total time includes -- (1) reading textual descriptions to find the specified issues, (2) copy and paste code segments to our IDE and (3) analysis of code to reproduce the reported issues by performing necessary modifications. 
    Fig. \ref{fig:issue-and-effort-classification} shows our sub-classification of the \emph{reproducible} category.
    
    - \emph{Reproducible without Modification (RWM):} The given code snippet is complete, stand-alone, and no explicit action is required to reproduce the issue. Fig. \ref{fig:exam-without-modification} shows a simple example of reproducible issue. Here, the user assigned \emph{null} to a \texttt{String} object and then added to another string. Not surprisingly, he found that the \emph{null} was also printed with the other string. To reproduce this issue, we just copy the code segment, paste it in the Eclipse IDE and then successfully reproduce the same output as noted in the question. Many of these issues are reported by new (e.g., reputation $ < 10$) and low reputed developers (e.g., $10 \le$ reputation $< 1K$).
    
    - \emph{Reproducible with Minor Modification (RMM):} The issue can be reproduced from the code by performing one or more modifications that are comparatively less complex and less time consuming. Table \ref{table:minor-modification} shows the low cost modifications added to the code. We spend about 15--30 minutes to reproduce each programming issue from this sub-category.
    
    - \emph{Reproducible with Major Modification (RMJM):}
    The reported issue can be reproduced from the code segment by performing one or more complex and time consuming modifications. Table \ref{table:major-modification} shows the high cost modifications. We spend about 30--60 minutes to reproduce each issue from this sub-category. Fig. \ref{fig:exam-question-to-perform-action} shows an example of reproducible issue where major modifications are performed on the code to reproduce the reported issue. 
    
    \textbf{Irreproducible (IREP)}: These issues are less likely to reproduce even after several minor and major code-level modifications. As mentioned above, two of the authors took part in the manual analysis. When one fails to reproduce the issue, the same question is analyzed by the other author. If both fail, we then conclude the issue as irreproducible. In some cases, we spent even a few hours on a single question. The developers who were hired for analyzing the reproducibility of Python questions also followed the same procedure. We make sure that both of the developers follow the same definition of minor/major effort level. Fig. \ref{example-issue} shows an example where the issue could not be reproduced since important details of the code/implementation are missing. Table \ref{table:reason-not-reproducible-java} and Table \ref{table:reason-not-reproducible-python} show our identified challenges that prevent a programming related issue from reproducing.

\begin{table}[!t]
	\centering
	\captionsetup{justification=centering, labelsep=newline}
	\caption{Minor Modifications}
	\label{table:minor-modification}
	\resizebox{4.5in}{!}{%
    \rowcolors{1}{}{lightgray!20}
    \begin{tabular}{p{10cm}} 
    \toprule
    \rowcolor{lightgray!60} \multicolumn{1}{c}{\textbf{Java}} \\ 
    \ding{59} {Add a demo class or a main method or a method definition}\\
    \ding{59} {Create constructor}\\
    \ding{59} {Declare identifier/object whose type can be easily determined}\\ 
    \ding{59} {Initialize identifier/object with default or sample value from the question} \\
    \ding{59} {Include import statements of native libraries} \\
    \ding{59} {Handle exception}\\
    \ding{59} {Resolve external library dependencies when import statements provided}\\
    \ding{59} {Resolve less complex compilation errors} \\
    \ding{59} {Create text/image/csv files} \\
    \ding{59} {Add sample values to the files from the question discussion} \\
    \ding{59} {Make minor changes to the code provided}\\
    \ding{59} {Invoke a method with no parameters or known parameters.}\\

    \rowcolor{lightgray!60} \multicolumn{1}{c}{\textbf{Python}}\\ 
    \ding{59} {Install and import library/package}\\
    \ding{59} {Create and initialize identifiers whose value can be easily determined}\\
    \ding{59} {Instantiate object of a class with appropriate parameters}\\
    \ding{59} {Fix indentation}\\
    \ding{59} {Resolve syntax error and warnings}\\
    \ding{59} {Function call with no parameters or known parameters}\\
    \ding{59} {Create folder structure}\\
    \ding{59} {Delete redundant code cleaning}\\
    \ding{59} {Instantiate object of a class with appropriate parameters}\\
    \ding{59} {Create simple test cases}\\
    \ding{59} {Reformat code (e.g., removal of space)}\\
    \bottomrule
    \end{tabular}
    }
\end{table}

\begin{table}[!t]
	\centering
	\captionsetup{justification=centering, labelsep=newline}
	\caption{Major Modifications}
	\label{table:major-modification}
	\resizebox{4.5in}{!}{%
	    \rowcolors{1}{}{lightgray!20}

    \begin{tabular}{p{10cm}} 
    \toprule
    \rowcolor{lightgray!60} \multicolumn{1}{c}{\textbf{Java}}\\
    \ding{59} {Resolve external library dependencies when import statements not provided}\\
    \ding{59} {Create database and table entries according to the problem description}\\
    \ding{59} {Make major changes to the code provided}\\
    \ding{59} {Declare/initialize identifier/object whose type cannot be easily determined}\\
    \ding{59} {Add sample input values to the files when not provided}\\
    \ding{59} {Invoke method with parameters when sample parameters not provided}\\
    \ding{59} {Debug code}\\

    \rowcolor{lightgray!60} \multicolumn{1}{c}{\textbf{Python}}\\
    \ding{59} {Install library/package when import statements are not provided}\\
    \ding{59} {Create file structure by predicting how the code uses files}\\
    \ding{59} {Resolve many syntax errors and warnings}\\
    \ding{59} {Create complex test cases}\\
    \ding{59} {Resolve logical errors by analyzing code or debugging}\\
    \ding{59} {Resolve many indentation errors}\\
    \ding{59} {Invoke function with parameters when sample parameters not provided}\\
    \bottomrule
    \end{tabular}
    }
\end{table}

\begin{figure}
	\centering
	\includegraphics[width=4in]{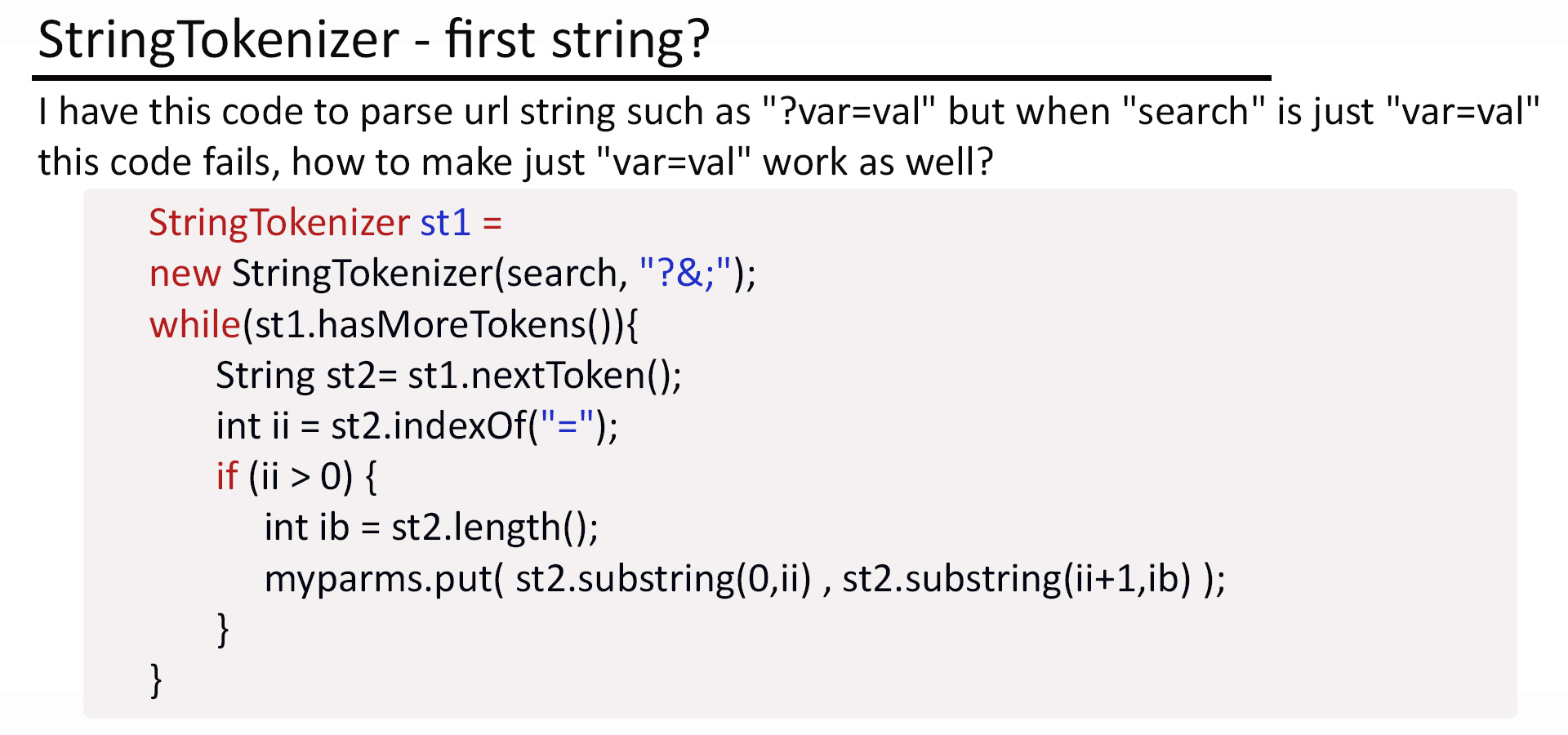}
	\caption{An example question with an inaccurate claim about the programming issue 
	(\url{https://stackoverflow.com/questions/750557}).
	}
	
	\label{fig:exam-inaccurate-claim}
\end{figure}

Besides the major classifications above, we also consider the appropriateness of developers' claims about issues during the question submission. We found the following two additional types of issues that could not be reproduced:

    \emph{Inaccurate Claim (IAC):}
    In some cases, the stated programming issues in the Stack Overflow questions might not be accurate. We call these issues Inaccurate Claim. The question shown in the Fig. \ref{fig:exam-inaccurate-claim} shows an example of an inaccurate claim about a programming issue. Here, the user raised an issue that they could parse the string  \texttt{?var=val} but could not parse the string  \texttt{var=val}. When we attempt to execute this code in our IDE, we find that both strings can be parsed successfully. Two programmers even commented, \emph{``Sorry, that code doesn't fail''} and \emph{``This works for me too.''} We find similar occurrences with run-time errors, compiler errors, and unexpected behaviour 
    reported by the users in Stack Overflow questions. Such anomalies could have several explanations. First, this might happen due to the difference in the development environment. Second, users might fail to identify the defective code, and hence, they submit the wrong code fragments.
    
    \emph{Ill-Defined Issue (IDEF)}: Ill-defined issues refer to the problems which cannot be reproduced consistently under any circumstances. That is, the question submitter does not specify the context precisely in which situation the alleged programming issue occurs. For instance, a user claimed that he was getting a run-time exception after clicking buttons. During code execution we find several buttons in the user interface and some of them trigger runtime exceptions. However, since the user does not specify a button, the reported issue could not be reproduced.

\subsection{Answering RQ$_2$}
\label{subsec:rq2}
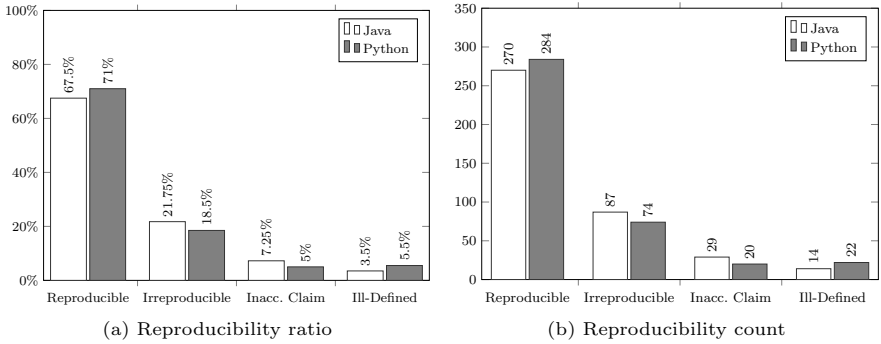
\begin{figure}[!htb]
\centering
    \pgfplotstableread{
    		1	67.5   71     270   284 
    		2	21.75  18.5   87    74
    		3	7.25   5      29    20
    		4	3.5    5.5    14    22
    }\datatable
    \subfloat[Reproducibility ratio]{
    \label{fig:reproducibility-status-ratio}
   	  \resizebox{2.25in}{!}{%
      \begin{tikzpicture}
        	\begin{axis}[
        	xtick=data,
        	xticklabels={Reproducible, Irreproducible, Inacc. Claim, Ill-Defined
        	},
        	enlarge y limits=false,
        	enlarge x limits=0.15,
        	ymin=0,ymax=100,
        	ybar,
        	bar width=0.75cm,
        	width=3.8in,
        	height = 2.85in,
        	ytick={0,20,...,100},
            yticklabels={0\%,20\%,40\%,60\%,80\%,100\%,},
        	ymajorgrids=false,
        	major x tick style = {opacity=0},
        	minor x tick num = 1,    
        	minor tick length=1ex,
        	legend style={
         	legend pos=north east,
        	legend cell align=left
            },
            nodes near coords style={rotate=90,  anchor=west}, 
        	nodes near coords =\pgfmathprintnumber{\pgfplotspointmeta}\%
        	]
        	\addplot[draw=black!80, fill=black!0] table[x index=0,y index=1] \datatable;
        	\addplot[draw=black!80, fill=black!50] table[x index=0,y index=2] \datatable;

            \legend	{Java,
            		 Python
            		 }
        	\end{axis}
    	\end{tikzpicture}
    	}
    	}
      \subfloat[Reproducibility count]{
      \label{fig:reproducibility-status-count}
   	  \resizebox{2.25in}{!}{%
      \begin{tikzpicture}
        	\begin{axis}[
        	xtick=data,
        	xticklabels={Reproducible, Irreproducible, Inacc. Claim, Ill-Defined
        	},
        	enlarge y limits=false,
        	enlarge x limits=0.15,
        	ymin=0,ymax=350,
        	ybar,
        	bar width=0.7cm,
        	width=3.8in,
        	height = 2.8in,
        	ytick={0,50,...,350},
        	ymajorgrids=false,
        	major x tick style = {opacity=0},
        	minor x tick num = 1,    
        	minor tick length=1ex,
        	legend style={
         	legend pos=north east,
        	legend cell align=left
            },
            nodes near coords style={rotate=90,  anchor=west}, 
        	nodes near coords =\pgfmathprintnumber{\pgfplotspointmeta}
        	]
        	\addplot[draw=black!80, fill=black!0] table[x index=0,y index=3] \datatable;
        	\addplot[draw=black!80, fill=black!50] table[x index=0,y index=4] \datatable;

            \legend	{Java,
            		 Python
            		 }
        	\end{axis}
    	\end{tikzpicture}
    	}
    	}
\caption{Issue reproducibility status.}
\label{fig:reproducibility-status}
\end{figure}

\textbf{What proportion of reported issues in Stack Overflow questions can be reproduced successfully?} We classify the issue reproducibility status into two major categories, which are again classified into further sub-categories. 
Fig. \ref{fig:reproducibility-status} shows the statistics on the reproducibility status of the questions. According to our analysis, 270 (67.5\%) issues among 400 can be reproduced using Java code segments. We can reproduce 284 (71.0\%) among 400 issues using Python code segments. On the contrary, 87 (21.8\%) issues for Java and 74 (18.5\%) issues for Python cannot be reproduced due to the challenges shown in Table \ref{table:reason-not-reproducible-java} and Table \ref{table:reason-not-reproducible-python}. We find inaccurate claims for 7.3\% of Java questions and 5.0\% of Python questions. This might be due to the lack of programming experience or proper analysis of the code during a question submission. We could not determine the reproducibility for 3.5\% of Java questions and 5.5\% of Python questions. The users often fail to provide the appropriate context in their questions, which are required to reproduce the reported issues. We thus call them ill-defined issues (IDEF).

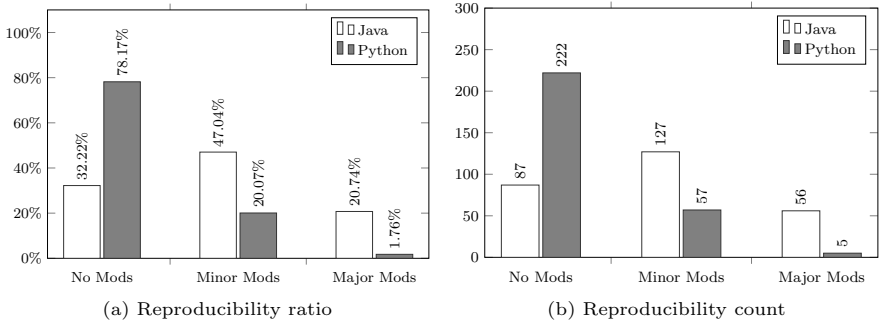
\begin{figure}[!htb]
\centering
   	\pgfplotstableread{
    		1	32.22   78.17   87   222  
    		2	47.04   20.07   127  57
    		3	20.74   1.76    56   5

    	}\datatable
      \subfloat[Reproducibility ratio]{
      \label{fig:reproducible-status-according-effort-level-ratio}
      \resizebox{2.25in}{!}{%
      \begin{tikzpicture}
        	\begin{axis}[
        	xtick=data,
        	xticklabels={No Mods, Minor Mods, Major Mods},
        	enlarge y limits=false,
        	enlarge x limits=0.2,
        	ymin=0,ymax=110,
        	ybar,
        	bar width=0.7cm,
        	width=3.5in,
        	height = 2.5in,
        	ytick={0,20,...,100},
            yticklabels={0\%,20\%,40\%,60\%,80\%,100\%,},
        	ymajorgrids=false,
        	major x tick style = {opacity=0},
        	minor x tick num = 1,    
        	minor tick length=1ex,
        	legend style={
         	legend pos=north east,
        	legend cell align=left
            },
            nodes near coords style={rotate=90,  anchor=west}, 
        	nodes near coords =\pgfmathprintnumber{\pgfplotspointmeta}\%
        	]
        	\addplot[draw=black!80, fill=black!0] table[x index=0,y index=1] \datatable;
        	\addplot[draw=black!80, fill=black!50] table[x index=0,y index=2] \datatable;

            \legend	{Java,
            		 Python
            		 }
        	\end{axis}
    	\end{tikzpicture}
     	}
     	}
      \subfloat[Reproducibility count]{
      \label{fig:reproducible-status-according-effort-level-count}
      \resizebox{2.25in}{!}{%
      \begin{tikzpicture}
        	\begin{axis}[
        	xtick=data,
        	xticklabels={No Mods, Minor Mods, Major Mods},
        	enlarge y limits=false,
        	enlarge x limits=0.2,
        	ymin=0,ymax=300,
        	ybar,
        	bar width=0.7cm,
        	width=3.5in,
        	height = 2.45in,
        	ytick={0,50,...,300},
        	ymajorgrids=false,
        	major x tick style = {opacity=0},
        	minor x tick num = 1,    
        	minor tick length=1ex,
        	legend style={
         	legend pos=north east,
        	legend cell align=left
            },
            nodes near coords style={rotate=90,  anchor=west}, 
        	nodes near coords =\pgfmathprintnumber{\pgfplotspointmeta}
        	]
        	\addplot[draw=black!80, fill=black!0] table[x index=0,y index=3] \datatable;
        	\addplot[draw=black!80, fill=black!50] table[x index=0,y index=4] \datatable;

            \legend	{Java,
            		 Python
            		 }
        	\end{axis}
    	\end{tikzpicture}
     	}
     	}
\caption{Reproducible issues and modification effort level.}
\label{fig:reproducible-status-according-effort-level}
\end{figure}

The ratio of reproducible status according to the effort level is shown in Fig. \ref{fig:reproducible-status-according-effort-level}. For Java, half of them require minor modification, whereas 20.7\% of issues require major modification to make the code snippets capable of reproducing the reported issues. Fortunately, 32.2\% snippets can reproduce the issues without any modifications. For Python, two-third of issues can be reproduced without any modifications, whereas only 2.0\% of issues require major modification to make the code snippets capable of reproducing the reported issues.

The changes (e.g., editing actions) needed to reproduce an issue using Java code could be different from that using Python code (e.g., Table \ref{table:minor-modification} \& \ref{table:major-modification}), which can lead to different effort levels.
For example, the addition of a container class and the main method was essential for Java code but not for python code. Besides, Java and Python are different in type - compiled vs. interpreted, which could warrant different editing operations. Thus, although we have a well-defined set of effort levels, different levels of effort were spent to reproduce the issues with Java code and Python code.
Fig. \ref{fig:reproducible-status-according-effort-level} shows that about 78\% of issues can be reproduced  without modifying Python code examples since Python is a dynamic and interpreted language. However, we often need to add a dummy class, the main method, and declare variables with types to compile/execute Java code examples when users add a couple of bare statements. Thus, the majority of Java code segments (about 47\%) require minor modifications to reproduce issues.

We then investigate the evolution aspect and attempt to determine whether the evolution aspect affects the reproducibility of issues. For example, the initial version of a code segment may not reproduce the reported issue. However, the code segment can be revised through edits and able to reproduce the reported issues. We thus randomly sampled 25\% of questions from our manually analyzed dataset. In particular, we select 91 Java questions (reproducible without modification 22 + minor modification 32 + major modification 14 + irreproducible 23) and 90 Python questions (reproducible  without modification 56 + minor modification 14 + major modification 1 + irreproducible 19) for further analysis. 
We not only extract and investigate the first version but also all the remaining versions of these questions (i.e., evolution) from the revision history to see the changes (i.e., edits) performed over the revisions. 
We used the same environmental setup as mentioned in Section \ref{envSetup} for analyzing the code segments. Our findings are summarized as follows.

\begin{figure}
	\centering
	\includegraphics[width=4.3in]{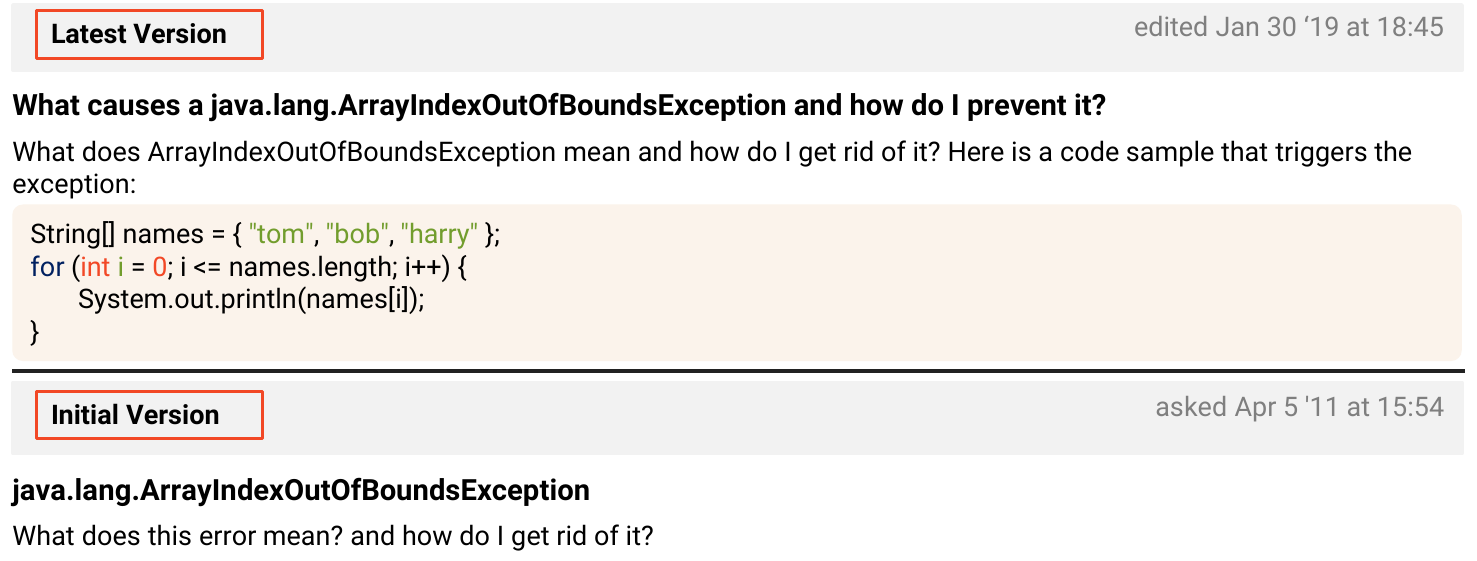}
	\caption{An example of question evolution. The initial version was asked to know the meaning of \texttt{java.lang.ArrayIndexOutOfBoundsException}. Then an example code segment was added to the question texts in the latest revision.
	(\url{https://stackoverflow.com/posts/5554734/revisions})
	}
	\label{fig:example-evolution}
\end{figure}

\begin{itemize}

    \item[] \textbf{Revision.} About 30.77\% (28 out of 91) of Java and 44.44\% (40 out of 90) of Python questions were not revised at all. Besides, 32.97\% (30 out of 91) Java and 28.89\%(26 out of 90) Python questions were revised (i.e., edited) only once (i.e., no of revision $=2$). However, the average number of revisions of our selected Java questions was 3.22. On the other hand, such average for Python questions was 2.92.
    
    \vspace{1mm}
    \item[] \textbf{Changes during evolution.}
    According to our investigation, 82.54\% (52 out of 63) of Java and 78\% (39 out of 50) Python code segments were either unchanged or revised only to improve their format (e.g., adding/removing space) during evolution. Among them, problem descriptions of 36.51\% (23 out of 63) Java and 46\% (23 out of 50) Python questions were also revised to improve only their format (e.g., bold/unbold text). Such revisions might improve the readability of text and code. However, they do not affect reproducibility. Other revisions include removing emotion (e.g., gratitude), title and tag edits, grammatical corrections (e.g., spelling mistakes), removing docstring, adding software versions, personal notes, import statements, solution code, and change API. Again, such changes did not affect much the reproducibility of the issues reported in the questions.
    
    \vspace{1mm}
    \item[] \textbf{Impact of question evolution on reproducibility.} Overall, the evolution aspect does not significantly affect the reproducibility of issues and, thus, our experimental findings. During the analysis, we found only two questions of Java (e.g., Fig. \ref{fig:example-evolution}) where clarifications and code segments were added to the question texts in the later revision. Thus, the first versions of those two questions did not satisfy our constraints since they did not include code segments. In addition, we found that the first versions of code segments of two Python questions were updated (e.g., added import statements) during evolution. For those two questions, the reproducible statuses were reproducible with minor modification and reproducible with major modification when we analyzed their first versions. In contrast, the statuses for their latest versions were reproducible without modification and minor modifications, respectively. Given the evidence of cosmetic, trivial changes, the impact of question evolution on the reproducibility of the issue might be negligible.
    
\end{itemize}


\begin{table}[htb]
\centering
    \caption{Issue Reproducibility Status vs. Presence of Accepted Answer}
	\label{table:relationship-reprostatus-accanswer}
	\resizebox{4.5in}{!}{%
    \begin{tabular}{l|c|c|c|c|c|c} \toprule
   \multicolumn{1}{l|}{\textbf{Reproducibility}} & \multicolumn{2}{c|}{\textbf{Accepted Answer}} & \multicolumn{2}{c|}{\textbf{No Accepted Answer}} &  \multicolumn{2}{c}{\textbf{Total}} \\ 
                                                    \multicolumn{1}{l|}{\textbf{Status}} & \textbf{Java}            & \textbf{Python}            & \textbf{Java}         & \textbf{Python}           & \textbf{Java}         & \textbf{Python} \\  \midrule
    \textbf{Reproducible}                            &  201 (74.4\%)           &  165 (58.1\%)              &  69 (25.6\%)         &  119 (41.9\%)             &   270              &      284       \\  \midrule
    \textbf{Irreproducible}                          &  18 (20.7\%)            &  19 (25.7\%)              &  69 (79.3\%)         &  55 (74.2\%)              &   87                &      74          \\ \bottomrule              
    \end{tabular}
    }
\end{table}






    
    


\begin{table}[htb]
\centering
\caption{Issue Reproducibility Subcategory vs. Presence of Accepted Answer}
	\label{table:relation-effort-level-accanswer}
	\resizebox{4.5in}{!}{%
    \begin{tabular}{l|c|c|c|c|c|c} \toprule
    \multicolumn{1}{l|}{\textbf{Reproducibility}} & \multicolumn{2}{c|}{\textbf{Accepted Answer}} & \multicolumn{2}{c|}{\textbf{No Accepted Answer}} &  \multicolumn{2}{c}{\textbf{Total}} \\ 
                                                        \multicolumn{1}{l|}{\textbf{Subcategory}}  & \textbf{Java}            & \textbf{Python}            & \textbf{Java}        & \textbf{Python}           & \textbf{Java}        & \textbf{Python} \\  \midrule
    \textbf{No Modifications}                         &  66 (75.9\%)           &  133 (59.9\%)              &  21 (24.1\%)        &  89 (40.1\%)             &   87               &      222        \\  \midrule
    \textbf{Minor Modifications}                           & 102 (80.3\%)           &  29 (50.9\%)              &  25 (19.7\%)         &  28 (49.1\%)             &   127               &      57       \\  \midrule
    \textbf{Major Modifications}                           &  33 (58.9\%)           &  3 (60.0\%)                  &  23 (58.9\%)         &  2 (40.0\%)                 &   56               &      5        \\  \midrule
    \textbf{Irreproducible}                               &  18 (20.69\%)           &  19 (25.68\%)              &  69 (79.31\%)        &  55 (74.32\%)             &   87                 &      74           \\ \bottomrule              
    \end{tabular}
    }
\end{table}

\subsection{Answering RQ$_3$}
\label{subsec:rq3}
\label{subsec:correlation-analysis}
Reproduction of the programming issue discussed in the question text allows one to experience the issue first hand, which could help users submit appropriate answers to the questions more promptly and more accurately. Therefore, we analyze the relationship between reproducibility status of the reported issues in Stack Overflow questions and the answer meta data such as presence of an accepted answer, time delay between question and accepted answer, and number of answers. In particular, we divide \emph{RQ3} into three sub-questions, and answer them with detailed statistics as follows:

\textbf{RQ$_3$(a): Does reproducibility of issues discussed in Stack Overflow questions encourage acceptable answers?}
Table \ref{table:relationship-reprostatus-accanswer} shows the confusion matrix where the rows represent the reproducibility status (\eg\ reproducible/irreproducible), and the columns represent the presence of an accepted answer (\eg\ present or absent). We note that 74.4\% (201 out of 270) Java and 58.1\% (165 out of 284) Python questions whose code segments are capable of reproducing the reported issues receive acceptable answers (i.e., solutions). On the contrary, only 20.7\% (18 out of 87) Java and 25.7\% (19 out of 74) Python questions with irreproducible issues receive acceptable answers. Thus, the reproducibility of issues increases the chance of getting a solution more than three times for Java and more than two times for Python. We examine the correlation between the two categorical variables -  reproducibility status and accepted answer presence. We use a statistical test, namely the \emph{Chi-Squared} test, to measure the independence of these two categorical variables shown in Table \ref{table:relationship-reprostatus-accanswer}. We find statistical significance \emph{p-value} ($p-value = 0 < 0.05$). Thus, there is a strong positive correlation between the issue reproducibility of questions and their chance of getting acceptable answers.

We also further analyze the reproducibility status based on effort level and determine their correlation with the accepted answer presence, as shown in Table \ref{table:relation-effort-level-accanswer}. From Table \ref{table:relation-effort-level-accanswer}, we see that there is a 75.9\% (66 out of 87) chance of getting an acceptable answer for Java questions when the issues can be directly reproduced from the verbatim code segment. On the contrary, the chance reduces to 58.9\% (33 out of 56) when the submitted code needs major modifications. Thus, the human efforts in reproducing the issues reported in questions are significantly correlated with the chance of getting the accepted answer. We also find statistically significant \emph{p-value} ($p-value = 0 < 0.05$) from the \emph{Chi-Squared} test. However, the difference in receiving accepted answers between reproducible without modification and minor modifications is not much high. For example, for Python questions, we see that there is a 59.9\% (133 out of 222) chance of getting an acceptable answer when the issues can be directly reproduced from the verbatim code segment. On the other hand, the chance is 50.9\% (29 out of 57) when the submitted code needs minor modifications. Such findings suggest that users can mostly answer questions appropriately even if the code segments require minor modifications to reproduce the reported issues. Surprisingly, the chance of getting an accepted answer is highest (80.3\%) for the questions whose code fragments require minor modifications to reproduce the issues. We thus further investigate this interesting scenario with more manual analysis. Three factors were identified that might explain such inconsistency on receiving accepted answers:

\begin{figure}[!htb]
 	\centering
	\includegraphics[width=4.5in]{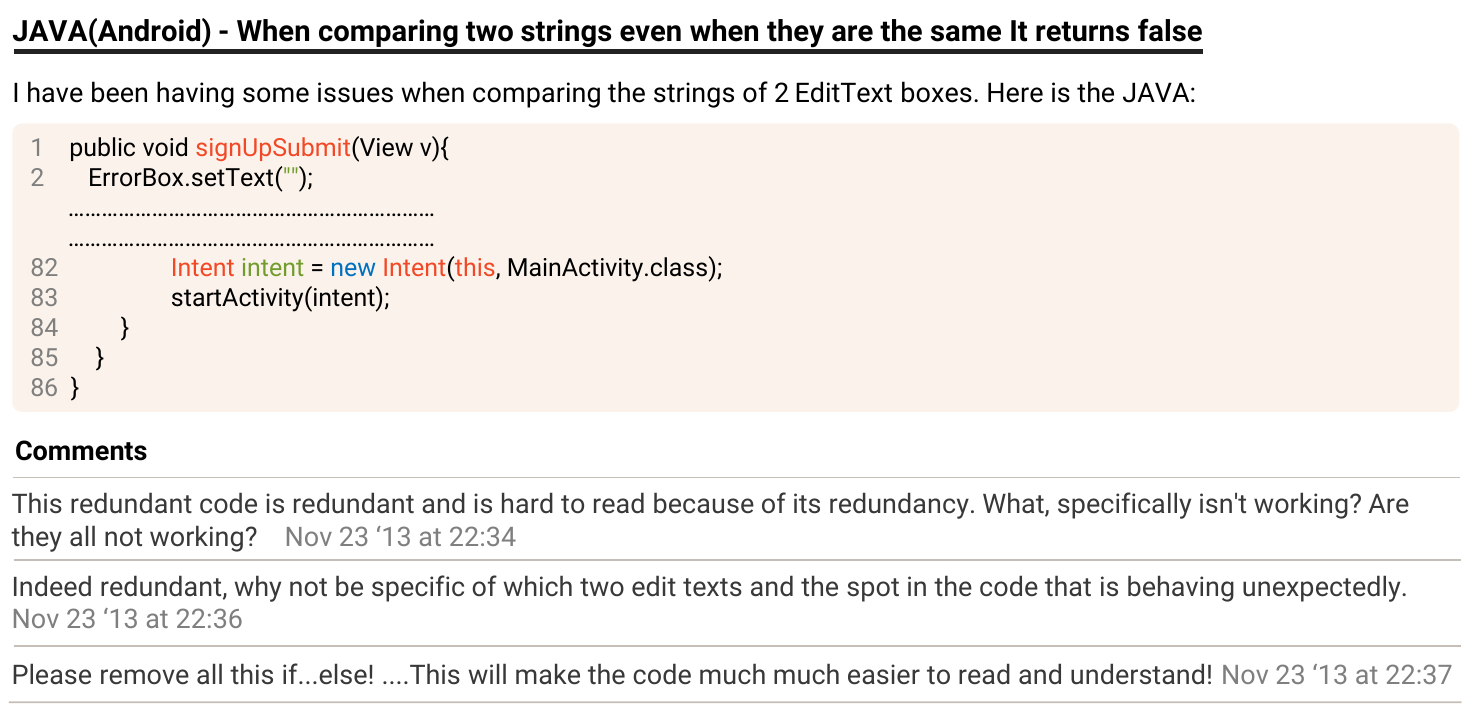}

	\caption{An example question where users complain against redundant code and suggest removing redundant statements.  (\url{https://stackoverflow.com/questions/20168726}).}
	\label{fig:exam-redundant-code}
\end{figure}

\begin{itemize}
    \item[] \textbf{Redundant and long code can prevent accepted answers.} The users of Stack Overflow often submit questions that contain unnecessarily long code. As a result, such questions often fail to get appropriate answers even though their code is capable of reproducing the reported issues without any modifications. We measure the length of each of the code segments included with questions whose issues could be reproduced without any modifications and did not receive accepted answers. According to our analysis, such code segments have more than $72$ lines of code (\texttt{LOC}) on average. Such a long code could place an extra burden on the other users during question answering.
    We then investigate to what extent \texttt{LOC} with more than 72 lines are less prone to get an accepted answer. According to our investigation, if the \texttt{LOC} of the code segments included with questions is less or equal to $72$, the chance of receiving accepted answers is 63.02\%. However, the chance reduces to 50\% when the \texttt{LOC} is more than $72$. We then attempt to see whether the code was actually needed or redundant. We thus eliminate the statements from the code segments that are not required to reproduce the reported issues. Interestingly, we can eliminate at least half of the code statements. Still, the issues can be reproduced. We also found that users commented against the long redundant code and suggested removing them (e.g., Fig. \ref{fig:exam-redundant-code}) during manual analysis. Thus, questions with long redundant code are less likely to receive accepted answers despite being reproducible without modifications.
    
    \vspace{1mm}
    \item[] \textbf{Identification of potential problem statements can help to receive accepted answers.} Users often fail to mention the potential problematic statements within the submitted code in their questions. This increases the analysis time for the other users of Stack Overflow while answering the question. Users also recommend spotting the potential code statements where the issue can occur. For example, one user commented on spotting the code that was behaving unexpectedly (e.g., Fig. \ref{fig:exam-redundant-code}). Thus, despite the verbatim code segments being able to reproduce the reported issues (without modifications), they might not receive the accepted answers.
    
    \vspace{1mm}
    \item[] \textbf{User reputation can affect questions getting accepted answers.} Questions submitted by new users (i.e., reputation $< 10$) are less likely to receive acceptable answers \citep{calefato2015mining, mondal2021early}. According to our analysis, the chance of getting accepted answers for questions submitted by new users is only 38.53\% (42 out of 109). On the contrary, such a chance is 71.37\% (177 out of 248) for the questions submitted by users except for new users. However, we see that about 50\% of questions whose issues can be reproduced without modifying the code segments but did not receive accepted answers were submitted by new users.
    %
    
\end{itemize}

We also see that the rate of receiving acceptable answers for Python questions when the submitted code needs major modification is almost similar when the issues can be directly reproduced. However, the number of questions is only three, so we did not go for a further manual investigation to expose the causes.

Besides accepted answers, we consider the answers with the highest score (a.k.a., best answer) since several questions do not have accepted answers. In particular, we attempt to see whether reproducibility of issues encourages high-scored answers. In Stack Overflow, the score usually estimates quality. Thus, a high score indicates high quality and vice versa.  However, in our dataset, a total of 40 questions (19 Java + 21 Python) did not receive any answers.
Furthermore, the highest score of 139 questions (Java 44 + 95 Python) is either negative or zero. An answer with a negative score could not be considered a high-quality answer. A score of zero usually indicates that the users did not assess the quality of the answer. Moreover, several questions have multiple answers with the highest score. Thus, selecting the answers other than the accepted ones could introduce noise in our analysis. However, we conduct a preliminary investigation using 293 answers (237 Reproducible + 56 Irreproducible) related to Java with the highest score. We were able to reproduce the result, which is aligned with when we considered the accepted answers. For example, the median score of the best answers to questions with reproducible issues is two times higher than those with irreproducible issues.  


\begin{figure}
	\centering
	\subfloat[Java]
	{
	\includegraphics[width=3in]{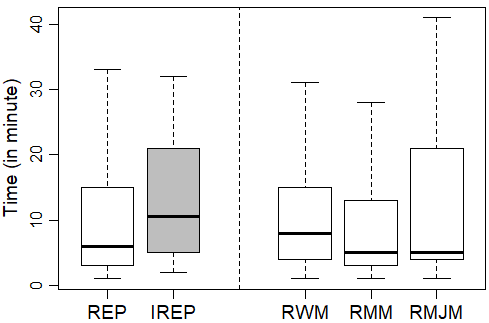}
	}
	
	\subfloat[Python]
	{
	\includegraphics[width=3in]{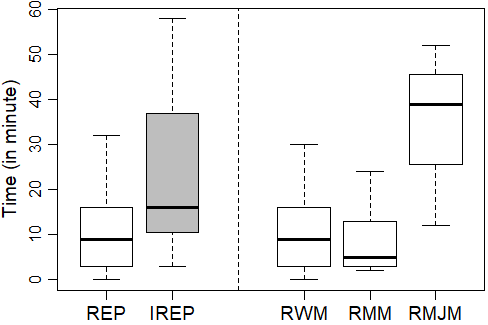}
	}

	\caption{Time delay in receiving accepted answers for reproducible and irreproducible issues (\small{\textbf{REP}=Reproducible; \textbf{IREP}=Irreproducible; \textbf{RWM}=Reproducible without modification; \textbf{RMM}=Reproducible with Minor Modifications; \textbf{RMJM}=Reproducible with Major Modifications).}}
	\label{fig:box-plot-accepted-answer}
\end{figure}

\textbf{RQ$_3$(b): Does reproducibility of issues discussed in Stack Overflow questions reduce the time to receive an accepted answer?}
According to RQ3a, there is a strong correlation between the reproducibility status of issues reported in questions and the chance of receiving an accepted answer. In fact, the chance is at least more than two times higher for reproducible issues. In this section, we investigate whether the delay in receiving an accepted answer could be influenced by the reproducibility status of the submitted issues in Stack Overflow questions. We determine the delay between the submission time of a question and that of the accepted answer and contrast that between reproducible and irreproducible issues using such delays. Fig. \ref{fig:box-plot-accepted-answer} shows the box plots for the delay of receiving accepted answers. We see that the median delay of getting an accepted answer for Java questions is about 5 minutes when the issue reported at the question is reproducible. On the contrary, such delay is almost double when the reported issue is not reproducible with the submitted code.
We also find similar results for Python questions. Here, the median delay of receiving an accepted answer for questions is about 9 minutes when the issue reported at the question is reproducible. On the contrary, such delay is about 16 minutes when the reported issue is not reproducible using the submitted code.

We find a significant difference in time delay for receiving an accepted answer between an issue being reproducible or irreproducible.  We use the \emph{Mann-Whitney-Wilcoxon} test, a non-parametric statistical significance test and find a statistical significance \emph{p-value} (\textbf{Java}: p-value $=$ $0.04$ $<$ $0.05$, \textbf{Python}: p-value $=$ $0.005$ $<$ $0.05$). We also examine the effect size using \emph{Cliff's delta} test and find a large \emph{effect size} (\textbf{Java}: Cliff's $|$d$|$ $= 0.9365857$ (large), \textbf{Python}: Cliff's $|$d$|$ $= 0.9590269$ (large)) with $95\%$ confidence. Given this evidence, the delay of receiving an acceptable answer is significantly higher for questions with irreproducible issues. Thus, the reproducibility of the reported issues is an important quality attribute for Stack Overflow questions, and such an attribute could help ensure accepted answers are submitted quickly, even within 5 minutes. Our further analysis shows that the time delay of getting accepted answers is slightly less for the questions whose code segments require minor modifications to reproduce the issues than those with no modifications. Surprisingly, the time delay of getting an acceptable answer for Python questions when the submitted code needs major modification is very high. However, we did not further analyze the causes since the number of questions belongs to that subcategory is only three.

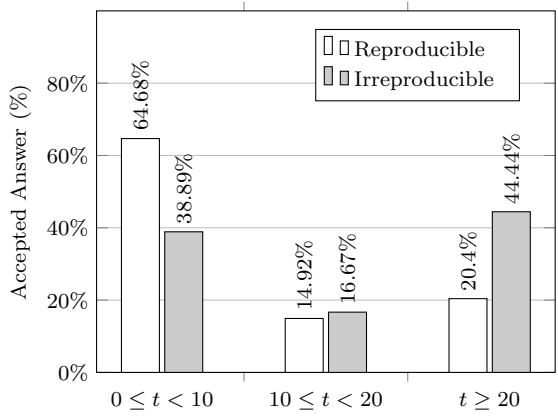
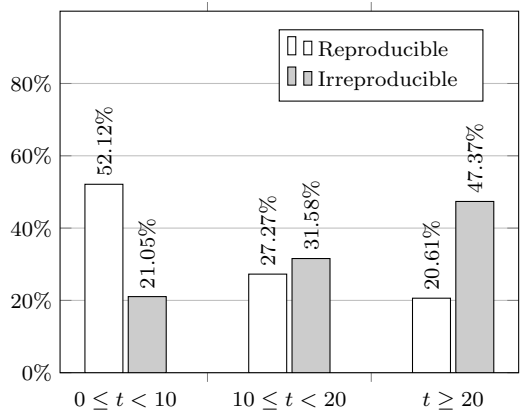
\begin{figure}
\centering
    \pgfplotstableread{
		1	64.68	38.89   52.12   21.05
		2	14.92	16.67   27.27   31.58
		3	20.40	44.44   20.61   47.37

	}\datatable
	\subfloat[Java]
	{
	\begin{tikzpicture}
	\begin{axis}[
	xtick=data,
	xticklabels={$0 \leq t<10$, $10 \leq t<20$, $t \geq 20$
	},
	enlarge y limits=false,
	enlarge x limits=0.2,
	ymin=0,ymax=100,
	ybar,
	ytick={0,20,...,80},
	yticklabels={0\%,20\%,40\%,60\%,80\%},
	bar width=0.5cm,
	width=3in,
	ylabel={Accepted Answer (\%)},
	height = 2.5in,
	ymajorgrids=true,
	xticklabel style={font=\small, /pgf/number format/fixed},	
	major x tick style = {opacity=0},
	minor x tick num = 1,    
	minor tick length=1ex,
	legend style={
		font=\small,
		cells={anchor=west},
		legend columns=1,
		at={(0.7,0.95)},
		anchor=north,
		text width=2cm,
		minimum height=0.4cm,
		/tikz/every even column/.append style={column sep=0.2cm}
	},
	nodes near coords style={rotate=90,  anchor=west}, 
    nodes near coords =\pgfmathprintnumber{\pgfplotspointmeta}\%
	]

	\addplot[draw=black!100, fill=black!0] table[x index=0,y index=1] \datatable;
	\addplot[draw=black!100, fill=black!20] table[x index=0,y index=2] \datatable;

    \legend	{Reproducible,
    		 Irreproducible			
    }
	\end{axis}
	\end{tikzpicture}
	}
	
	\subfloat[Python]
	{
	\begin{tikzpicture}
	\begin{axis}[
	xtick=data,
	xticklabels={$0 \leq t<10$, $10 \leq t<20$, $t \geq 20$
	},
	enlarge y limits=false,
	enlarge x limits=0.2,
	ymin=0,ymax=100,
	ybar,
	ytick={0,20,...,80},
	yticklabels={0\%,20\%,40\%,60\%,80\%},
	bar width=0.5cm,
	width=3in,
	height = 2.5in,
	ymajorgrids=true,
	xticklabel style={font=\small, /pgf/number format/fixed},	
	major x tick style = {opacity=0},
	minor x tick num = 1,    
	minor tick length=1ex,
	legend style={
		cells={anchor=west},
		legend columns=1,
		at={(0.7,0.95)},
		anchor=north,
		text width=2cm,
		minimum height=0.4cm,
		/tikz/every even column/.append style={column sep=0.2cm}
	},
	nodes near coords style={rotate=90,  anchor=west}, 
    nodes near coords =\pgfmathprintnumber{\pgfplotspointmeta}\%
	]

	\addplot[draw=black!100, fill=black!0] table[x index=0,y index=3] \datatable;
	\addplot[draw=black!100, fill=black!20] table[x index=0,y index=4] \datatable;

    \legend	{Reproducible,
    		 Irreproducible			
    }
	\end{axis}
	\end{tikzpicture}
	}
	\caption{Accepted answer fraction vs. the time delay between question and accepted answer submission}
	\label{fig:accepted-answer-ratio-vs-time-delay}
\end{figure}


\begin{table}[htb]
\centering
    \caption{Reproducibility Status vs. Time Delay in Receiving Accepted Answers of Questions}
	\label{table:relation-acc-ans-delay}
	\resizebox{4.5in}{!}{%
    \begin{tabular}{l|c|c|c|c|c|c} \toprule
    \textbf{Reproducibility} & \multicolumn{2}{c|}{\textbf{$ t < 10$}} & \multicolumn{2}{c|}{\textbf{$ t \geq 10$}} &  \multicolumn{2}{c}{\textbf{Total}} \\ 
                                                    \textbf{Status} & \textbf{Java}            & \textbf{Python}            & \textbf{Java}         & \textbf{Python}           & \textbf{Java}        & \textbf{Python} \\  \midrule
    \textbf{Reproducible}                            &  130 (64.7\%)           &  86 (52.1\%)              &  71 (35.3\%)         &  79 (47.9\%)             &   201                &   165         \\  \midrule
    \textbf{Irreproducible}                          &  7 (38.9\%)             &  4 (21.1\%)                  &  11 (61.1\%)         &  15 (78.9\%)                &   18                 &   20           \\ \bottomrule              
    \end{tabular}
    }
\end{table}



    
    

Although the above box plots demonstrate the benefits of issue reproducibility, we further classify the delay of getting accepted answers into three intervals: $0 \leq t<10$, $10 \leq t<20$, $t \geq 20$. Fig. \ref{fig:accepted-answer-ratio-vs-time-delay} shows the percentage of the accepted answers for reproducible and irreproducible issues against these intervals. For Java, we see that questions with reproducible issues receive about 65\% of their accepted answers within only 10 minutes. Such percentage is only 39\% for the questions with irreproducible issues. It also should be noted that 44\% of these questions require more than 20 minutes on average to receive the accepted answers. In Python, questions with reproducible issues receive about 52\% of their accepted answers within only 10 minutes. Such percentage is only 21\% for the questions with irreproducible issues. It also should be noted that 47\% of these questions require more than 20 minutes on average to receive the accepted answers. Table \ref{table:relation-acc-ans-delay} provides the raw statistics on the accepted answers and time interval. We examine the relation between these two categorical variables using a statistical test, namely \emph{Chi-Squared} test. Thus, given all the evidence above the tables and plots, issue reproducibility status is very likely to influence the time delay for getting the acceptable answers on Stack Overflow.

\begin{table}[htb]
\centering
    \caption{Issue Reproducibility Status vs. Presence of Answers}
	\label{table:relationship-reprostatus-answered-questions}
	\resizebox{4.5in}{!}{%
    \begin{tabular}{l|c|c|c|c|c|c} \toprule
   \multicolumn{1}{l|}{\textbf{Reproducibility}} & \multicolumn{2}{c|}{\textbf{Answered Question}} & \multicolumn{2}{c|}{\textbf{Unanswered Question}} &  \multicolumn{2}{c}{\textbf{Total}} \\ 
                \multicolumn{1}{l|}{\textbf{Status}} & \textbf{Java}    & \textbf{Python}          & \textbf{Java}         & \textbf{Python}           & \textbf{Java}      & \textbf{Python} \\  \midrule
    \textbf{Reproducible}                            &  264 (97.78\%)   &  283 (99.65\%)            &  6 (2.22\%)          &  1 (0.35\%)               &   270              &      284       \\  \midrule
    \textbf{Irreproducible}                          &  74 (85.06\%)    &  54 (72.97\%)             &  13 (14.94\%)        & 20 (27.03\%)              &   87               &      74          \\ \bottomrule              
    \end{tabular}
    }
\end{table}

\begin{figure}
	\centering
	\subfloat[Java]
	{
	\includegraphics[width=3in]{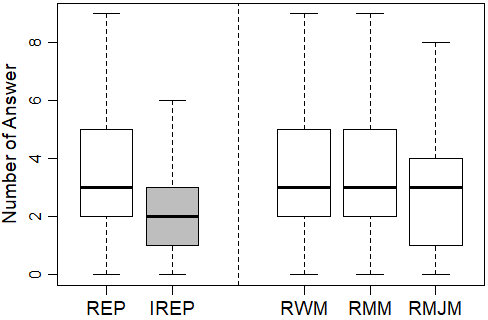}
	}
	
	\subfloat[Python]{
	\includegraphics[width=3in]{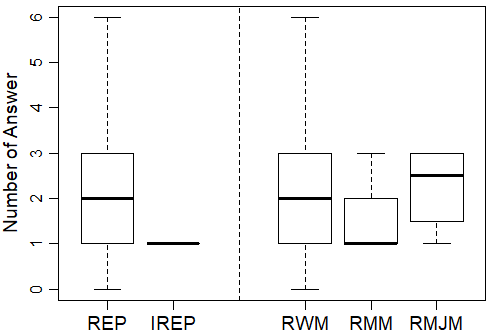}
	}

	\caption{Number of answers for questions with reproducible and irreproducible issues (\small{\textbf{REP}=Reproducible; \textbf{IREP}=Irreproducible; \textbf{RWM}=Reproducible without modification; \textbf{RMM}=Reproducible with Minor Modifications; \textbf{RMJM}=Reproducible with Major Modifications).}}
	\label{fig:box-plot-no-of-answer}
\end{figure}

\textbf{RQ$_3$(c): Does reproducibility of issues discussed in Stack Overflow questions encourage answers?}
According to RQ3a and RQ3b, the reproducibility of reported issues in Stack Overflow questions might encourage quick and high-quality responses from the users. In this section, we further investigate whether such reproducibility also encourages answers on Stack Overflow. 
First, we attempt to see whether the reproducibility of issues helps questions to be answered. Table \ref{table:relationship-reprostatus-answered-questions} shows the confusion matrix where the rows represent the reproducibility status (e.g.,  reproducible/irreproducible) and the columns represent the presence of answers (e.g., answered/unanswered). We see that 97.78\% (264 out of 270) Java and 99.65\% (283 out of 284) Python questions whose code segments can reproduce the reported issues receive one or more answers. Such percentages reduce to 85.06\% (74 out of 87) for Java and 72.97\% (54 out of 74) for Python when the code segments fail to reproduce the reported issues.  Thus, the reproducibility of issues reported in questions increases the chance of receiving answers. We then use the \emph{Chi-Squared} test to statistically measure the independence of these two categorical variables shown in Table \ref{table:relationship-reprostatus-answered-questions}. We find statistical significance \emph{p-value} ($p-value = 0 < 0.05$). Thus, there is a strong positive correlation between the issue reproducibility of questions and their chance of receiving answers. The percentage of answered questions is very high (e.g., 99.65\% for Python). Thus, we did not further analyze the correlation between the reproducibility status based on effort level and the presence of answers.

Next, we investigate whether such reproducibility encourages more answers.
Fig. \ref{fig:box-plot-no-of-answer} shows the box plots for the answer count of our selected questions against their issue reproducibility status. We clearly see that questions with issue reproducibility receive more answers on average than the counterpart.
We also find a significant difference in the number of answers between the two types of questions. 
\emph{Mann-Whitney-Wilcoxon} test shows \emph{$p-value = 0 < 0.05$} whereas  \emph{Cliff's  $|d|$ = 0.57 (large)} for Java questions and \emph{Cliff's  $|d|$ = 0.43 (medium)} for Python questions with $95\%$ confidence. Given these statistical findings, we suggest that the reproducibility status of a reported issue in a Stack Overflow question has a significant impact on its chance of getting more answers.

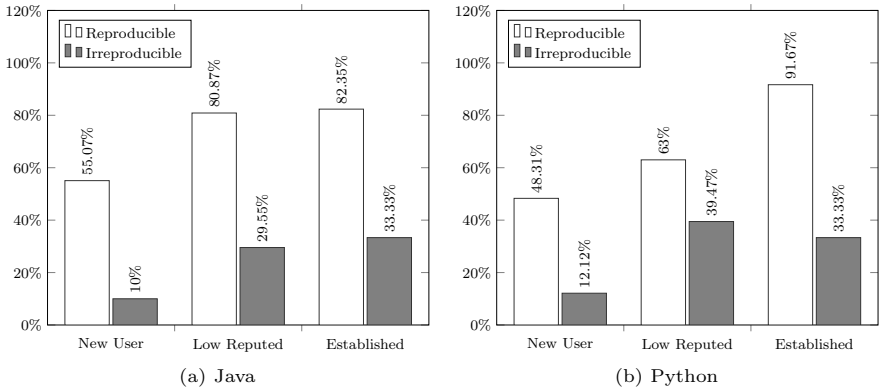
\begin{figure}[!htb]
\centering
    \pgfplotstableread{
    		1	55.07  10      48.31    12.12
    		2	80.87  29.55   63       39.47
    		3	82.35  33.33   91.67    33.33
    }\datatable
    \subfloat[Java]{
    \label{fig:compounding-variable-reputation-accepted-answer-java}
   	  \resizebox{2.25in}{!}{%
      \begin{tikzpicture}
        	\begin{axis}[
        	xtick=data,
        	xticklabels={New User, Low Reputed, Established, Trusted},
        	enlarge y limits=false,
        	enlarge x limits=0.25,
        	ymin=0,ymax=120,
        	ybar,
        	bar width=0.85cm,
        	width=3.5in,
        	height = 3in,
        	ytick={0,20,...,140},
            yticklabels={0\%,20\%,40\%,60\%,80\%,100\%,120\%,140\%},
        	ymajorgrids=false,
        	major x tick style = {opacity=0},
        	minor x tick num = 1,    
        	minor tick length=1ex,
        	legend style={
         	legend pos=north west,
        	legend cell align=left
            },
            nodes near coords style={rotate=90,  anchor=west}, 
        	nodes near coords =\pgfmathprintnumber{\pgfplotspointmeta}\%
        	]
        	\addplot[draw=black!80, fill=black!0] table[x index=0,y index=1] \datatable;
        	\addplot[draw=black!80, fill=black!50] table[x index=0,y index=2] \datatable;

            \legend	{Reproducible,
            		 Irreproducible
            		 }
        	\end{axis}
    	\end{tikzpicture}
    	}
    	}
      \subfloat[Python]{
      \label{fig:compounding-variable-reputation-accepted-answer-python}
   	  \resizebox{2.25in}{!}{%
      \begin{tikzpicture}
        	\begin{axis}[
        	xtick=data,
        	xticklabels={New User, Low Reputed, Established, Trusted},
        	enlarge y limits=false,
        	enlarge x limits=0.25,
            ymin=0,ymax=120,
        	ybar,
        	bar width=0.85cm,
        	width=3.5in,
        	height = 3in,
        	ytick={0,20,...,140},
            yticklabels={0\%,20\%,40\%,60\%,80\%,100\%,120\%,140\%},
        	ymajorgrids=false,
        	major x tick style = {opacity=0},
        	minor x tick num = 1,    
        	minor tick length=1ex,
        	legend style={
         	legend pos=north west,
        	legend cell align=left
            },
            nodes near coords style={rotate=90,  anchor=west}, 
        	nodes near coords =\pgfmathprintnumber{\pgfplotspointmeta}\%
        	]
        	\addplot[draw=black!80, fill=black!0] table[x index=0,y index=3] \datatable;
        	\addplot[draw=black!80, fill=black!50] table[x index=0,y index=4] \datatable;

            \legend	{Reproducible,
            		 Irreproducible
            		 }
        	\end{axis}
    	\end{tikzpicture}
    	}
    	}
\caption{Impact of question submitters' reputation in receiving accepted answers. \small{(\textbf{New user} (reputation score $< 10$); \textbf{Low Reputed user} ($10 \le$ reputation score $< 1K$); \textbf{Established user} ($1K \le$ reputation score $< 20K$)}.}
\label{fig:compounding-variable-reputation-accepted-answer}
\end{figure}

\subsection{Answering RQ\textsubscript4}
\label{subsec:rq4}

\textbf{What factors can affect questions receiving answers besides reproducibility of issues?}
In Section \ref{subsec:rq3}, we investigate the correlation between issue reproducibility and answer meta-data, such as the presence of an accepted answer. However, several factors can be associated with reproducibility that can affect questions receiving answers and hurt the correlation. In this section, we thus investigate three potential factors: \emph{user reputation, number of posts} and \emph{question submission time}, and determine their impact on questions receiving accepted answers, time delay in receiving accepted answers and number of answers.

\begin{table}[htb]
\centering
    \caption{Issue Reproducibility Status vs. Presence of Accepted Answer According to the Reputation of the Question Submitter \small{(\textbf{New user} (reputation score $< 10$); \textbf{Low Reputed user} ($10 \le$ reputation score $< 1K$); \textbf{Established user} ($1K \le$ reputation score $< 20K$); \textbf{Trusted user} (reputation score $\geq 20K$))}}
	\label{table:compounding-variable-reputation-accepted-answer}
	
	\subfloat[Java]{
	\label{table:compounding-variable-reputation-accepted-answer-java}
	\resizebox{4.5in}{!}{%
    \begin{tabular}{l|c|c|c|c}
    \toprule
    \multirow{2}{*}{\textbf{Repro. Status}} & \multicolumn{4}{c}{\textbf{User Category}} \\ 
                                            & \textbf{New}      & \textbf{Low Reputed} & \textbf{Established} & \textbf{Trusted}  \\ \midrule
    \textbf{Reproducible}                   &  38/69 (55.07\%)  & 148/183 (80.87\%)    & 14/17 (82.35\%)      & 1/1 (100\%)        \\ \midrule
    \textbf{Irreproducible}                 &  4/40 (10\%)      & 13/44 (29.55\%)      & 1/3 (33.33\%)        & --                \\ \bottomrule
    \end{tabular}
    }
    }
    
    \subfloat[Python]{
	\label{table:compounding-variable-reputation-accepted-answer-python}
	\resizebox{4.5in}{!}{%
   \begin{tabular}{l|c|c|c|c}
    \toprule
    \multirow{2}{*}{\textbf{Repro. Status}} & \multicolumn{4}{c}{\textbf{User Category}} \\ 
                                          & \textbf{New}       & \textbf{Low Reputed} & \textbf{Established} & \textbf{Trusted}  \\ \midrule
    \textbf{Reproducible}                 &  57/118 (48.31\%)  &  97/154 (63\%)       & 11/12 (91.67\%)      & --          \\ \midrule
    \textbf{Irreproducible}               &  4/33 (12.12\%)    &   15/38 (39.47\%)    & 1/3 (33.33\%)        & --          \\ \bottomrule
    \end{tabular}
    }
    }
    
\end{table}

\vspace{2mm}
\subsubsection{User Reputation}

Stack Overflow designs a reputation system of registered users to incentivize their contributions and to allow assessment from the users \citep{calefato2015mining, mondal2021early}. Several studies argue that questions submitted by users with a higher reputation are more likely to be answered and resolved \citep{asaduzzaman2013answering, Masud-InsightUnresolvedQuestionsSO-MSR2015}. We thus consider the users' reputation as a potential factor that could affect how questions receive accepted answers, the delay in receiving accepted answers, and the number of answers.
The Stack Overflow data dump only reports the latest reputation scores of the users, which might not be appropriate for our analysis. However, Stack Overflow stores all the user activities (e.g., votes, acceptances, bounties) required to estimate users' reputations. We thus use the snapshot of the activities of users to determine their reputation during a question submission. We used a standard equation provided by the Stack Overflow to determine reputation \citep{HDR-StackOverflow-2020}. We divide the users into four categories based on their reputation score \citep{Calefato-HowToAskForTechnicalHelp-IST2018}. They are -- \emph{New} user (score $< 10$), \emph{Low Reputed} user ($10 \le$ score $< 1K$), \emph{Established} user ($1K \le$ score $< 20K$) and \emph{Trusted} user (score $\geq 20K$). Then we investigate how user reputation affects questions in receiving accepted answers, time delay in receiving accepted answers and number of answers.
%
%

Fig. \ref{fig:compounding-variable-reputation-accepted-answer} shows how user reputation affects questions receiving accepted answers besides the issue reproducibility of questions. We see that questions submitted by users with a higher reputation (e.g., established users) have a higher chance of receiving accepted answers than those submitted by new users. 
For example, questions from Python with reproducible issues receive 91.67\% of accepted answers if they were submitted by established users, as opposed to 48.31\% for the new users.
However, the impact of reputation is less visible
to Java questions with reproducible issues when they were submitted by low reputed and established users. According to our analysis, the rate of receiving accepted answers is more than 80\% for both the user categories.
We also see that the rate of receiving accepted answers by questions with irreproducible issues is very close if submitted by either low reputed or established users. However, regardless of reputation, the chance of receiving accepted answers by questions with reproducible issues is \emph{consistently} higher than those with irreproducible issues. In fact, it is 2 to 5 times higher for the questions with reproducible issues, which indicates the importance of issue reproducibility.
For example, from Table \ref{table:compounding-variable-reputation-accepted-answer}, we see that there is an 80.87\% (148 out of 183) chance of getting an acceptable answer for Java questions submitted by low reputed users when the issues can be reproduced. On the contrary, the chance reduces to only 29.55\% (13 out of 44) when the submitted code cannot reproduce the reported issues. We did not consider the trusted user since a trusted user submitted only one Java question (e.g., Table \ref{table:compounding-variable-reputation-accepted-answer}). Unfortunately, there was no Python question in our dataset submitted by Trusted users.


\begin{table}[htb]
\centering
    \caption{Issue Reproducibility Status vs. Time Delay of Receiving Accepted Answer According to the Reputation of the Question Submitter \small{(\textbf{New user} (reputation score $< 10$); \textbf{Low Reputed user} ($10 \le$ reputation score $< 1K$); \textbf{Established user} ($1K \le$ reputation score $< 20K$); \textbf{Trusted user} (reputation score $\geq 20K$))}}
	\label{table:compounding-variable-reputation-time-delay}
	
	\subfloat[Java]{
	\label{table:compounding-variable-reputation-time-delay-java}
	\resizebox{4.5in}{!}{%
    \begin{tabular}{l|c|c|c|c|c|c|c|c}
    \toprule
    \multirow{3}{*}{\textbf{Repro. Status}} & \multicolumn{8}{c}{\textbf{User Category}}                                                                                                                                                                                                                                                               \\ 
                                            & \multicolumn{2}{c|}{\textbf{New}} & \multicolumn{2}{c|}{\textbf{Low Reputed}} & \multicolumn{2}{c|}{\textbf{Established}}  & \multicolumn{2}{c}{\textbf{Trusted}}                                        \\ 
                                            & \textbf{Avg}  & \textbf{Med} & \textbf{Avg} & \textbf{Med} & \textbf{Avg} & \textbf{Med} & \textbf{Avg} & \textbf{Med} \\ \midrule
                                            
    \textbf{Reproducible}                   & 17.03 & 11.50 & 9.30  & 5  & 11.57 & 5.5 & 4  & 4  \\ \midrule
    \textbf{Irreproducible}                 & 12.50 & 13.50 & 15.38 & 10 & 5     & 5   & -- & -- \\ \bottomrule
    \end{tabular}
    }
    }
    
    \subfloat[Python]{
	\label{table:compounding-variable-reputation-time-delay-python}
	\resizebox{4.5in}{!}{%
    \begin{tabular}{l|c|c|c|c|c|c|c|c}
    \toprule
    \multirow{3}{*}{\textbf{Repro. Status}} & \multicolumn{8}{c}{\textbf{User Category}}                                                                                                                                                                                                                                                               \\ 
                                            & \multicolumn{2}{c|}{\textbf{New}} & \multicolumn{2}{c|}{\textbf{Low Reputed}} & \multicolumn{2}{c|}{\textbf{Established}}  & \multicolumn{2}{c}{\textbf{Trusted}}                                        \\ 
                                             & \textbf{Avg}  & \textbf{Med} & \textbf{Avg} & \textbf{Med} & \textbf{Avg} & \textbf{Med} & \textbf{Avg} & \textbf{Med} \\ \midrule
                                            
    \textbf{Reproducible}                   & 11.40 & 7  & 10.70 & 7  & 6.45 & 3  & --  & -- \\ \midrule
    \textbf{Irreproducible}                 & 33.25 & 37 & 19.33 & 15 & 43   & 43 & -- & -- \\ \bottomrule
    \end{tabular}
    }
    }
    
\end{table}

Next, we attempt to see whether reputation affects the time delay of receiving accepted answers. Table \ref{table:compounding-variable-reputation-time-delay} shows the average and median time delay (in minutes) of receiving accepted answers for each user category. We see in many cases that questions submitted by highly reputed users receive accepted answers with a of minimum delay. For example, the median time delay of receiving accepted answers to Java questions with reproducible issues is five and half minutes when submitted by established users, but $11.50$ minutes for new users. However, the median time delay of receiving accepted answers for Python questions with reproducible issues is the same (i.e., seven minutes) for new and low reputed users (e.g., Table \ref{table:compounding-variable-reputation-time-delay-python}). Surprisingly, the median time delay of receiving accepted answers for Java questions is slightly higher for established users than low reputed users (e.g., Table \ref{table:compounding-variable-reputation-time-delay-java}).
However, more interestingly, the time delay is controlled by the reproducibility status of the reported issues. The time delay of receiving accepted answers to questions with reproducible issues is significantly lower than those with irreproducible issues. For example, for the low reputed user category, the median time delay for receiving accepted answers to Java questions is five minutes and seven minutes for Python questions whose code segments can reproduce the reported issues. On the contrary, such median time delays are as high as 10 and 15 minutes for the Java and Python questions, whose code segments fail to reproduce the issues. We see the only exception for the Java questions submitted by established users, where the time delay of receiving accepted answers is slightly higher for the questions with reproducible issues. However, in our dataset, the number of questions with the irreproducible issue submitted by established users is only one. Thus, their impact on our findings about time delay might be negligible.



\begin{table}[htb]
\centering
    \caption{Issue Reproducibility Status vs. Number of Answers According to the Reputation of the Question Submitter \small{(\textbf{New user} (reputation score $< 10$); \textbf{Low Reputed user} ($10 \le$ reputation score $< 1K$); \textbf{Established user} ($1K \le$ reputation score $< 20K$); \textbf{Trusted user} (reputation score $\geq 20K$))}}
	\label{table:compounding-variable-reputation-number-of-answers}
	
	\subfloat[Java]{
	\label{table:compounding-variable-reputation-number-of-answers-java}
	\resizebox{4.5in}{!}{%
    \begin{tabular}{l|c|c|c|c|c|c|c|c}
    \toprule
    \multirow{3}{*}{\textbf{Repro. Status}} & \multicolumn{8}{c}{\textbf{User Category}}                                                                                                                                                                                                                                                               \\ 
                                            & \multicolumn{2}{c|}{\textbf{New}} & \multicolumn{2}{c|}{\textbf{Low Reputed}} & \multicolumn{2}{c|}{\textbf{Established}}  & \multicolumn{2}{c}{\textbf{Trusted}}                                        \\ 
                                            & \textbf{Avg}  & \textbf{Med} & \textbf{Avg} & \textbf{Med} & \textbf{Avg} & \textbf{Med} & \textbf{Avg} & \textbf{Med} \\ \midrule
                                            
    \textbf{Reproducible}                   & 1.57 & 2 & 1.83 & 3 & 2.06 & 3 & 4 & 4  \\ \midrule
    \textbf{Irreproducible}                 & 2.25 & 2 & 2.66 & 2 & 0.67 & 1 & -- & -- \\ \bottomrule
    \end{tabular}
    }
    }
    
    \subfloat[Python]{
	\label{table:compounding-variable-reputation-number-of-answers-python}
	\resizebox{4.5in}{!}{%
    \begin{tabular}{l|c|c|c|c|c|c|c|c}
    \toprule
    \multirow{3}{*}{\textbf{Repro. Status}} & \multicolumn{8}{c}{\textbf{User Category}}                                                                                                                                                                                                                                                               \\ 
                                            & \multicolumn{2}{c|}{\textbf{New}} & \multicolumn{2}{c|}{\textbf{Low Reputed}} & \multicolumn{2}{c|}{\textbf{Established}}  & \multicolumn{2}{c}{\textbf{Trusted}}                                        \\ 
                                             & \textbf{Avg}  & \textbf{Med} & \textbf{Avg} & \textbf{Med} & \textbf{Avg} & \textbf{Med} & \textbf{Avg} & \textbf{Med} \\ \midrule
                                            
    \textbf{Reproducible}                   & 1.77 & 2  & 1.90 & 2 & 1.58 & 1.5 & --  & -- \\ \midrule
    \textbf{Irreproducible}                 & 1    & 1  & 1.23 & 1 & 1.33 & 1   & --  & --   \\ \bottomrule
    \end{tabular}
    }
    }
    
\end{table}

Finally, we then attempt to see whether reputation affects the number of answers besides the reproducibility of issues. As shown in Table \ref{table:compounding-variable-reputation-number-of-answers}, reputation has almost no impact on questions in receiving answers. For example, the median number of answers for Python questions with irreproducible issues is only one for all the user categories. The number of answers is even lower for the questions with reproducible issues submitted by the established users than new and low reputed users (e.g., Table \ref{table:compounding-variable-reputation-number-of-answers-java}). However, the number of answers to questions with reproducible issues is consistently higher than those with irreproducible issues regardless of reputation-based user categories.

\begin{figure}[!htb]
\centering
  \subfloat[Java]{
  \label{fig:compounding-variable-yearwiseposts-accepted-answer-java}
  \resizebox{3.25in}{!}{%
    \begin{tikzpicture}
    \begin{axis}[
        width=4.25in,
        height=3.25in,
        axis lines=left,
        grid=both,
        legend style={draw=none, font=\large},
        legend style={
        at={(1,-0.15)},
        legend cell align=left,
        legend columns= 1
        },
        unbounded coords=jump,
        label style={font=\small},
        ticklabel style={font=\small},
        ymin=0,
        xmin=0,
        ymax=1.1,
        xmax=12,
        xtick={1,2,3,4,5,6,7,8,9,10,11,12},
        xticklabels={2008,2009,2010,2011,2012,2013,2014,2015,2016,2017,2018,2019},
        ytick={0,0.1,...,1.1},
        xlabel=Years,
        legend entries={Percentage of Java posts,
        Percentage of acc. ans. (\emph{repro. status = reproducible}),
        Percentage of acc. ans. (\emph{repro. status = irreproducible})
        },
        cycle list name = auto 
    ]
    \addplot
    coordinates {(1,0.15) (2,0.21) (3,0.33) (4,0.39) (5,0.46) (6,0.52) (7,0.61) (8,0.60) (9,0.55) (10,0.50) (11,0.47) (12,0.42)};
    \addplot
    coordinates {(1,0.83) (2,0.79) (3,0.84) (4,0.90) (5,0.92) (6,0.78) (7,0.67) (8,0.67) (9,0.77) (10,0.61) (11,0.47) (12,nan)};
    \addplot
    coordinates {(1,1) (2,0.3) (3,0.14) (4,0) (5,0.25) (6,0.2) (7,0.17) (8,0.20) (9,0.11) (10,0.14) (11,0) (12,nan)};
    
    \end{axis}
    \end{tikzpicture}
    }
    }

  \subfloat[Python]{
  \label{fig:compounding-variable-yearwiseposts-accepted-answer-python}
  \resizebox{3.25in}{!}{%
    \begin{tikzpicture}
    \begin{axis}[
        width=4.25in,
        height=3.25in,
        axis lines=left,
        grid=both,
        legend style={draw=none, font=\large},
        legend cell align=left,
        legend style={
        at={(1,-0.15)},
        legend cell align=left
        },
        unbounded coords=jump,
        label style={font=\small},
        ticklabel style={font=\small},
        ymin=0,
        xmin=0,
        ymax=1.1,
        xmax=12,
        xtick={1,2,3,4,5,6,7,8,9,10,11,12},
        xticklabels={2008,2009,2010,2011,2012,2013,2014,2015,2016,2017,2018,2019},
        ytick={0,0.1,...,1.1},
        xlabel=Years,
        legend entries={Percentage of Python posts,
        Percentage of acc. ans. (\emph{repro. status = reproducible}),
        Percentage of acc. ans. (\emph{repro. status = irreproducible}),
        },
        cycle list name = auto 
    ]
    \addplot
    coordinates {(1,0) (2,0.05) (3,0.09) (4,0.08) (5,0.13) (6,0.21) (7,0.28) (8,0.36) (9,0.46) (10,0.64) (11,0.82) (12,1)};
    \addplot
    coordinates {(1,nan) (2,1) (3,0.8) (4,1) (5,0.79) (6,0.83) (7,0.59) (8,0.46) (9,0.50) (10,0.57) (11,0.59) (12,0.48)};
    \addplot
    coordinates {(1,nan) (2,nan) (3,1) (4,0.33) (5,0.50) (6,0.29) (7,0.67) (8,0.33) (9,0.3) (10,0.3) (11,0.08) (12,0.06)};
    
    \end{axis}
    \end{tikzpicture}
    }
    }
    
\caption{Issue Reproducibility Status vs. Presence of Accepted Answer According to the Number of Posts Over Years}
\label{fig:compounding-variable-yearwiseposts-accepted-answer}
\end{figure}
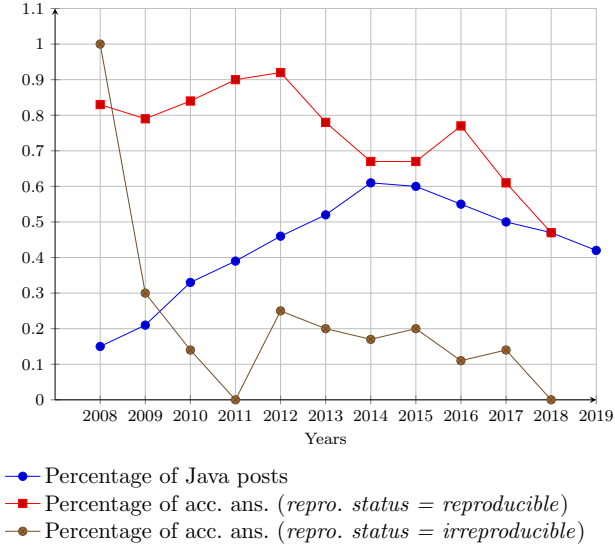
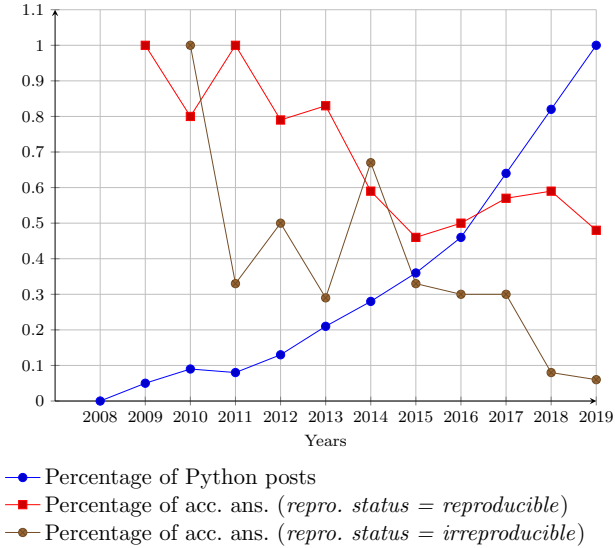

\subsubsection{Number of Posts} 

In Stack Overflow, the popularity of a discussed technology in questions might affect them how they
receive answers. For example, Python is becoming more popular than Java over the last few years. As a result, the Python community is growing and thus attracting more questions than Java in Stack Overflow. Therefore, we investigate whether the popularity of technology affects questions receiving answers. In particular, we compute the ratio of posts (e.g., questions and answers) in terms of the total number of posts related to Java and Python submitted in Stack Overflow from 2008 to 2019 and then normalized the ratio from 0 to 1. Next, we find how the post ratio affects the questions receiving answers.

Fig. \ref{fig:compounding-variable-yearwiseposts-accepted-answer} shows how posts ratio (i.e., number of posts) affects questions receiving accepted answers. We see that the number of posts does not significantly impact questions with reproducible issues receiving accepted answers over the first few years (e.g., from 2008 to 2013). For example, more than 80\% of questions receive acceptable answers in all those years. However, after 2014, the rate of accepted answers decreases for questions related to Java and Python, although Python posts continue to grow. Overall, we find a negative correlation between the number of posts and the rate of receiving accepted answers using \emph{Pearson correlation coefficient}. Such a correlation suggests that the rate of receiving accepted answers might decrease when the number of posts increases. However, from Fig. \ref{fig:compounding-variable-yearwiseposts-accepted-answer}, we see that the rate of receiving accepted answers for the questions whose code segments can reproduce the issues is significantly higher than those with irreproducible issues.


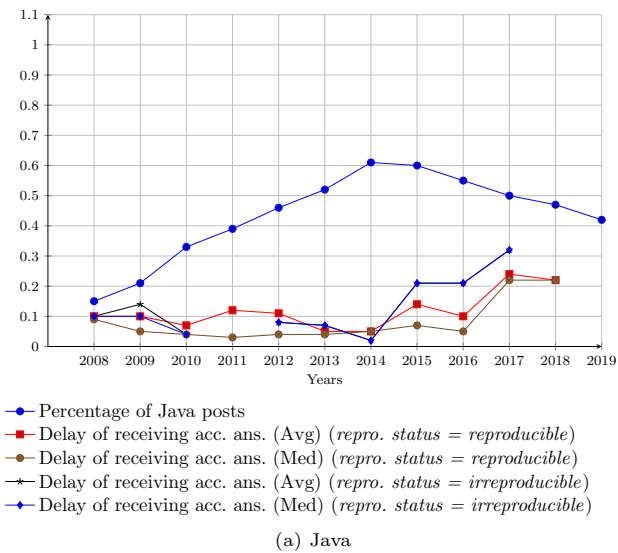
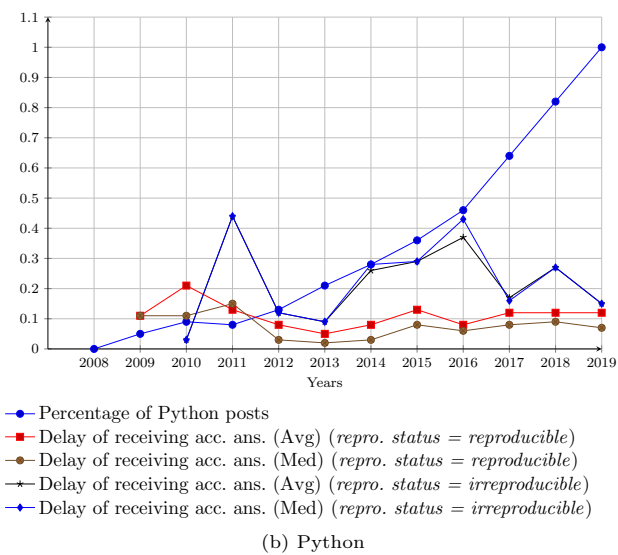
\begin{figure}[!htb]
\centering
  \subfloat[Java]{
  \label{fig:compounding-variable-yearwiseposts-delay-java}
  \resizebox{3.25in}{!}{%
    \begin{tikzpicture}
    \begin{axis}[
        width=5in,
        height=3.25in,
        axis lines=left,
        grid=both,
        legend style={draw=none, font=\large},
        legend style={
        at={(1,-0.15)},
        legend cell align=left,
        legend columns= 1
        },
        unbounded coords=jump,
        label style={font=\small},
        ticklabel style={font=\small},
        ymin=0,
        xmin=0,
        ymax=1.1,
        xmax=12,
        xtick={1,2,3,4,5,6,7,8,9,10,11,12},
        xticklabels={2008,2009,2010,2011,2012,2013,2014,2015,2016,2017,2018,2019},
        ytick={0,0.1,...,1.1},
        xlabel=Years,
        legend entries={Percentage of Java posts,
        Delay of receiving acc. ans. (Avg) (\emph{repro. status = reproducible}),
        Delay of receiving acc. ans. (Med) (\emph{repro. status = reproducible}),
        Delay of receiving acc. ans. (Avg) (\emph{repro. status = irreproducible}),
        Delay of receiving acc. ans. (Med) (\emph{repro. status = irreproducible})
        },
        cycle list name = auto 
    ]
    \addplot
    coordinates {(1,0.15) (2,0.21) (3,0.33) (4,0.39) (5,0.46) (6,0.52) (7,0.61) (8,0.60) (9,0.55) (10,0.50) (11,0.47) (12,0.42)};
     \addplot
    coordinates {(1,0.1) (2,0.1) (3,0.07) (4,0.12) (5,0.11) (6,0.05) (7,0.05) (8,0.14) (9,0.10) (10,0.24) (11,0.22) (12,nan)};
     \addplot
    coordinates {(1,0.09) (2,0.05) (3,0.04) (4,0.03) (5,0.04) (6,0.04) (7,0.05) (8,0.07) (9,0.05) (10,0.22) (11,0.22) (12,nan)};
     \addplot
    coordinates {(1,0.1) (2,0.14) (3,0.04) (4,nan) (5,0.08) (6,0.07) (7,0.02) (8,0.21) (9,0.21) (10,0.32) (11,nan) (12, nan)};
     \addplot
    coordinates {(1,0.1) (2,0.1) (3,0.04) (4,nan) (5,0.08) (6,0.07) (7,0.02) (8,0.21) (9,0.21) (10,0.32) (11,nan) (12,nan)};
    \end{axis}
    \end{tikzpicture}
    }
    }
    
  \subfloat[Python]{
  \label{fig:compounding-variable-yearwiseposts-delay-python}
  \resizebox{3.25in}{!}{%
    \begin{tikzpicture}
    \begin{axis}[
        width=5in,
        height=3.25in,
        axis lines=left,
        grid=both,
        legend style={draw=none, font=\large},
        legend cell align=left,
        legend style={
        at={(1,-0.15)},
        legend cell align=left
        },
        unbounded coords=jump,
        label style={font=\small},
        ticklabel style={font=\small},
        ymin=0,
        xmin=0,
        ymax=1.1,
        xmax=12,
        xtick={1,2,3,4,5,6,7,8,9,10,11,12},
        xticklabels={2008,2009,2010,2011,2012,2013,2014,2015,2016,2017,2018,2019},
        ytick={0,0.1,...,1.1},
        xlabel=Years,
        legend entries={Percentage of Python posts,
        Delay of receiving acc. ans. (Avg) (\emph{repro. status = reproducible}),
        Delay of receiving acc. ans. (Med) (\emph{repro. status = reproducible}),
        Delay of receiving acc. ans. (Avg) (\emph{repro. status = irreproducible}),
        Delay of receiving acc. ans. (Med) (\emph{repro. status = irreproducible})
        },
        cycle list name = auto 
    ]
    \addplot
    coordinates {(1,0) (2,0.05) (3,0.09) (4,0.08) (5,0.13) (6,0.21) (7,0.28) (8,0.36) (9,0.46) (10,0.64) (11,0.82) (12,1)};
    \addplot
    coordinates {(1,nan) (2,0.11) (3,0.21) (4,0.13) (5,0.08) (6,0.05) (7,0.08) (8,0.13) (9,0.08) (10,0.12) (11,0.12) (12,0.12)};
    \addplot
    coordinates {(1,nan) (2,0.11) (3,0.11) (4,0.15) (5,0.03) (6,0.02) (7,0.03) (8,0.08) (9,0.06) (10,0.08) (11,0.09) (12,0.07)};
    \addplot
    coordinates {(1,nan) (2,nan) (3,0.03) (4,0.44) (5,0.12) (6,0.09) (7,0.26) (8,0.29) (9,0.37) (10,0.17) (11,0.27) (12,0.15)};
    \addplot
    coordinates {(1,nan) (2,nan) (3,0.03) (4,0.44) (5,0.12) (6,0.09) (7,0.28) (8,0.29) (9,0.43) (10,0.16) (11,0.27) (12,0.15)};
    \end{axis}
    \end{tikzpicture}
    }
    }
    
\caption{Issue Reproducibility Status vs. Time Delay of Receiving Accepted Answers According to the Number of Posts Over Years}
\label{fig:compounding-variable-yearwiseposts-delay}
\end{figure}

Fig. \ref{fig:compounding-variable-yearwiseposts-delay} shows how the number of posts affects the time delay of receiving accepted answers. We see that number of posts does not affect the time delay in receiving acceptable answers to the Java questions with reproducible issues. However, there is a negative correlation between posts ratio and time delay of receiving accepted answers of the Python questions with reproducible issues. That is, question submitters may have to wait a bit more time to receive acceptable answers even after the code segments can reproduce the issue when more questions are submitted. However, the time delay in receiving accepted answers is still consistently less than the questions with irreproducible issues. For example, the median time delay for the Java questions with reproducible issues posted in 2009 was only five minutes. On the contrary, such delay was 10 minutes for questions with irreproducible issues. Thus, regardless of the number of posts, the reproducibility status of reported issues controls the time delay in receiving accepted answers.

\begin{figure}[!htb]
\centering
  \subfloat[Java]{
  \label{fig:compounding-variable-yearwiseposts-no-of-answers-java}
  \resizebox{3.25in}{!}{%
    \begin{tikzpicture}
    \begin{axis}[
        width=4.25in,
        height=3in,
        axis lines=left,
        grid=both,
        legend style={draw=none, font=\large},
        legend style={
        at={(1,-0.15)},
        legend cell align=left,
        legend columns= 1
        },
        unbounded coords=jump,
        label style={font=\small},
        ticklabel style={font=\small},
        ymin=0,
        xmin=0,
        ymax=1.1,
        xmax=12,
        xtick={1,2,3,4,5,6,7,8,9,10,11,12},
        xticklabels={2008,2009,2010,2011,2012,2013,2014,2015,2016,2017,2018,2019},
        ytick={0,0.1,...,1.1},
        xlabel=Years,
        legend entries={Percentage of Java posts,
        No. of ans. (Avg) (\emph{repro. status = reproducible}),
        No. of ans. (Med) (\emph{repro. status = reproducible}),
        No. of ans. (Avg) (\emph{repro. status = irreproducible}),
        No. of ans. (Med) (\emph{repro. status = irreproducible})
        },
        cycle list name = auto 
    ]
    \addplot
    coordinates {(1,0.15) (2,0.21) (3,0.33) (4,0.39) (5,0.46) (6,0.52) (7,0.61) (8,0.60) (9,0.55) (10,0.50) (11,0.47) (12,0.42)};
     \addplot
    coordinates {(1,0.08) (2,0.02) (3,0.02) (4,0.02) (5,0.02) (6,0.01) (7,0.02) (8,0.02) (9,0.02) (10,0.01) (11,0.02) (12,nan)};
     \addplot
    coordinates {(1,0.06) (2,0.04) (3,0.05) (4,0.06) (5,0.03) (6,0.02) (7,0.02) (8,0.03) (9,0.02) (10,0.02) (11,0.02) (12,nan)};
     \addplot
    coordinates {(1,0.04) (2,0.04) (3,0.03) (4,0.02) (5,0.02) (6,0.02) (7,0.01) (8,0.01) (9,0.02) (10,0.01) (11,0.01) (12,nan)};
     \addplot
    coordinates {(1,0.04) (2,0.04) (3,0.01) (4,0.02) (5,0.01) (6,0.02) (7,0.01) (8,0.01) (9,0.01) (10,0.01) (11,0.01) (12,nan)};
    
    \end{axis}
    \end{tikzpicture}
    }
    }
    
  \subfloat[Python]{
  \label{fig:compounding-variable-yearwiseposts-no-of-answers-python}
  \resizebox{3.25in}{!}{%
    \begin{tikzpicture}
    \begin{axis}[
        width=4.25in,
        height=3in,
        axis lines=left,
        grid=both,
        legend style={draw=none, font=\large},
        legend cell align=left,
        legend style={
        at={(1,-0.15)},
        legend cell align=left
        },
        unbounded coords=jump,
        label style={font=\small},
        ticklabel style={font=\small},
        ymin=0,
        xmin=0,
        ymax=1.1,
        xmax=12,
        xtick={1,2,3,4,5,6,7,8,9,10,11,12},
        xticklabels={2008,2009,2010,2011,2012,2013,2014,2015,2016,2017,2018,2019},
        ytick={0,0.1,...,1.1},
        xlabel=Years,
        legend entries={Percentage of Python posts,
        No. of ans. (Avg) (\emph{repro. status = reproducible}),
        No. of ans. (Med) (\emph{repro. status = reproducible}),
        No. of ans. (Avg) (\emph{repro. status = irreproducible}),
        No. of ans. (Med) (\emph{repro. status = irreproducible})
        },
        cycle list name = auto 
    ]
    \addplot
    coordinates {(1,0) (2,0.05) (3,0.09) (4,0.08) (5,0.13) (6,0.21) (7,0.28) (8,0.36) (9,0.46) (10,0.64) (11,0.82) (12,1)};
    \addplot
    coordinates {(1,nan) (2,0.04) (3,0.02) (4,0.03) (5,0.02) (6,0.02) (7,0.02) (8,0.02) (9,0.02) (10,0.02) (11,0.02) (12,0.01)};
    \addplot
    coordinates {(1,nan) (2,0.04) (3,0.03) (4,0.02) (5,0.02) (6,0.02) (7,0.02) (8,0.02) (9,0.01) (10,0.02) (11,0.02) (12,0.01)};
    \addplot
    coordinates {(1,nan) (2,nan) (3,0.04) (4,0.03) (5,0.02) (6,0.01) (7,0.02) (8,0.01) (9,0.01) (10,0.01) (11,0.01) (12,0.01)};
    \addplot
    coordinates {(1,nan) (2,nan) (3,0.04) (4,0.03) (5,0.02) (6,0.01) (7,0.01) (8,0.01) (9,0.01) (10,0.01) (11,0.01) (12,0)};
    
    \end{axis}
    \end{tikzpicture}
    }
    }
    
\caption{Issue Reproducibility Status vs. Number of Answers According to the Number of Posts Over Years}
\label{fig:compounding-variable-yearwiseposts-no-of-answers}
\end{figure}

We then analyze the impact of the number of posts on questions receiving answers. According to our analysis, the number of posts does not significantly impact the number of Java answers for the first few years. Later, the number of answers decreases for questions with reproducible/irreproducible issues when the number of posts increases (i.e., negative correlation). However, we see that questions with reproducible issues encourage more answers than those of irreproducible issues (e.g., Fig. \ref{fig:compounding-variable-yearwiseposts-no-of-answers}). For example, the median number of answers to the Java questions with reproducible issues posted in 2011 was six as opposed to only two for the questions with irreproducible issues.

\begin{figure}[!htb]
\centering
    \pgfplotstableread{
    		1	73.80  22.03   57.89    35.71
    		2	75.90  17.86   58.51    15.63
    		3	75.36  19.12   59.36    29.63
    		4	71.42  26.31   53.85    20
    }\datatable
    \subfloat[Java]{
    \label{fig:compounding-variable-question-submission-time-accepted-answer-java}
   	  \resizebox{2.25in}{!}{%
      \begin{tikzpicture}
        	\begin{axis}[
        	xtick=data,
        	xticklabels={Day, Night, Weekday, Weekend},
        	enlarge y limits=false,
        	enlarge x limits=0.2,
        	ymin=0,ymax=120,
        	ybar,
        	bar width=0.8cm,
        	width=4in,
        	height = 3in,
        	ytick={0,20,...,140},
            yticklabels={0\%,20\%,40\%,60\%,80\%,100\%,120\%,140\%},
        	ymajorgrids=false,
        	major x tick style = {opacity=0},
        	minor x tick num = 1,    
        	minor tick length=1ex,
        	legend style={
         	legend pos=north west,
        	legend cell align=left
            },
            nodes near coords style={rotate=90,  anchor=west}, 
        	nodes near coords =\pgfmathprintnumber{\pgfplotspointmeta}\%
        	]
        	\addplot[draw=black!80, fill=black!0] table[x index=0,y index=1] \datatable;
        	\addplot[draw=black!80, fill=black!50] table[x index=0,y index=2] \datatable;

            \legend	{Reproducible,
            		 Irreproducible
            		 }
        	\end{axis}
    	\end{tikzpicture}
    	}
    	}
      \subfloat[Python]{
      \label{fig:compounding-variable-question-submission-time-accepted-answer-python}
   	  \resizebox{2.25in}{!}{%
      \begin{tikzpicture}
        	\begin{axis}[
        	xtick=data,
        	xticklabels={Day, Night, Weekday, Weekend},
        	enlarge y limits=false,
        	enlarge x limits=0.2,
        	ymin=0,ymax=120,
        	ybar,
        	bar width=0.8cm,
        	width=4in,
        	height = 3in,
        	ytick={0,20,...,140},
            yticklabels={0\%,20\%,40\%,60\%,80\%,100\%,120\%,140\%},
        	ymajorgrids=false,
        	major x tick style = {opacity=0},
        	minor x tick num = 1,    
        	minor tick length=1ex,
        	legend style={
         	legend pos=north west,
        	legend cell align=left
            },
            nodes near coords style={rotate=90,  anchor=west}, 
        	nodes near coords =\pgfmathprintnumber{\pgfplotspointmeta}\%
        	]
        	\addplot[draw=black!80, fill=black!0] table[x index=0,y index=3] \datatable;
        	\addplot[draw=black!80, fill=black!50] table[x index=0,y index=4] \datatable;

            \legend	{Reproducible,
            		 Irreproducible
            		 }
        	\end{axis}
    	\end{tikzpicture}
    	}
    	}
\caption{Impact of question submission time in receiving accepted answers.}
\label{fig:compounding-variable-question-submission-time-accepted-answer}
\end{figure}
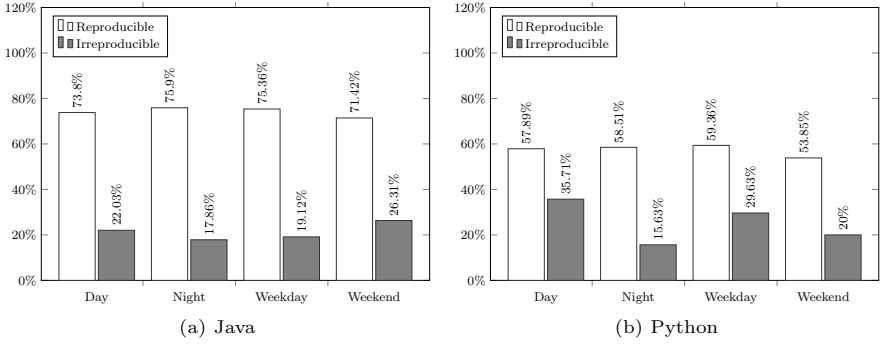

\subsubsection{Question Submission Time} 

Several studies suggest that question submission time might affect questions receiving answers \citep{bosu2013building,calefato2018ask}. We thus investigate the impact of the question submission time on questions receiving answers besides reproducibility of issues. We first convert the question submission time to Universal Time Coordinated (UTC) and then divide the submission time into four time frames based on working hours and day. They are - \emph{day}, \emph{night}, \emph{weekday} and \emph{weekend}. Then we attempt to see how such submission time frames affect questions receiving their accepted answers, the time delay of receiving accepted answers, and the number of answers.

\begin{table}[htb]
\centering
    \caption{Issue Reproducibility Status vs. Presence of Accepted Answer According to the Question Submission Time}
	\label{table:compounding-variable-question-submission-time-accepted-answer}
	
	\subfloat[Java]{
	\label{table:compounding-variable-question-submission-time-accepted-answer-java}
	\resizebox{4.5in}{!}{%
    \begin{tabular}{l|c|c|c|c}
    \toprule
    \multirow{2}{*}{\textbf{Repro. Status}} & \multicolumn{2}{c|}{\textbf{Working Hour}}  & \multicolumn{2}{c}{\textbf{Working Day}} \\ 
                                            & \textbf{Day}      & \textbf{Night} & \textbf{Weekday} & \textbf{Weekend}  \\ \midrule
    \textbf{Reproducible}                   & 138/187 (73.80\%) & 63/83 (75.90\%) & 156/207 (75.36\%) & 45/63 (71.42\%)       \\ \midrule
    \textbf{Irreproducible}                 & 13/59 (22.03\%)   & 5/28 (17.86\%)  & 13/68 (19.12\%) & 5/19 (26.31\%)           \\ \bottomrule
    \end{tabular}
    }
    }
    
    \subfloat[Python]{
	\label{table:compounding-variable-question-submission-time-accepted-answer-python}
	\resizebox{4.5in}{!}{%
   \begin{tabular}{l|c|c|c|c}
    \toprule
    \multirow{2}{*}{\textbf{Repro. Status}} & \multicolumn{2}{c|}{\textbf{Working Hour}}  & \multicolumn{2}{c}{\textbf{Working Day}} \\ 
                                          & \textbf{Day}      & \textbf{Night} & \textbf{Weekday} & \textbf{Weekend}  \\ \midrule
    \textbf{Reproducible}                 & 110/190 (57.89\%) & 55/94 (58.51\%)& 130/219 (59.36\%) & 35/65 (53.85\%)    \\ \midrule
    \textbf{Irreproducible}               & 15/42 (35.71\%)   & 5/32 (15.63\%) & 16/54 (29.63\%)   & 4/20 (20\%)   \\ \bottomrule
    \end{tabular}
    }
    }
    
\end{table}

Fig. \ref{fig:compounding-variable-question-submission-time-accepted-answer} shows the rate of receiving accepted answers for different time slots. We see that question submission time does not affect questions significantly in receiving acceptable answers. For example, the percentage of the accepted answers for the Java questions with reproducible issues is about 71\%--75\% (e.g., Fig. \ref{fig:compounding-variable-question-submission-time-accepted-answer-java}). Such percentage is about 53\%--59\% for questions related to Python (e.g., Fig. \ref{fig:compounding-variable-question-submission-time-accepted-answer-python}). However, submission time negatively impacts receiving acceptable answers if questions are submitted at night, and their code segments could not reproduce the issues. This is because the experts could be less active in Stack Overflow during the night than daytime. Anyway, regardless of the submission time, the main difference in receiving accepted answers is due to the reproducibility status of issues. According to our investigation, questions with reproducible issues receive about $2-5$ times higher chance of receiving accepted answers (e.g., Table. \ref{table:compounding-variable-question-submission-time-accepted-answer}). For example, 76\% of Java questions with reproducible issues that were submitted at daytime receive accepted answers. On the contrary, such a rate is only about 18\% when the code segments included with questions fail to reproduce the issues.

\begin{table}[htb]
\centering
    \caption{Issue Reproducibility Status vs. Time Delay of Receiving Accepted Answer According to the Question Submission Time}
	\label{table:compounding-variable-question-submission-time-delay}
	
	\subfloat[Java]{
	\label{table:compounding-variable-question-submission-time-delay-java}
	\resizebox{4.5in}{!}{%
    \begin{tabular}{l|c|c|c|c|c|c|c|c}
    \toprule
    \multirow{3}{*}{\textbf{Repro. Status}} & \multicolumn{4}{c|}{\textbf{Working Hour}} & \multicolumn{4}{c}{\textbf{Working Day}}                                                                                                                                                                                                                                                               \\ 
                                            & \multicolumn{2}{c|}{\textbf{Day}} & \multicolumn{2}{c|}{\textbf{Night}} & \multicolumn{2}{c|}{\textbf{Weekday}}  & \multicolumn{2}{c}{\textbf{Weekend}}                                        \\ 
                                            & \textbf{Avg}  & \textbf{Med} & \textbf{Avg} & \textbf{Med} & \textbf{Avg} & \textbf{Med} & \textbf{Avg} & \textbf{Med} \\ \midrule
                                            
    \textbf{Reproducible}                   & 10.08 & 5  & 12.67 & 7 & 10.20 & 5  & 13.29 & 7  \\ \midrule
    \textbf{Irreproducible}                 & 13.84 & 10 & 15    & 7 & 12.38 & 10 & 18.80 & 11 \\ \bottomrule
    \end{tabular}
    }
    }
    
    \subfloat[Python]{
	\label{table:compounding-variable-question-submission-time-delay-python}
	\resizebox{4.5in}{!}{%
    \begin{tabular}{l|c|c|c|c|c|c|c|c}
    \toprule
    \multirow{3}{*}{\textbf{Repro. Status}} & \multicolumn{4}{c|}{\textbf{Working Hour}} & \multicolumn{4}{c}{\textbf{Working Day}}                                                                                                                                                                                                                                                             \\ 
                                            & \multicolumn{2}{c|}{\textbf{Day}} & \multicolumn{2}{c|}{\textbf{Night}} & \multicolumn{2}{c|}{\textbf{Weekday}}  & \multicolumn{2}{c}{\textbf{Weekend}}                                        \\ 
                                             & \textbf{Avg}  & \textbf{Med} & \textbf{Avg} & \textbf{Med} & \textbf{Avg} & \textbf{Med} & \textbf{Avg} & \textbf{Med} \\ \midrule
                                            
    \textbf{Reproducible}                   & 10.42 & 6.5 & 11.15 & 8  & 10.56 & 7  & 11.03 & 6 \\ \midrule
    \textbf{Irreproducible}                 & 18.8  & 15  & 36.8  & 43 & 23.63 & 20 & 22 & 20   \\ \bottomrule
    \end{tabular}
    }
    }
    
\end{table}

\begin{table}[htb]
\centering
    \caption{Issue Reproducibility Status vs. Number of Answers According to the Question Submission Time}
	\label{table:compounding-variable-question-submission-number-of-answers}
	
	\subfloat[Java]{
	\label{table:compounding-variable-question-submission-number-of-answers-java}
	\resizebox{4.5in}{!}{%
    \begin{tabular}{l|c|c|c|c|c|c|c|c}
    \toprule
    \multirow{3}{*}{\textbf{Repro. Status}} & \multicolumn{4}{c|}{\textbf{Working Hour}} & \multicolumn{4}{c}{\textbf{Working Day}}                                                                                                                                                                                                                                                               \\ 
                                            & \multicolumn{2}{c|}{\textbf{Day}} & \multicolumn{2}{c|}{\textbf{Night}} & \multicolumn{2}{c|}{\textbf{Weekday}}  & \multicolumn{2}{c}{\textbf{Weekend}}                                        \\ 
                                            & \textbf{Avg}  & \textbf{Med} & \textbf{Avg} & \textbf{Med} & \textbf{Avg} & \textbf{Med} & \textbf{Avg} & \textbf{Med} \\ \midrule
                                            
    \textbf{Reproducible}                   & 1.81 & 3 & 1.67 & 3 & 1.85 & 3 & 1.52 & 2  \\ \midrule
    \textbf{Irreproducible}                 & 2.42 & 1 & 2.36 & 2 & 2.51 & 2 & 2 & 1 \\ \bottomrule
    \end{tabular}
    }
    }
    
    \subfloat[Python]{
	\label{table:compounding-variable-question-submission-number-of-answers-python}
	\resizebox{4.5in}{!}{%
    \begin{tabular}{l|c|c|c|c|c|c|c|c}
    \toprule
    \multirow{3}{*}{\textbf{Repro. Status}} & \multicolumn{4}{c|}{\textbf{Working Hour}} & \multicolumn{4}{c}{\textbf{Working Day}}                                                                                                                                                                                                                                                             \\ 
                                            & \multicolumn{2}{c|}{\textbf{Day}} & \multicolumn{2}{c|}{\textbf{Night}} & \multicolumn{2}{c|}{\textbf{Weekday}}  & \multicolumn{2}{c}{\textbf{Weekend}}                                        \\ 
                                             & \textbf{Avg}  & \textbf{Med} & \textbf{Avg} & \textbf{Med} & \textbf{Avg} & \textbf{Med} & \textbf{Avg} & \textbf{Med} \\ \midrule
                                            
    \textbf{Reproducible}                   & 1.83 & 2 & 1.84 & 2 & 1.85 & 2 & 1.78 & 1 \\ \midrule
    \textbf{Irreproducible}                 & 1.31 & 1 & 0.91 & 1 & 1.13 & 1 & 1.15 & 1  \\ \bottomrule
    \end{tabular}
    }
    }
    
\end{table}

Next, we investigate how the time frames affect the delay between question submission and receiving accepted answers. According to our analysis, developers not only receive less accepted answers but also have to wait for more to receive them if they submit their questions at night. For example, the median time delay of receiving acceptable answers to Python questions with reproducible issues is six and half minutes if they were submitted at daytime (e.g., Fig. \ref{table:compounding-variable-question-submission-time-delay-python}). On the contrary, such delay is eight minutes during the night. Surprisingly, the delay in receiving accepted answers to Python questions is as high as 15 minutes in the daytime and 43 minutes at night when the issues reported at questions cannot be reproduced. From Table \ref{table:compounding-variable-question-submission-time-delay}, we see that the time delay in receiving accepted answers is significantly less for the questions with reproducible issues, no matter when the question is submitted.

Finally, we investigate whether the question submission time affects questions receiving answers. Table \ref{table:compounding-variable-question-submission-number-of-answers} shows issue reproducibility status vs. the number of answers based on the question submission time. We see that the impact of questions submission time on the number of answers is negligible. As shown in Table \ref{table:compounding-variable-question-submission-number-of-answers}, the median number of answers is very close whenever the questions were submitted. For example, the median number of answers is three for Java and two for Python for the questions with reproducible issues. However, the difference in receiving answers to questions is mainly due to the reproducibility status. For example, the median number of the answer is three for Java questions with reproducible issues if they were submitted during the daytime. However, such a number reduces to only one when the issues reported in questions cannot be reproduced. Thus, reproducibility of issues encourages more answers to questions regardless of their submission time.

\section{Key Findings \& Discussion}
\label{discussion}

\subsection{Findings of Issue Reproducibility: Java vs Python}

We compare the findings of reproducibility of issues between two popular programming languages -- Java and Python, where Java is statically typed, and Python is a dynamically typed programming language. In particular, we attempt to see whether the correlation between reproducibility of issues and answer meta-data for both the programming languages is consistent. 


\begin{table}[htb]
\centering
    \caption{Issue Reproducibility Status vs. Answer Meta-data}
	\label{table:overall-findings}
	\resizebox{4.65in}{!}{%
    \begin{tabular}{p{3cm}|p{2.5cm}|p{2.5cm}|p{2.5cm}} \toprule
    \multicolumn{1}{c|}{\multirow{2}{*}{\textbf{Dimension}}}        & \multicolumn{3}{c}{\textbf{Findings}}                                                                          \\ 
    & \multicolumn{1}{c|}{\textbf{Java}} & \multicolumn{1}{c|}{\textbf{Python}} & \multicolumn{1}{c}{\textbf{Overall}} \\ \midrule
    
    \textbf{Accepted Answers}    & \textbf{More than three times higher} for the questions with reproducible issues that the questions with irreproducible issues. &   \textbf{More than two times higher} for the questions with reproducible issues that the questions with irreproducible issues. & Question submitters could enhance the chance of getting acceptable answers to their questions \textbf{at least two times more} if their code segments included with the questions can reproduce the questions' issues. \\ \midrule
    
    \textbf{Delay in Receiving Accepted Answer} & The question submitters could get acceptable answers to their questions with reproducible issues \textbf{two times faster} than the questions with irreproducible issues. & We get a similar time delay for Python questions, like Java. & Issue reproducibility success makes a question receiving acceptable answers \textbf{two times faster} than the questions with irreproducible issues. \\ \midrule
    
    \textbf{Number of Answers} & Questions with reproducible issues receive \textbf{one and half times more} answers than the questions with irreproducible issues. & Questions with reproducible issues receive \textbf{two times more} answers than the questions with irreproducible issues. &  Reproducibility of questions' issues helps to receive \textbf{at least one and a half times more} answers than the questions with irreproducible issues. \\ \bottomrule
    \end{tabular}
    }
\end{table}

Table \ref{table:overall-findings} summarizes the correlation between questions with reproducible and irreproducible issues and answer meta-data. We see that reproducibility of issues reported in questions significantly -- (1) increases the chance of receiving acceptable answers, (2) decreases the delay of receiving such acceptable answers, and also (3) increases the number of answers for both the programming languages -- Java and Python. Such findings establish issue reproducibility as a general question quality paradigm of Stack Overflow. However, about two-thirds (e.g., 78\%) of issues discussed in Python questions can be reproduced without any modification of the given code segments. Such a percentage is about 32\% for Java code segments. Nevertheless, the rate of receiving accepted answers to Python questions is lower than those of Java questions. We thus further investigate such discrepancies between Java and Python and attempt to expose the causes as follows.

\textbf{Python is dynamic and interpreted.} Users often include incomplete code segments with questions while discussing a programming-related issue. However, to reproduce the reported issues, first, we had to compile/execute the code segments. Since Python is a dynamic and an interpreted language, the code segments of Python can be executed even if they consist of a couple of statements. On the contrary, adding a container class and the main method was often essential for Java statements to compile/execute them. Thus, the majority of Java code segments (about 47\%) require minor modifications, whereas Python code segments require no such modifications to reproduce the issues.

\textbf{New users submitted more Python questions than Java.} From Section \ref{subsec:rq4}, we see that questions submitted by \emph{new} users (reputation score $< 10$) have a lower chance of receiving accepted answers even if their issues can be reproduced. Since Python is a promising language, easy to use and learn, it attracts more beginners and newcomers. In our dataset, 53.17\% of Python questions (151 out of 284) whose code segments can reproduce the issues were submitted by new users. However, their rate of receiving accepted answers is below 50\%. On the contrary, the percentage of Java questions submitted by new users was only 25.56\%. Thus, overall, the rate of receiving accepted answers to Java questions is higher than those of Python questions, even if their issues can be reproduced.

\textbf{More posts might reduce the chance of receiving accepted answers.} From Section \ref{subsec:rq4}, we find a negative correlation between the number of posts and the rate of receiving accepted answers. In particular, the rate of accepted answers declined significantly after 2014 when Python posts grew exponentially.
Our dataset contains 73.60\% of Python questions (209 out of 284) with reproducible issues that were submitted after 2014. However, at that time, the rate of accepted answers reduced to about 52\%. On the other hand, only 40\% of Java questions (108 out of 270) were submitted after 2014, and thus, their rate of getting accepted answers was affected less than Python.

\subsection{Key Findings \& Guidelines}
\label{keyFindings}

When submitting a question to Stack Overflow, it is recommended to add a bit of a code fragment so that the reported program issues can be reproduced easily \citep{squire2014bit}. Expert users of Stack Overflow also suggest adding complete and standalone code in the question \citep{JonSkeet2010}. However, our study delves further into the submitted code, and delivers more in-depth insights on the question of issue reproducibility using such code. 

\begin{inparaenum}[(a)]
    
    \item \textbf{Redundancy Hurts.} Only those statements that are required to reproduce the reported issue should be added, and redundant code should be avoided. Long and redundant code unnecessarily wastes the developer's time, which might also hurt the question's chance of getting an accepted answer. During the manual analysis of Java questions, we find comments from answerers against the long redundant code and suggested removing such redundant statements. For example, users commented against the inclusion of a long redundant code (e.g., $LOC = 86$) that the code was hard to read and understand because of its redundancy (see Fig. \ref{fig:exam-redundant-code}).
    
    \item \textbf{Dependency Matters.} Including  import statements for the external libraries is very important. This is one of the major difficulties that we faced during the reproduction of issues reported in questions with the submitted code. For example, missing external libraries was the third major challenge for Java that prevents reproducibility of issues. The question text also should point to the external libraries (if used) so that the users can include them in the IDE during issue reproduction.
    
    \item \textbf{Executable Code for Debugging.} The submitted code segment should compile and run if the reported issue requires debugging to reproduce. This is especially required when the program shows stochastic or unexpected behavior. Without a reproducible code segment, such questions are generally hard to answer effectively.
    
    
    \item \textbf{Inconsistent Problem Description Confusion.} Inconsistency between the problem description and the problematic code segment often confuses the users. The description of a question should specify the programming problem accurately. Sometimes questions describe issues that are different from the issues found when investigating the code segments. Such a situation confuses the users in deciding whether to consider the description or the issue found in the code segment. Our dataset contains 10.75\% (43 out of 400) Java and 10.50\% (42 out of 400) Python questions that discuss the issues inaccurately/inappropriately. According to our investigation, such inconsistent (e.g., inaccurate) discussion prevents reproducibility and discourages acceptable answers. Thus, the problem description must be consistent with the problem of the submitted code segment.

    \begin{figure}[!htb]
 	\centering
	\includegraphics[width=4.5in]{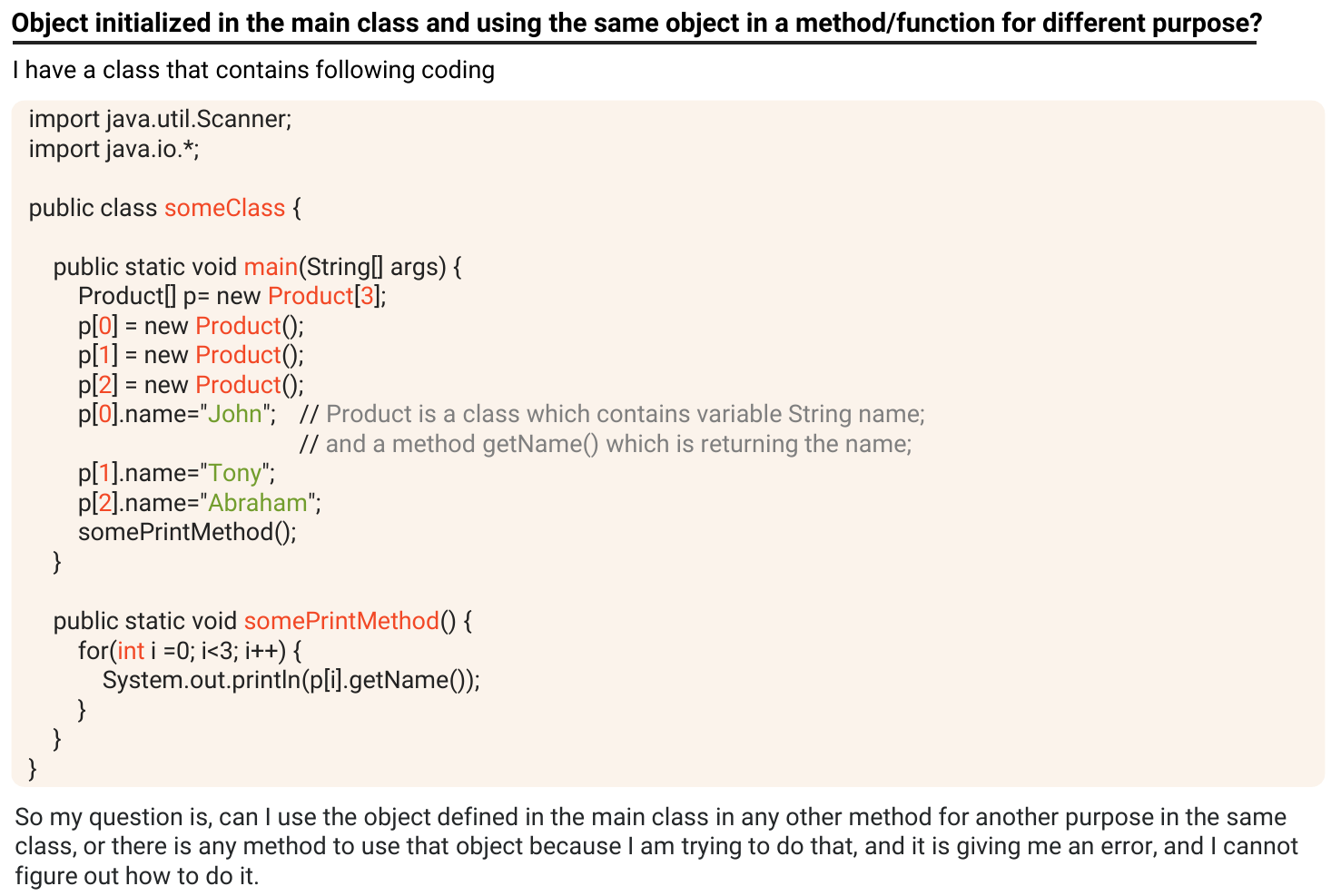}

	\caption{An example question where code comments help define a missing class to reproduce the reported issue.  (\url{https://stackoverflow.com/questions/17779601}).}
	\label{fig:exam-code-comment}
    \end{figure}
    
    \item \textbf{Code Comments Support Readability and Reproducibility.} Code comments not only enhance code readability but also sometimes support the issue description of questions. They help users identify the potential part of code where the reported issue might occur. They also hint at external dependencies (e.g., libraries, API, class) that assist often users in making the code segment complete, executable, and reproducible. Fig. \ref{fig:exam-code-comment} shows an example where the class Product was missing, which was required to execute the code and reproduce the reported issue. Interestingly, the code comments hint at the name of the variable (e.g., \texttt{String name}), method (e.g., \texttt{getName()}) and return value of the class. Such comments helped us define the class, execute the code, and reproduce the issue. Our investigation identified 13 (7\%) Java questions where code comments helped us reproduce the issues. However, unnecessary comments (e.g., declaration of an identifier) can make the code segment noisy and thus hurt the readability of the code. Thus, the question submitter should comment only on the part of code that is not self-explanatory or is difficult to understand.
    
    \item \textbf{Details of Environment Setup Helps.} Programming issues are often related to the environment setup. Therefore, missing environment setup details might hinder reproducibility or demand more human effort. Such environment setup includes Operating Systems (e.g., Windows), software versions (e.g., Python 2), IDEs (e.g., Eclipse) and application software (e.g., Mozilla Firefox). Sometimes, the question submitter does not specify the context precisely in which environment setup the alleged programming issue is encountered.  In that cases, users face difficulties reproducing the reported issues and answering questions. During our qualitative analysis, we found that users asked for which environment the question submitters had encountered the issues. Consider the example where one user commented, \textit{``It would be better to know your target device and API version, environment setup etc. before deep diving.''} The question received answers when the question submitter edited it and added the environment details (e.g., Python 3.6 64 bit).
    However, we could not reproduce about 8\% (6 out of 74) of Python issues due to the lack of environment setup specifications. Even some programming issues are addressed and resolved in advanced versions. Thus, it might help analyze the code segments and reproduce their issues if the environment setup or system dependency is specified in the description.
    
    \item \textbf{Renaming Library Confuses.} Users often rename Python libraries when importing them (e.g., import matplotlib.pyplot as \emph{plt}). Then they use the short name in their code to call the functions from the libraries. There are some familiar short names for some libraries (e.g., import numpy as \emph{np}) that are  commonly used and recognized by other developers. However, developers sometimes use short names for external or their own libraries in code segments without including the import statements. Such libraries are not easy to find using short names. Thus, the question submitter should include the import statements when they rename libraries.
    
    \item \textbf{Error Log or Stack Trace Helps.} Error logs help developers determine what is happening (i.e., the root cause of errors) in code segments and suggest solutions. Besides the error log, a stack trace supports debugging. It shows the call stack when an uncaught exception is thrown. Thus, submission of the error log or stack trace with the problem description (when necessary) might help users to understand the problem and answer the question correctly.
    
    \item \textbf{It Is Acceptable If Code Requires Minor Modifications to Reproduce Issues.} Reproducibility of issues encourages more accepted answers for Java and Python. However, our further analysis based on effort level shows that the difference in the accepted answers for the categories reproducible without modification and minor modifications is not high (e.g., Table \ref{table:relation-effort-level-accanswer}). For example, the difference between that two categories in receiving accepted answers is only about 5\% for Java and 9\% for Python. Such a finding suggests that users are able to answer questions appropriately even if the code segments require minor modifications to reproduce the discussed issues.

\end{inparaenum}

\subsection{Impact of Environment Setup on Reproducibility}

We used the latest release of the Java Development Kit (\texttt{JDK}) to execute the Java code segments and reproduce the programming-related issues reported in questions. In particular, we used \texttt{Eclipse Oxygen.3a} and \texttt{NetBeans 8.2} with \texttt{JDK}-1.8. However, Java is backward compatible. We could not reproduce only one issue due to an incompatible class/API with \texttt{JDK} 1.8. We also could not find any equivalent class/API to replace the incompatible ones. However, we used two major releases of Python interpreters (e.g., \texttt{Python 2.7.17} and \texttt{3.6.9}) to address the version incompatibility problem while executing Python code segments. Primarily, the \texttt{Python 2.7.17} interpreter was used to execute the code segments and reproduce the issues. Upon failure, we used the \texttt{Python 3.6.9} interpreter. Unfortunately, we could not reproduce five issues due to version incompatibility. However, we did not find any issues that were reproducible using the first version of the code segments and irreproducible in the latest version while analyzing the evolution aspects of questions. Thus, according to our analysis, the environmental setup might impact the reproducibility of a negligible number of issues (e.g., one Java + five Python) and thus does not affect our major findings.
    
\section{Threats to Validity} \label{threat}

Threats to \emph{external validity} relate to the generalizability of a technique. Our results may not generalize to all the questions of Stack Overflow. We analyze statically significant samples from two popular programming languages -- Java and Python to mitigate this threat. Java is statically typed, whereas Python is a dynamically typed programming language. We see that the results from both languages are consistent. Thus, we believe that our insights can generalize to other programming languages. Moreover, we investigate a wide variety of questions of different types of issues in order to combat potential bias in our results. But we caution readers not to over-generalize our results.

Threats to \emph{internal validity} relate to experimental errors
and biases \citep{tian2014automated}. The reproducibility status (\eg\ reproducible, irreproducible) of the reported issue is threatened by the subjectivity of our classification approach. Thus, we cross-validate our results when an issue cannot be reproduced. We finalize the reproducibility status as irreproducible only if both examiners fail to reproduce the reported issue.

Threats to \emph{construct validity} relate to suitability of evaluation metrics.
We use the \emph{Mann-Whitney-Wilcoxon} test, which is a widely used non-parametric test for evaluating the difference between two sample sets. However, the significance level might suffer due to the limited size of the samples. We thus consider the effect size along with the \emph{p-value}. To see the correlation between two categorical variables we use the \emph{Chi-square} test. This statistical test of independence works well when there is a  small number of categories ($\le 20$) \citep{chiSquareTest}.

\section{Related Work} \label{relatedwork}

There have been various studies on the reproducibility of programming issues \citep{querytousablecode, gistable, mu2018understanding, rahman2020why, Mondal-SOIssueReproducability-MSR2019, crashdroid,crashscope,guided-ga-crash,yakusu}, computational reproducibility \citep{marwick2017computational, freire2012computational, goecks2010galaxy, dalle2012reproducibility, liu2019successes}, experimental reproducibility \citep{dit2015supporting, teran2014toward}, issues in reproducing research results \citep{boettiger2015introduction, cito2016using, poldrack2015publication, jimenez2017popper, playford2016administrative, scheitle2017towards, walters2013modeling, de2015initiative, gruning2018practical}, and reproducibility issues in software engineering \citep{rodriguez2018reproducibility, crick2014can}. However, to the best of our knowledge, we first investigate the reproducibility of issues reported in Stack Overflow questions in two popular programming languages -- Java and Python. This section overviews the reproducibility challenges of various research areas.

\subsection{Reproducibility of Programming Issues}

Several studies investigate usability (e.g., parsability, compilability) and executability challenges of code segments posted at crowd-sourced developer forums (e.g., GitHub, SO) \citep{querytousablecode, gistable}.
Yang et al. \citep{querytousablecode} analyze the \emph{usability} of 914,974 Java code snippets on Stack Overflow and report that only 3.89\% are parsable and 1.00\% are compilable. They analyze the code segments found in only the accepted answers of Stack Overflow and employ automated tools such as Eclipse JDT and ASTParser for the parsing and compiling the code. Then they report the errors that prevent the code segments from parsing and compiling without human involvement. Similarly, we analyze the Java and Python code snippets from the questions of Stack Overflow. However, unlike their approach, which is completely automated, our approach is a combination of automatic and manual analysis. Not only we make the code parsable, compilable, and runnable using appropriate modifications to the code, we also overcome the challenges to make them reproduce the reported issues on Stack Overflow. \citeauthor{gistable} investigate the executability of Python code found on the GitHub Gist system. Their primary focus was the execution of the Python snippets. However, we go beyond code execution and manually investigate the reproducibility of issues using Java and Python code snippets submitted with Stack Overflow questions. They also report the types of execution failures encountered while running Python gists. Similarly, we categorize the reproducibility status and identify the reasons why the issues could not be reproduced. Interestingly some reasons are common between ours and their study, such as import error, syntax error. However, the executability of a code does not always guarantee the reproducibility of an issue reported in the Stack Overflow question. Reproducibility requires testing and debugging, which warrant manual analysis, and this was not done by any of the earlier works. Besides, our research context differs from theirs since they examine gists shared on GitHub, whereas we deal with code snippets found in Stack Overflow questions.

Several researchers investigate the reproducibility challenges of software bugs and security vulnerabilities \citep{erfani2014works, rahman2020why, mu2018understanding}. Joorabchi et al. \citep{erfani2014works} analyze 1,643 irreproducible bug reports and investigate the causes of their irreproducibility. They reveal six root causes, such as environmental differences, insufficient information that prevent bug-reports' reproducibility. Rahman et al.~\citep{rahman2020why} conduct a study to understand the irreproducibility of software bugs. They investigate 576 irreproducible bug reports from two popular software systems (e.g., Firefox, Eclipse) and identify 11 challenges (e.g., bug duplication, missing information, ambiguous specifications) that might prevent bug reproducibility. The authors then survey 13 developers to investigate how the developers cope with irreproducible bugs. According to the study findings, developers either close these bugs or solicit further information. Mu et al.~\citep{mu2018understanding} analyze 368 security vulnerabilities to quantify their reproducibility. Their study suggests that individual vulnerability reports are insufficient to reproduce the reported vulnerabilities due to missing information. Besides, many vulnerability reports do not include details of software installation options and configurations, or the affected operating system (OS) that could hinder reproducibility. Then they survey hackers, researchers, and engineers who have domain expertise in software security. Survey findings suggest that apart from Internet-scale crowd-sourcing and some interesting heuristics, manual efforts (e.g., debugging) based on experience are the sole way to retrieve missing information from reports. Our study relates to the above studies in terms of research methodologies and problem aspects. However, our research context differs from theirs since we attempt to reproduce the issues reported in Stack Overflow questions using the code segments included with them.

Tahaei et al.~\citet{tahaei2020understanding} analyze 1,733 privacy-related questions of SO to understand the challenges and confusion that developers face while dealing with privacy-related topics. Ebert et al.~\citep{ebert2019confusion} investigate the reasons (e.g., missing rationale) and impacts (e.g., merge decision is delayed) of confusion in code reviews. Their study analyzes how developers cope with confusion during code reviews. For example, developers attempt to deal with confusion by requesting information, improving the familiarity with existing code, and discussing off-line the code review tool. However, our research goals are different. They attempt to find reasons and impacts of confusion in code reviews. We find the challenges that prevent the reproducibility of issues using the code segments with the questions. We also find the impacts of reproducibility of issues on receiving acceptable answers to questions, the time delay of receiving acceptable answers, and the number of answers. Finally, we provide evidence-based guidelines to write effective code examples for Stack Overflow questions.

Ford et al.~\citep{ford2018we} deploy a month-long, just-in-time mentorship program to SO to guide the novices with formative feedback of their questions. Such mentorship reduces the negative experience caused by delays in getting answers or adverse feedback. However, human mentorship is costly. Thus, it is hard to sustain such mentorship. Horton and Parnin~\citep{horton2019dockerizeme} present DockerizeMe, a system for inferring the dependencies required to execute a Python code snippet without import errors. However, our research findings might inspire automated tool support to assist the reproducibility of issues. Terragni et al.~\citep{terragni2016csnippex} propose a technique CSnippEx to automatically convert Java code snippets into compilable Java source code files. We plan to analyze SO code segments included with questions and then suggest users improve the code segments to support reproducibility.

Due to the growing popularity and importance of the Stack Overflow Q\&A site, several studies focus on question/answer quality analysis. Duijn  et  al. \citep{qclassification} collect the Java code segments found in Stack Overflow questions and suggest that several code-level constructs (\eg\  code length, keywords) are correlated to the quality of a code fragment. A number of studies investigate the quality of a code segment by measuring its readability \citep{5332232, daka2015modeling, posnett2011simpler, scalabrino2016improving,buse2008metric, understanding} and understandability \citep{trockman2018automatically,lin2008evaluation, scalabrino2017automatically}. Unfortunately, their capability of reproducing the issues reported in Stack Overflow questions was not investigated by any of the early studies. This makes our work novel. As our empirical findings suggest, like readability and understandability, the reproducibility of issues could be considered as one of the major quality aspects of Stack Overflow questions.

Several other studies \citep{wang2018understanding, squire2014bit, calefato2015mining, calefato2018ask} investigate how to get a fast answer, create a high-quality post, or mine a successful answer. They suggest that information presentation, code-text ratio, and question posting time are the key factors behind getting high-quality answers. Similarly, our findings show that the reproducibility of an issue discussed in the question is likely to encourage high-quality responses, including the acceptable answers. \citet{rahman2015insight} investigate why questions on Stack Overflow remain unresolved. However, they also do not consider the issue of reproducibility in their study.

\subsection{Computational Reproducibility}

Several studies investigate the challenges of computational reproducibility \citep{marwick2017computational, freire2012computational, goecks2010galaxy, dalle2012reproducibility, liu2019successes}. Marwick~\citep{marwick2017computational} investigates the computational reproducibility in archaeological researches. According to his investigation, such studies often miss important information that prevents reproducibility of statistical results. Furthermore, the inaccessibility and unavailability of raw data prevent reproducing the computational results. Liu and Salganik~\citep{liu2019successes} also claim that the unavailability of data and source code makes computational reproducibility challenging to achieve in practice. 
Finally, Marwick offers a compendium that incorporates a collection of files (e.g., raw data, R scripts) and a Docker file that recreates the computational environment to enhance computational reproducibility. 
We analyze the reproducibility of programming-related issues discussed in Stack Overflow. However, we also encountered similar challenges, such as missing important code statements and sample datasets.
Freire et al.~\citep{freire2012computational} argue that encapsulating all the components (e.g., raw data, source code) is also challenging. Users of Stack Overflow also face challenges uploading the supporting data and figures required to reproduce the issues reported in questions. However, the study by Freire et al.~\citep{freire2012computational} demonstrates several tools (e.g., Madagascar\footnote{http://www.reproducibility.org/wiki/Tutorial}) that could support computational reproducibility.

Goecks et al.~\citep{goecks2010galaxy} investigate computational reproducibility challenges in life science researches (e.g., genomic research). Their study finds that managing large datasets or multiple data sources and complex computational tools are the major challenges. To mitigate such challenges, they introduce an open web-based platform named Galaxy\footnote{http://usegalaxy.org}. It automatically tracks and manages data provenance, captures descriptive information about datasets, tools, and their invocations, and enables users to apply tools to datasets to support reproducible computational research.

\subsection{Experimental Reproducibility and Simulation-based Experiments}

Dit et al.~\citep{dit2015supporting} investigate experimental reproducibility in software maintenance. They find several challenges that make the reproducibility of such studies difficult. According to them, the lack of datasets, appropriate tools, and implementation details (e.g., parameter values) prevent others from reproducing such studies. To enhance reproducibility, they introduce several maintenance techniques and approaches as a set of experiments. They also develop a library of components (a.k.a., component library) to create a body of actionable knowledge to facilitate reproducibility in future research and allow the research community to contribute to it as well.

Dalle~\citep{dalle2012reproducibility} investigates the issues that prevent reproducibility in simulation-based experiments. They classify the issues that could hinder reproducibility into two main categories: human factors and technical issues. Human factors include hidden parameters, insufficient detail of publications, business-related limitations, and manipulation errors. On the other hand, technical issues include software bugs, software unavailability, floating points errors, operating system dependency. Teran-Somohano et al.~\citep{teran2014toward} investigate the reproducibility of experiments with simulation models and exposes several challenges. In particular, they report three challenges that might hamper reproducibility. They are (i) underutilization of the experimental design, (ii) limited transparency in the result analysis, and (ii) ad-hoc adaptation of experiments. They then introduce a conceptual framework to manage simulated experiments to improve reproducibility and replicability. Our study relates to the above studies in terms of research methodologies. We identify the challenges that prevent reproducibility of issues by manually analyzing 800 questions of Stack Overflow. We also provide evidence-based guidelines to improve the code segments include with questions to promote reproducibility.

\subsection{Issues in Reproducing Research Results}

Boettiger~\citep{boettiger2015introduction} investigates why one research project's code cannot be executed successfully by subsequent researchers and fails to replicate the original results. The author exposes four technical challenges that pose substantial barriers to reproduce the original results. They are (i) software dependency, (ii) imprecise documentation, (iii) code rot, and (iv) barriers to adoption and reuse in existing solutions (e.g., workflow software, virtual machines). 
We compile/execute the code segments included with 800 questions related to Java and Python to reproduce issues discussed in the questions. However, we expose several challenges such as database dependency and outdated code that prevent reproducibility of issues.
Boettiger suggests Docker, which can combine multiple areas (e.g., operating system virtualization, modular re-usable elements, versioning) and a `DevOps' philosophy to address those challenges. 

Playford et al.~\citep{playford2016administrative} study the challenges of reproducibility with micro-level administrative, social science data. Such challenges include data access, data retention, working with dynamic data, and difficulties in undertaking exploratory data analysis. Their investigation suggests that sharing both data and code is vital for reproducible research. Poldrack and Poline~\citep{poldrack2015publication} also argue that only sharing data is not sufficient to reproduce the research result. Reproducible research demands sharing analysis methods with detailed descriptions, essential resources (e.g., software, computing resources) to the reviewers, and data analysis code. Unfortunately, the analysis code is still rarely available and easily readable (due to insufficient comments), often scattered across many different files in different languages that prevent research results' reproducibility.

Scheitle et al.~\citep{scheitle2017towards} investigate the reproducibility of research results in computer networking. They argue that most of the results are not reproducible due to four main challenges. They are (i) author unavailability, (ii) artifact unavailability, (iii) lack of detail, and (iv) unclear terminology and expectations. They introduce an ecosystem that incentivizes authors and reproducers to contribute to reproducible research. In particular, they create a reproduction bundle that consists of code, data, and other relevant artifacts, along with the initial submission to a venue.

Walters~\citep{walters2013modeling} attempts to discover the difficulties in reproducing the study results in molecular modeling and cheminformatics papers. According to the study, software unavailability and inaccessibility of supporting information hinder the reproducibility of study results. Walters suggests that clear and easy descriptions of any new methods and their source code could support reproducibility.

Grüning et al.~\citep{gruning2018practical} investigate why many research areas suffer from poor reproducibility. They find that traditional publication approaches do not well capture computationally intensive domains where results rely on a series of complex methodological decisions. Several guidelines have emerged for achieving reproducibility. However, implementing them is difficult due to assembling software tools and associated libraries, connecting tools into pipelines, and specifying parameters. Finally, the authors attempt to make computational reproducibility practical by implementing a Python library called galaxy-lib. This library provides utilities to create a container with all dependencies for a given analysis.


\subsection{Reproducibility Challenges in Software Engineering}

Rodríguez-Pérez et al.~\citep{rodriguez2018reproducibility} investigate how reproducibility and credibility in Empirical Software Engineering (ESE) have been addressed. In particular, they target studies where the SZZ algorithm \citep{sliwerski2005changes} has been applied to detect software bugs. They analyze 187 papers. According to their analysis, 39\% of the papers do not provide a replication package or detailed description of the methodology and data that prevent reproducibility. They warn researchers about the risk of every assumption of such an algorithm. They ask to provide a manual analysis of the results of every assumption. They also recommend researchers who modify the SZZ algorithm to publish the software implementation.

Crick et al.~\citep{crick2014can} investigate reproducibility and replicability challenges of scientific software engineering results. The two major issues that hinder reproducibility are (i) algorithm descriptions are too high-level and (ii) too obscure. Thus, they recommend explaining the algorithm in such a way that any reader could implement it. They also introduce an open platform for scientific software development to share scientific software engineering results.

Cito and Gall~\citep{cito2016using} study the reproducibility issues in software engineering research. They find that published code and data come with many undocumented assumptions, dependencies, and configurations that make reproducibility hard to achieve. According to them, Docker containers can overcome these issues, aid the reproducibility of research artifacts in software engineering.

Neto et al.~\citep{de2015initiative} investigate reproducible research issues with software testing techniques (STT). They argue that reproducible research requires executable artifacts. Such artifacts include source code or test cases, description of studies, and methods. They also initiate a collaborative effort to define, execute, and deploy experiments with improved description, accessibility, and availability of required artifacts to reproduce the experimental findings.

Liem and Panichella~\citep{liem2020run} investigate the reproducibility of outcomes of Machine Learning (ML) models applied in software engineering tasks. They expose three potential threats that might hinder the reproducibility. They are (i) the sensitivity of ML methods to randomized resamplings from datasets, (ii) the randomized nature of ML learning procedures, and (iii) inconsistency between different implementations of the same ML method in different libraries. They survey 45 recent papers related to predictive tasks (e.g., defect prediction, code smell detection). About 50\% of the papers address the threats related to randomized data sampling through multiple repetitions. Only 8\% of the papers address the random nature of ML, and only 18\% of the papers report parameter values. Finally, they provide several guidelines to cope with the threats. For example, they suggest reporting parameter values, library/function names, and library versions.

We investigate the reproducibility challenges of programming-related issues discussed in Stack Overflow questions. However, the above studies are related to our study in terms of methodology and research goal. Our investigation -- (1) reports the reproducibility status of programming-related issues discussed in 800 questions of Stack Overflow, (2) addresses the challenges that hinder reproducibility and (3) guides users to improve the example code segments included with questions to support reproducibility. 

\subsection{Miscellaneous}

Ferro \citep{ferro2017reproducibility} studies the reproducibility challenges in Information Retrieval (IR) evaluation. He discovers three main concerns for reproducibility. They are system runs (i.e., the outputs of the execution), experimental collections (e.g., documents, topics), and meta-evaluation studies (e.g., the robustness of the experimental collections, reliability of the adopted evaluation measures). This study introduces a model of the entities involved in IR evaluation \citep{silvello2016data}, based on semantic Web and Linked Open Data (LOD) technologies, and by making (a subset of) the Conference and Labs of the Evaluation Forum (CLEF) experimental data accessible through a running prototype.

Cohen-Boulakia et al.~\citep{cohen2017scientific} investigate the barriers of reproducibility-friendly scientific workflow systems. They find that (i) a short description of the methodology, (ii) insufficient data, (iii) incompatibility of tools, (iv) tracing difficulty of the workflow history and reusing parts of executed workflows, and (v) the maintenance of workflow runtime environment are the major reproducibility issues.

\section{Conclusion and Future Work}\label{conclusion}

We manually analyzed 800 randomly selected questions from Stack Overflow related to Java and Python and investigated the reproducibility of their reported issues using the code segments provided. We answered four research questions in our work.  We classified the status of reproducibility into two major categories - reproducible and irreproducible. We find that 68\% of issues can be reproduced using the code submitted with Java questions, and 
71\% 
of issues can be reproduced using the code submitted with Python questions after performing minor or major modifications as necessary.  We also investigated and reported on why several issues could not be reproduced using the provided code segments. We then examined the correlation between the reproducibility of programming issues in questions and answer meta-data. Our findings suggest that the reproducibility 
is likely to encourage more high-quality responses, including acceptable answers.
We also investigate several confounding factors (e.g., user reputation) that affect questions receiving answers. However, such factors do not hurt the correlation between reproducibility status and answer meta-data.
Thus, programming issue reproducibility
can be treated as a novel metric of question quality for Stack Overflow, which was previously unrecognized in earlier studies.
In the future, we plan to develop automated tool support for predicting the reproducibility of issues discussed in questions and turning irreproducible issues into reproducible issues.

\begin{acknowledgements}
This research is supported by the Natural Sciences and Engineering Research Council of Canada (NSERC), Canada First Research Excellence Fund (CFREF) grant coordinated by the Global Institute for Food Security (GIFS), and Tenure-track Startup Fund, Dalhousie University.
\end{acknowledgements}

\bibliographystyle{spbasic}      

\bibliography{reference}

%
%

\end{document}